\documentclass[%
preprint,
nofootinbib,
 amsmath,amssymb,
 aps,
]{revtex4-1}

\usepackage{graphicx}
\usepackage{dcolumn}
\usepackage{bm}
\usepackage{hyperref}

\usepackage{comment}
\allowdisplaybreaks

\newcommand\Ref[1] {Ref.\,\cite{#1}}
\newcommand\Refs[1] {Refs.\,\cite{#1}}
\newcommand\eqn[1] {Eq.\,(\ref{#1})}
\newcommand\eqns[2] {Eqs.\,(\ref{#1}) and~(\ref{#2})}

\newcommand\Tab[1] {Tab.\,{\ref{#1}}}

\def\bsp#1\esp{\begin{split}#1\end{split}}

\newcommand{\order}[1]{{\cal O}\left(#1\right)}

\newcommand{\ymin}{\ensuremath{y_{\min}}}
\newcommand{\as}{\ensuremath{\alpha_\textrm{s}}}
\newcommand{\cusp}{\text{cusp}}
\newcommand{\mMDT}{\text{mMDT}}
\newcommand{\sing}{\text{sing}}
\newcommand{\rL}{\text{L}}
\newcommand{\LP}{\text{LP}}
\newcommand{\rR}{\text{R}}
\newcommand{\rd}{\text{d}}

\newcommand{\rg}{\text{g}}
\newcommand{\zcut}{\ensuremath{z_\text{cut}}}

\DeclareRobustCommand{\Sec}[1]{Sec.~\ref{#1}}

\DeclareRobustCommand{\App}[1]{App.~\ref{#1}}
\DeclareRobustCommand{\Tab}[1]{Table~\ref{#1}}

\DeclareRobustCommand{\Fig}[1]{Fig.~\ref{#1}}

\DeclareRobustCommand{\Eq}[1]{Eq.~(\ref{#1})}

\DeclareRobustCommand{\Ref}[1]{Ref.~\cite{#1}}
\DeclareRobustCommand{\Refs}[1]{Refs.~\cite{#1}}

\usepackage{color}
\definecolor{darkblue}{rgb}{0,0,0.5}
\definecolor{darkred}{rgb}{0.5,0,0}
\definecolor{darkgreen}{rgb}{0,0.5,0}

\begin{document}


\title{Two- and Three-Loop Data for Groomed Jet Mass}

\author{Adam Kardos}
 \email{kardos.adam@science.unideb.hu}
 \affiliation{University of Debrecen, 4010 Debrecen, PO Box 105, Hungary}
\author{Andrew J.~Larkoski}%
 \email{larkoski@reed.edu}
\affiliation{Physics Department, Reed College, Portland, OR 97202, USA}%

\author{Zolt\'an Tr\'ocs\'anyi}
 \email{Zoltan.Trocsanyi@cern.ch}
 \homepage{http://pppheno.elte.hu}
\affiliation{
 Institute for Theoretical Physics, ELTE E\"otv\"os Lor\'and University, 
P\'azm\'any P\'eter 1/A, H-1117 Budapest, Hungary
\\ and
MTA-DE Particle Physics Research Group, University of Debrecen, 
4010 Debrecen, PO Box 105, Hungary
}%

\date{\today}

\begin{abstract}
We discuss the status of resummation of large logarithmic contributions to groomed event shapes of hadronic final states in electron-positron annihilation. We identify the missing ingredients needed for
next-to-next-to-next-to-leading logarithmic (NNNLL) resummation of the mMDT groomed jet mass in $e^+e^-$ collisions: the low-scale collinear-soft constants at two-loop accuracy, $c_{S_c}^{(2)}$, and the three-loop non-cusp anomalous dimension of the global soft function, $\gamma_S^{(2)}$. We present a method for extracting those constants  using fixed-order codes: the {\tt EVENT2} program to obtain the color coefficients of $c_{S_c}^{(2)}$, and {\tt MCCSM} for extracting $\gamma_S^{(2)}$. We present all necessary formulae for  resummation of the mMDT groomed heavy jet mass distribution at NNNLL accuracy.
\end{abstract}

\maketitle

\tableofcontents

\section{\label{sec:introduction}Introduction}

At the Large Hadron Collider (LHC), quantum chromodynamics (QCD) has been studied at an unprecedented precision in both experiment and theory.  Major advances have especially recently been made in the field of jet substructure in which the internal dynamics of jets are exploited to learn information about the short-distance physics; see \Refs{Larkoski:2017jix,Asquith:2018igt} for recent reviews.  Because jets are identified as restricted regions of phase space, particles that land in a jet can come from numerous sources from outside of the jet that contaminate a simple theoretical picture of the jet's dynamics.  To mitigate this contamination radiation, which may come from initial-state radiation, secondary parton scatterings, or secondary proton scatterings (pile-up), analysis techniques broadly referred to as jet grooming have been introduced.  The jet groomers systematically identify emissions in the jet that likely came from such contamination and remove them from consideration.  The best theoretically understood jet groomers are the modified mass mass drop tagger (mMDT) \cite{Dasgupta:2013ihk,Dasgupta:2013via} and soft drop \cite{Larkoski:2014wba} algorithms, due to their unique feature of elimination of non-global logarithms (NGLs) \cite{Dasgupta:2001sh}, leading correlations between in-jet and out-of-jet scales.

The elimination of NGLs of mMDT and soft drop groomers means that calculations of observables can be meaningfully extended beyond leading-logarithmic accuracy.  \Refs{Frye:2016okc,Frye:2016aiz} derived an all-orders factorization theorem for observables like the jet mass measured on mMDT/soft drop groomed jets, in the limit that the jet mass is small with respect to the jet energy.  This factorization theorem enabled calculation to next-to-next-to-leading logarithmic (NNLL) accuracy for jet observables, resulting in predictions with controlled theoretical uncertainties over a wide, and experimentally-relevant, region of phase space.  Many other calculations for mMDT/soft drop groomed jet mass (and related observables) have followed \cite{Marzani:2017mva,Marzani:2017kqd,Makris:2017arq,Kang:2018jwa,Chay:2018pvp,Baron:2018nfz,Makris:2018npl,Kang:2018vgn,Lee:2019lge,Marzani:2019evv,Gutierrez-Reyes:2019msa,Cal:2019gxa,Kang:2019prh}, and both ATLAS and CMS have measured these groomed jet masses in collisions \cite{Aaboud:2017qwh,Sirunyan:2018xdh,Aad:2019vyi} and compared directly to theoretical predictions.  Further calculations have been performed of groomed observables that are sensitive to additional structure in jets \cite{Salam:2016yht,Larkoski:2017iuy,Larkoski:2017cqq,Napoletano:2018ohv}, and related grooming procedures have been introduced \cite{Chien:2019osu} that exploit the nice theoretical properties of these groomers.

With the high theoretical precision achievable with mMDT/soft drop and the continual experimental measurements, a realistic goal of this program is a first extraction of underlying physical quantities from jet substructure.  First studies have demonstrated that distributions of observables on these groomed jets are very sensitive to the value of $\as$ \cite{Bendavid:2018nar}.  
Further, grooming promises to provide an unambiguous theoretical definition of the top quark mass, and a universal robustness to its value \cite{Hoang:2017kmk}.  To reach the necessary theoretical precision for these measurements requires not only high logarithmic accuracy, but also high fixed-order accuracy and well-understood non-perturbative corrections.  Next-to-next-to-leading fixed order (NNLO) predictions for mMDT and soft drop groomed events have been produced using the {\tt MCCSM} (Monte Carlo for the CoLoRFulNNLO Subtraction Method) code \cite{Kardos:2016pic,Kardos:2018kth} for jet production in $e^+e^-$ collisions, and a detailed analysis of non-perturbative corrections within the context of the factorization theorem for groomed jets has been presented \cite{Hoang:2019ceu}.  Further, advances in NNLO multi-jet production in hadron collisions like at the LHC have been steadily made over the past few years \cite{Badger:2017jhb,Abreu:2017hqn,Chawdhry:2018awn,Abreu:2018jgq,Badger:2018enw,Abreu:2018zmy,Chicherin:2018old,Abreu:2019odu,Badger:2019djh,Hartanto:2019uvl}.

For precision extraction of the strong coupling, for example, logarithmic accuracy should extend to next-to-next-to-next-to-leading logarithmic order (NNNLL), the first order at which simple additive matching to NNLO fixed-order is possible.  NNNLL is the order to which thrust \cite{Becher:2008cf,Abbate:2010xh} and $C$-parameter \cite{Hoang:2014wka,Hoang:2015hka} are resummed for precision extraction of $\as$ from LEP data.  Such high precision is possible because improving logarithmic accuracy requires calculation of universal anomalous dimensions and constants for the functions that appear in the factorization theorem for these observables.  In the case of mMDT/soft drop groomed jet mass, the factorization theorem for the process $e^+e^-\to $ hadrons reads
\begin{equation}
\frac{1}{\sigma_0}\frac{\rd^2\sigma}{\rd\tau_\rL\, \rd\tau_\rR} = H(Q^2)S(\zcut) \left[
J(\tau_\rL)\otimes S_c(\tau_\rL,\zcut)
\right]  \left[
J(\tau_\rR)\otimes S_c(\tau_\rR,\zcut)
\right]\,,
\end{equation}
where $\sigma_0$ is the leading-order $e^+e^-\to q\bar q$ cross section, $\zcut$ is a parameter of the mMDT/soft drop groomer, and $\tau_\rL$, $\tau_\rR$ are the left- and right-hemisphere masses, respectively.  Extending this factorization theorem beyond NNLL requires the two-loop constants for each of the functions on the right, as well as their three-loop anomalous dimensions.  The hard $H(Q^2)$ and jet functions $J(\tau)$ are known completely through three loops, and the global soft function $S(\zcut)$ is known completely through two loops \cite{Bell:2018oqa,Bell:2018vaa,Bell:2020yzz}.  Consistency of the factorization theorem requires the sum of anomalous dimensions of the functions to vanish so that the cross section is renormalization-group invariant.  Therefore, for NNNLL resummation, we only need to calculate the two-loop constants of the collinear-soft function $S_c(\tau,\zcut)$ and the three-loop global soft function anomalous dimension.

In this paper, we use numerical fixed-order event generators to extract these necessary ingredients.  We will restrict our analysis to jet masses groomed with mMDT for simplicity and compactness.  In this case, we use {\tt {\tt EVENT2}} \cite{Catani:1996vz} to find the two-loop constants $c_{S_c}^\text{mMDT}$ of the collinear-soft function (in Laplace conjugate space) to be
\begin{align}
c_{S_c}^{\mMDT} &=\left(\frac{\as}{4\pi}\right)^2\left[ C_F^2 \left(22\pm 4\right)+C_F C_A\left(41\pm 1\right)+C_F T_R n_f \left(14.4\pm 0.1\right)\right]\,,    
\end{align}
separated into distinct color channels where $C_F = 4/3$, $C_A = 3$, and $T_R=1/2$ in QCD, and $n_f$ is the number of active quark flavors.  Using {\tt MCCSM}, we find the three-loop anomalous dimension of the global soft function $\gamma_S^{\mMDT}$ to be
\begin{equation}
\gamma_S^{\mMDT} = \left(\frac{\as}{4\pi}\right)^3\left[-11600 \pm 2000\right] \qquad (n_f = 5)\,,
\end{equation}
where we fix the number of active quarks to $n_f = 5$.  These results enable resummation to NNNLL accuracy for jet substructure observables for the first time, which we present in a companion paper \cite{Kardos:2020gty}.

The outline of this paper is as follows.  In \Sec{sec:grooming}, we review the mMDT grooming algorithm and its all-orders factorization theorem.  In \Sec{sec:twoloop}, we discuss our procedure for extracting the two-loop constants of the collinear-soft function.  In \Sec{sec:threeloop}, we extract the three-loop anomalous dimension of the global soft function using numerical results from {\tt MCCSM}.  We conclude in \Sec{sec:conclusions}.  Several appendices provide the analytical and numerical details of the results provided in the main text.

\section{\label{sec:grooming}mMDT Grooming and Factorization Theorem}

In this paper, we restrict our analysis to hadronic final states in electron-positron annihilation where high fixed-order results are available and can be used to extract currently-unknown data for resummation.  As such, we will only discuss the grooming algorithm and observables for $e^+e^-$ collisions, and we point readers interested in other applications to the cited literature.  In this section, we will review the definition of the grooming algorithm, as well as the all-orders factorization theorem that we exploit.

The mMDT groomer \cite{Dasgupta:2013ihk}, or soft drop with angular exponent $\beta = 0$ \cite{Larkoski:2014wba}, proceeds as follows:
\begin{enumerate}

\item Divide the final state of an $e^+e^-\to$ hadrons event into two hemispheres in any infrared and collinear safe way.  This could include separation with thrust \cite{Brandt:1964sa,Farhi:1977sg}, or in a recoil-free manner using broadening axes \cite{Georgi:1977sf,Larkoski:2014uqa}.

\item In each hemisphere, run the Cambridge/Aachen jet algorithm \cite{Dokshitzer:1997in,Wobisch:1998wt} to produce an angular-ordered pairwise clustering history of particles.  The clustering metric appropriate for $e^+e^-$ collisions is
\begin{equation}
d_{ij} = 1-\cos\theta_{ij}\,,
\end{equation}
where $\theta_{ij}$ is the angle between particles $i$ and $j$ in the same hemisphere.

\item Starting with the left hemisphere and at widest angle, step through the Cambridge/Aachen particle branching tree.  At each branching in the tree, test if
\begin{equation}
\frac{\min[E_i,E_j]}{E_i+E_j} > \zcut\,,
\end{equation}
where $i$ and $j$ are the daughter particles at that branching and $\zcut$ is some fixed numerical value where $0\leq \zcut < 0.5$.  If this is true, stop and return all particles that remain in the left hemisphere.  If it is false, remove the lower energy branch, and continue to the next branching at smaller angle.  Repeat for the right hemisphere.

\item Once the groomer has terminated, any observable can be measured on the particles that remain in the two hemispheres.

\end{enumerate}
We will refer to this procedure as the mMDT groomer or simply mMDT throughout this paper.

From this procedure, \Ref{Frye:2016aiz} proved an all-orders factorization theorem for the cross section differential in the groomed hemisphere masses $\tau_\rL$ and $\tau_\rR$, where 
\begin{equation}
\tau_i = \frac{m_i^2}{E_i^2}\,,
\end{equation}
for mass $m_i$ and energy $E_i$ of hemisphere $i$.  For $\tau_\rL,\tau_\rR \ll\zcut \ll 1$, the cross section factorizes into:
\begin{equation}
\frac{1}{\sigma_0}\frac{\rd^2\sigma}{\rd\tau_\rL\, \rd\tau_\rR} = H(Q^2)S(\zcut) \left[
J(\tau_\rL)\otimes S_c(\tau_\rL,\zcut)
\right]  \left[
J(\tau_\rR)\otimes S_c(\tau_\rR,\zcut)
\right]\,,
\end{equation}
where $\sigma_0$ is the leading-order cross section for $e^+e^-\to q\bar q$, $H(Q^2)$ is the hard function for quark--anti-quark production in $e^+e^-$ collisions, $S(\zcut)$ is the global soft function for mMDT grooming, $J(\tau_i)$ is the quark jet function for hemisphere mass $\tau_i$, $S_c(\tau_i,\zcut)$ is the collinear-soft function for hemisphere mass $\tau_i$ with mMDT grooming, and we suppress the dependence on the renormalization scale $\mu$ in all functions.  The symbol $\otimes$ denotes convolution over the hemisphere mass $\tau_i$.   In most of the expressions that follow, for simplicity we will set $\sigma_0 = 1\,$GeV$^{-2}$.  While we will calculate data relevant for the functions in this factorization theorem in this paper, we will do so through the single-differential cross section of the groomed heavy hemisphere mass, defined as
\begin{equation}
\frac{1}{\sigma_0}\frac{\rd\sigma_\rg}{\rd\rho} \equiv \int \rd\tau_\rL \, \rd\tau_\rR\, \frac{1}{\sigma_0}\frac{\rd^2\sigma}{\rd\tau_\rL\, \rd\tau_\rR}\left[
\Theta(\tau_\rL - \tau_\rR)\,\delta(\rho - \tau_\rL)+
\Theta(\tau_\rR - \tau_\rL)\,\delta(\rho - \tau_\rR)
\right]\,.
\end{equation}
The subscript g on the cross section denotes that this is groomed.  Additionally, we note that this definition of the heavy hemisphere mass differs from the standard definition in the ungroomed case when the heavy hemisphere mass is defined as:
\begin{equation}
\rho = \frac{\max[m_\rL^2,m_\rR^2]}{Q^2}\,,
\end{equation}
with $Q$ being the center-of-mass energy.  When hemispheres are groomed, the grooming eliminates their dominant correlations, and so it is more natural to define the groomed mass with respect to the hemisphere energy, and not the center-of-mass energy.

The factorization theorem is simplest to analyze in Laplace space, where we Laplace transform in both $\tau_\rL$ and $\tau_\rR$ to eliminate the convolutions.  In Laplace space, the cross section becomes a simple product:
\begin{equation}\label{eq:laplacefact_text}
\frac{\sigma(\nu_\rL,\nu_\rR)}{\sigma_0} = H(Q^2)S(\zcut)\tilde J(\nu_\rL) \tilde S_c(\nu_\rL,\zcut)\tilde J(\nu_\rR) \tilde S_c(\nu_\rR,\zcut)\,,
\end{equation}
where $\nu_\rL$ ($\nu_\rR$) is the Laplace conjugate of $\tau_\rL$ ($\tau_\rR$).  In this product form, each function in the factorization theorem satisfies a simple renormalization group equation (RGE),
\begin{equation}
\mu\frac{\partial \tilde F}{\partial \mu} = \left(d_F \Gamma_{\cusp} \log\frac{\mu^2}{\mu_F^2}+\gamma_F\right) \tilde F\,,
\quad (\tilde F = H\,,
\:S\,,\:\tilde J\,,\:\tilde S_c)
\end{equation}
where $d_F$ is a constant, $\mu_F$ is the canonical scale, and $\gamma_F$ is the non-cusp anomalous dimension particular to $\tilde F$.  $\Gamma_{\cusp}$ is the cusp anomalous dimension for back-to-back light-like Wilson lines in the fundamental representation of color SU(3).  Large logarithms of hemisphere masses can be resummed to all orders in $\as$ through this renormalization group equation.  It can be solved exactly, and we present its explicit solution through $\as^3$ in \App{app:foexpansion}.

\begin{table}
\begin{center}
\begin{tabular}{c| c c c c c c}
 & \ \ $\Gamma_{\cusp}$\ \  & \ \ $\gamma_F$ \ \ &\ \  $\beta$ \ \ & \ \ $c_F$ \ \ &\ \  Matching\\
 \hline
 LL& $\as$ & -  & $\as$ & -  & - \\
 NLL & $\as^2$ & $\as$  & $\as^2$ &  - & $\as$ \\
 NNLL & $\as^3$ & $\as^2$ & $\as^3$ & $\as$ & $\as^2$\\
  NNNLL & $\as^4$ & $\as^3$ & $\as^4$ & $\as^2$ & $\as^3$\\
\end{tabular}
\end{center}
\caption{
$\as$-order of ingredients needed for resummation to the logarithmic accuracy given.  $\Gamma_{\cusp}$ is the cusp anomalous dimension, $\gamma_F$ is the non-cusp anomalous dimension for function $\tilde F$, $\beta$ is the QCD $\beta$-function, and $c_F$ are the low-scale constants for function $\tilde F$.  The final column shows the relative order to which the resummed cross section can be additively matched to fixed-order.
}\label{tab:logtab}
\end{table}

The order to which logarithms can be resummed through this renormalization group equation depends on the accuracy to which its components are calculated.  For the canonical definition of logarithmic accuracy \cite{Catani:1992ua}, \Tab{tab:logtab} shows the order in $\as$ to which the components of the renomalization group equation are needed.  In this table, ``LL'' denotes leading logarithmic accuracy, ``NLL'' is next-to-leading logarithmic accuracy, etc.  The factorization theorem successfully resums all logarithms of both $\rho$ and $\zcut$ simultaneously, to leading power in the limit where $\rho\ll\zcut\ll 1$.  Thus ``NLL'', for example, refers to the resummation of all terms up through those of the form $\alpha_s^n\log^n \rho$, $\alpha_s^n\log^n \zcut$, and $\alpha_s^n\log^n \frac{\rho}{\zcut}$ in the exponent of the cross section cumulative in the mass $\rho$. \Ref{Frye:2016aiz} resummed the mMDT groomed mass distribution to NNLL accuracy, requiring the new calculation of one-loop constants and two-loop non-cusp anomalous dimensions.  The goal of this paper is to extend the order of the known components to accomplish NNNLL resummation.


The cusp anomalous dimension is now known to four-loop order \cite{Gracey:1994nn,Beneke:1995pq,Davies:2016jie,Lee:2016ixa,Henn:2016men,Moch:2017uml,Grozin:2018vdn,Moch:2018wjh,Bruser:2019auj,Henn:2019rmi,vonManteuffel:2019wbj,Lee:2019zop,Henn:2019swt,vonManteuffel:2020vjv}, and the QCD $\beta$-function is as well \cite{vanRitbergen:1997va}.  The hard and jet functions are known through three-loop order \cite{Moch:2005id,Gehrmann:2005pd,Becher:2006mr,Baikov:2009bg,Lee:2010cga,Gehrmann:2010ue,vanRitbergen:1997va,Becher:2008cf}, as they are relevant for resummation of a broad class of observables.  The global mMDT soft function is now completely known to two-loops \cite{Frye:2016aiz,Bell:2018vaa,Bell:2018oqa,Bell:2020yzz}, while only the two-loop non-cusp anomalous dimension of the collinear-soft function is known.  Thus, to achieve NNNLL accurate resummation of mMDT groomed mass, we need the two-loop collinear-soft function constants, and the three-loop non-cusp anomalous dimensions of the soft and collinear-soft functions.  Actually, because the cross section is renormalization-group invariant, the sum of non-cusp anomalous dimensions of the functions in the factorization theorem must vanish:
\begin{equation}
0=\gamma_H+ \gamma_S+2\gamma_J+2\gamma_{S_c}\,.
\end{equation}
Hence, only the two-loop constants of the collinear-soft function and the three-loop non-cusp anomalous dimension of the global soft function are needed to accomplish resummation at NNNLL accuracy.  In the following sections, we numerically extract these values from fixed-order codes.  All results needed for solving the renomalization group equations to ${\cal O}(\as^3)$ are provided in the appendices.

Before moving to the numerical extraction, it is useful to comment on the region of phase space in which mMDT grooming acts non-trivially.  Assuming that the grooming parameter is small, $\zcut \ll 1$ formally, then only low-energy particles in each hemisphere are groomed away.  To leading power in a soft particle's energy $E_s$, the mMDT constraint is
\begin{equation}
E_s > E_H\zcut = \frac{Q}{2} \zcut\,,
\end{equation}
where $E_H = Q/2$ is the hemisphere energy, which is half of the center-of-mass collision energy in the soft limit.  For such a soft emission that just passes mMDT, its contribution to the hemisphere mass is
\begin{align}
\rho = \frac{m_H^2}{E_H^2} = \frac{2E_H E_s (1-\cos\theta_s)}{E_H^2} \leq 2\frac{E_s}{E_H}\,,
\end{align}
where the inequality corresponds to a soft emission right at the hemisphere boundary.  Taking the upper bound as the parametric scaling of the soft particle's energy, we have
\begin{equation}
E_s\sim \frac{\rho E_H}{2}\,.
\end{equation}
Then, using the mMDT constraint, the value of the hemisphere mass for such an emission is
\begin{equation}
\rho \gtrsim 2\zcut\,,
\end{equation}
and mMDT strongly affects the cross section for values of $\rho$ smaller than this.

\section{\label{sec:twoloop}Extraction of Two-Loop Constants}

In this section, we will present a method for numerical extraction of the two-loop constant terms of the collinear-soft function $c_{S_c}^{(2)}$ for the factorization theorem formulated in Laplace space, \Eq{eq:laplacefact_text}.  Our procedure for doing so will be to relate the cross section differential in the groomed hemisphere mass $\rho$ to the total cross section.  We can express the leading-power (LP) differential cross section for $\rho\to 0$ as a sum of terms with support exclusively at $\rho = 0$ and terms with support away from 0:
\begin{equation}
\frac{\rd\sigma_\text{g,\LP}}{\rd\rho} = D_{\delta,\rg}\,\delta(\rho) + \frac{\rd\sigma^{\sing}_\rg}{\rd\rho}\,.
\end{equation}
The terms that have support away from $\rho = 0$ are defined to integrate to 0 on $\rho \in[0,1]$.  The total cross section can then be written as:
\begin{equation}
\sigma_{\text{tot}} = D_{\delta,\rg}+\int_0^1\!\rd\rho\, \left(
\frac{\rd\sigma_\rg}{\rd\rho}-\frac{\rd\sigma^{\sing}_\rg}{\rd\rho}
\right)\,,
\label{eq:stot1}
\end{equation}
where $\rd\sigma_\rg/\rd\rho$ is the full differential cross section of the groomed hemisphere mass for all $\rho > 0$.

The desired unknown two-loop constants are contained in the $\delta$-function coefficient, $D_{\delta,\rg}$.  
One approach to determining them is to numerically evaluate the integral that remains using the known total cross section for $e^+e^-\to $ hadrons and a fixed-order code.  
Performing this particular numerical integral is somewhat complicated by the fact that grooming introduces multiple regions and so the integrand exhibits cusps and has support all the way out to the maximal value of $\rho$.
We will significantly rewrite this expression into a more directly useful form so that the numerical integrals we must perform are restricted to the region where grooming turns on, $\rho\sim \zcut$.  First, we separate the integral into two parts, 
\begin{align}
\int_0^1\!\rd\rho\, \left(
\frac{\rd\sigma_\rg}{\rd\rho}-\frac{\rd\sigma^{\sing}_\rg}{\rd\rho}
\right) &= \int_0^{2\zcut}\!\rd\rho\, \left(
\frac{\rd\sigma_\rg}{\rd\rho}-\frac{\rd\sigma^{\sing}_\rg}{\rd\rho}
\right) + \int_{2\zcut}^1\!\rd\rho\, 
\frac{\rd\sigma_\rg}{\rd\rho}- \int_{2\zcut}^1\!\rd\rho\, 
\frac{d\sigma^{\sing}_\rg}{\rd\rho}\,.
\label{eq:step1}
\end{align}
Below 2\zcut\ mMDT grooming does, while above 2\zcut\ it does not, dominantly enforce constraints on the radiation.
Then we add and subtract the singular differential cross section for ungroomed heavy hemisphere mass to the second integral on the right of \eqn{eq:step1}:
\begin{equation}
\int_{2\zcut}^1\!\rd\rho\, 
\frac{d\sigma_\rg}{\rd\rho}=\int_{2\zcut}^4\!\rd\rho\, \left(
\frac{\rd\sigma_\rg}{\rd\rho}\,\Theta(1-\rho)-\frac{\rd\sigma}{\rd\rho}
\right) +\int_{2\zcut}^4\!\rd\rho\, 
\frac{\rd\sigma}{\rd\rho}\,.
\label{eq:step2}
\end{equation}
For $\rho \sim 1 \gg \zcut$, mMDT grooming has no effect, so the first integral in this expression is dominated by the region near $\rho \simeq \zcut$, and we can drop the step function $\Theta$. Integrals on the right extend to 4, rather than 1 as simple consequence of the factor of 4 difference in definition of the groomed versus ungroomed heavy hemisphere mass, as discussed in \Sec{sec:grooming}.
For the integral exclusively of the ungroomed heavy hemisphere mass cross section, we can write it as
\begin{align}
\int_{2\zcut}^4\!\rd\rho\, 
\frac{\rd\sigma}{\rd\rho} = \int_0^4\!\rd\rho\, \left(
\frac{\rd\sigma}{\rd\rho}-\frac{\rd\sigma^{\sing}}{\rd\rho}\right)+\order{\zcut}+\int_{2\zcut}^4\rd\rho\, \frac{\rd\sigma^{\sing}}{\rd\rho}\,.
\end{align}
Assuming $\zcut\ll 1$, the lower bound of the first integral can be set to 0, neglecting power corrections that vanish as $\zcut \to 0$.

After implementing all these identities, the total cross section can be written as
\begin{equation}
\bsp
\sigma_\text{tot}&=D_{\delta,\rg}+\int_0^{2\zcut}\!\rd\rho\, \left(
\frac{\rd\sigma_\rg}{\rd\rho}-\frac{\rd\sigma^{\sing}_\rg}{\rd\rho}
\right) + \int_{2\zcut}^1\!\rd\rho\, \left(
\frac{\rd\sigma_\rg}{\rd\rho}-\frac{\rd\sigma}{\rd\rho}
\right)\\
&+\int_0^4\!\rd\rho\, \left(
\frac{\rd\sigma}{\rd\rho}-\frac{\rd\sigma^{\sing}}{\rd\rho}\right)+\int_{2\zcut}^1\!\rd\rho\, \left(\frac{\rd\sigma^{\sing}}{\rd\rho}- \frac{\rd\sigma^{\sing}_\rg}{\rd\rho}\right)+\int_1^4\!\rd\rho\, \frac{\rd\sigma^{\sing}}{\rd\rho}
+\order{\zcut}\,.
\esp
\label{eq:stot2}
\end{equation}
Further progress can be made by replacing the total cross section $\sigma_\text{tot}$ with its expression as an integral over the ungroomed hemisphere mass cross section. Writing the ungroomed hemisphere mass cross section at LP as
\begin{equation}
\frac{\rd\sigma_{\LP}}{\rd\rho} = D_{\delta}\,\delta(\rho) + \frac{\rd\sigma^{\sing}}{\rd\rho}\,,
\end{equation}
the total cross section can also be expressed as
\begin{equation}
\sigma_\text{tot} = D_\delta+\int_0^4\!\rd\rho\, \left(
\frac{\rd\sigma}{\rd\rho}-\frac{\rd\sigma^{\sing}}{\rd\rho}\right)\,.
\label{eq:stot3}
\end{equation}
Using \eqns{eq:stot2}{eq:stot3}, we find the relationship
\begin{align}\label{eq:const_rel}
D_\delta&=D_{\delta,\rg}+\int_0^{2\zcut}\!\rd\rho\, \left(
\frac{\rd\sigma_\rg}{\rd\rho}-\frac{\rd\sigma^{\sing}_\rg}{\rd\rho}
\right) + \int_{2\zcut}^1\!\rd\rho\, \left(
\frac{\rd\sigma_\rg}{\rd\rho}-\frac{\rd\sigma}{\rd\rho}
\right)\\
&
\hspace{2cm} +\int_{2\zcut}^1\!\rd\rho\, \left(\frac{\rd\sigma^{\sing}}{\rd\rho}- \frac{\rd\sigma^{\sing}_\rg}{\rd\rho}\right)+\int_1^4\!\rd\rho\, \frac{\rd\sigma^{\sing}}{\rd\rho}\,.\nonumber
\end{align}
Through two-loop order, the only unknowns in this equation are the collinear-soft function constants $c_{S_c}^{(2)}$ within the $D_{\delta,\rg}$ term and the first two integrals.  However, those integrals can be numerically evaluated with a fixed-order code.  We will first validate this relationship at one-loop, and then use it at two-loops to determine $c_{S_c}^{(2)}$.

\subsection{Test at One-Loop}

At one-loop order, \Eq{eq:const_rel} simplifies significantly.  For $\rho > 2\zcut$, the groomed and ungroomed differential cross sections are identical, and so the third term in \Eq{eq:const_rel} is identically zero.  Hence, the one-loop relationship between the groomed and ungroomed cross sections can be expressed as
\begin{align}\label{eq:oneloop_fit}
&D_\delta^{(1)}-D_{\delta,\rg}^{(1)}-\int_{2\zcut}^1\!\rd\rho\, \left(\frac{\rd\sigma^\text{sing,(1)}}{\rd\rho}- \frac{\rd\sigma^\text{sing,(1)}_\rg}{\rd\rho}\right)-\int_1^4\!\rd\rho\, \frac{\rd\sigma^\text{sing,(1)}}{\rd\rho}\\
&
\hspace{8cm}=\int_0^{2\zcut}\!\rd\rho\, \left(
\frac{\rd\sigma_\rg^{(1)}}{\rd\rho}-\frac{\rd\sigma^\text{sing,(1)}_\rg}{\rd\rho}
\right) \,.\nonumber
\end{align}
This equation has been written so that everything on the left is known, and the right side can be evaluated numerically.  Using the formulas from \App{app:mmdt_res}, the left side evaluates to
\begin{align}\label{eq:oneloop_intext}
D_\delta^{(1)}-D_{\delta,\rg}^{(1)}-\int_{2\zcut}^1\!\rd\rho\, \left(\frac{\rd\sigma^\text{sing,(1)}}{\rd\rho}- \frac{\rd\sigma^\text{sing,(1)}_\rg}{\rd\rho}\right)-\int_1^4\!\rd\rho\, \frac{\rd\sigma^\text{sing,(1)}}{\rd\rho}&= \frac{\as C_F}{2\pi}\left(
\frac{\pi^2}{3}-2\log^2 2
\right)\nonumber\\
&\simeq 2.32896\,\frac{\as C_F}{2\pi}\,.
\end{align}

\begin{figure}
    \centering
    \includegraphics[width=8cm]{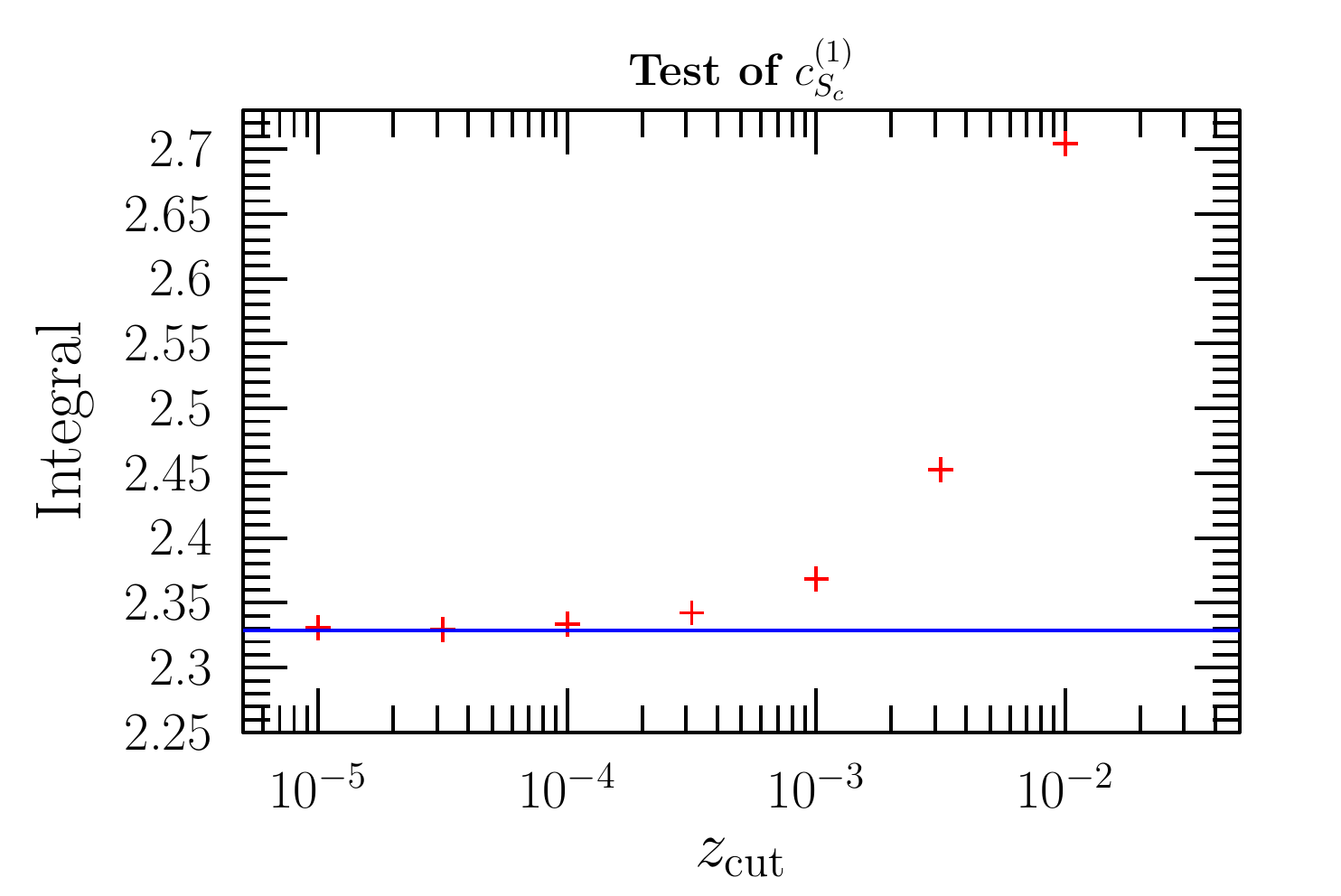}
    \caption{Plot of the numerical integral of \Eq{eq:oneloop_numint} as a function of $\zcut$ as evaluated in {\tt {\tt EVENT2}}.  The exact value for $\zcut \to 0$ is plotted as the solid line.}
    \label{fig:one-loop_integral}
\end{figure}

To evaluate the right side of \Eq{eq:oneloop_fit}, we generate about $10^{13}$ $e^+e^-\to q\bar q g$ events in {\tt EVENT2} \cite{Catani:1996vz}.  The events are groomed with mMDT with a range of $\zcut$ values from $10^{-5}$ to $10^{-2}$, in powers of $\sqrt{10}\simeq 3.16$.  On the groomed events, we then measure the heavy hemisphere mass.  For each value of $\zcut$, we calculate the integrand of the integral on the right side of \Eq{eq:oneloop_fit} as a table of values ranging from $\rho \in [e^{-20},2\zcut]$, in steps of powers of $e^{0.1}$.  To compute the integral, we then fit a smooth interpolating function to the table and numerically integrate.  The result of this procedure is shown in \Fig{fig:one-loop_integral}.  Here, we plot the result of the integral divided by the coupling factor:
\begin{equation}\label{eq:oneloop_numint}
\text{Integral} = \frac{2\pi}{\as C_F}\int_0^{2\zcut}\!\rd\rho\, \left(
\frac{\rd\sigma_\rg^{(1)}}{\rd\rho}-\frac{\rd\sigma^\text{sing,(1)}_\rg}{\rd\rho}
\right)\,.
\end{equation}
As $\zcut$ decreases, the integral values are seen to converge to the exact expected value computed in \Eq{eq:oneloop_intext}.

\subsection{Fit at Two-Loops}

Having validated the procedure at one-loop, we move on to using it at two-loops to determine the constant terms of the collinear-soft function, $c_{S_c}^{(2)}$.  Unlike at one-loop, all terms in \Eq{eq:const_rel} are generically non-zero, so we have the equality:
\begin{align}\label{eq:twoloop_ints}
&D_\delta^{(2)}-D_{\delta,\rg}^{(2)}-\int_{2\zcut}^1\!\rd\rho\, \left(\frac{\rd\sigma^\text{sing,(2)}}{\rd\rho}- \frac{\rd\sigma^\text{sing,(2)}_\rg}{\rd\rho}\right)-\int_1^4\!\rd\rho\, \frac{\rd\sigma^\text{sing,(2)}}{\rd\rho}\\
&
\hspace{4cm}
=\int_0^{2\zcut}\!\rd\rho\, \left(
\frac{\rd\sigma_\rg^{(2)}}{\rd\rho}-\frac{\rd\sigma^\text{sing,(2)}_\rg}{\rd\rho}
\right) + \int_{2\zcut}^1\!\rd\rho\, \left(
\frac{\rd\sigma^{(2)}_\rg}{\rd\rho}-\frac{\rd\sigma^{(2)}}{\rd\rho}
\right)\,.\nonumber
\end{align}
The right side of the equality is unknown and must be evaluated numerically.  The left side is completely known, except for the two-loop constant of the collinear-soft function, $c_{S_c}^{(2)}$.  Using the results of \App{app:mmdt_res} and the singular ungroomed heavy hemisphere mass distribution calculated in \Refs{Chien:2010kc,Kelley:2011ng,Monni:2011gb}, the left side can be expressed as:
\begin{align}
&D_\delta^{(2)}-D_{\delta,\rg}^{(2)}-\int_{2\zcut}^1\!\rd\rho\, \left(\frac{\rd\sigma^\text{sing,(2)}}{\rd\rho}- \frac{\rd\sigma^\text{sing,(2)}_\rg}{\rd\rho}\right)-\int_1^4\!\rd\rho\, \frac{\rd\sigma^\text{sing,(2)}}{\rd\rho}\\
&
\hspace{0.5cm}
=\left(
\frac{\as}{2\pi}
\right)^2 C_F \left[
C_F\left(
-2\left(
\frac{\pi^2}{3}-2\log^2 2
\right)\log^2\zcut
+\left(
-\frac{9}{4}+6\log^2 2-\pi^2-8\log^3 2\right.\right.\right.\nonumber\\
&
\hspace{1cm}
\left.+\frac{8}{3}\pi^2\log 2+8\zeta_3
\right)\log\zcut+\frac{9}{4}\log 2+2\log^2 2 -\frac{\pi^2}{3}-6\log^3 2+2\pi^2\log 2+6\zeta_3\nonumber\\
&
\hspace{1cm}
\left.\left.+2\log^4 2-\frac{4}{3}\pi^2\log^2 2+\frac{61}{360}\pi^4-8\zeta_3\log 2+\frac{\log 2}{4}\gamma_{S,C_F}^{(1)}- \frac{c_{S,C_F}^{(2)}}{4}- \frac{c_{S_c,C_F}^{(2)}}{2}
\right)\right.\nonumber\\
&
\hspace{0.5cm}
+C_A\left(
-\frac{11}{3}\left(
\frac{\pi^2}{3}-2\log^2 2
\right)\log\zcut-\frac{508}{81}-\frac{202}{27}\log 2-\frac{67}{9}\log^2 2-\frac{67}{24}\pi^2-\frac{11}{3}\log^3 2\right.\nonumber\\
&
\hspace{1cm}
\left.
+\frac{11}{36}\pi^2\log 2+\frac{88}{9}\zeta_3+\frac{\pi^2}{3}\log^2 2+\frac{11}{30}\pi^4+4\zeta(\bar 3,\bar 1)+\frac{\log 2}{4}\gamma_{S,C_A}^{(1)}-\frac{c_{S,C_A}^{(2)}}{4}-\frac{c_{S_c,C_A}^{(2)}}{2}
\right)\nonumber\\
&
\hspace{0.5cm}
+T_R n_f\left(
\frac{4}{3}\left(
\frac{\pi^2}{3}-2\log^2 2
\right)\log\zcut-\frac{34}{81}+\frac{56}{27}\log 2+\frac{20}{9}\log^2 2+\frac{19}{18}\pi^2+\frac{4}{3}\log^3 2 \right.\nonumber\\
&
\hspace{1cm}
\left.\left.
-\frac{\pi^2}{9}\log 2-\frac{32}{9}\zeta_3+\frac{\log 2}{4}\gamma_{S,n_f}^{(1)}-\frac{c_{S,n_f}^{(2)}}{4} -\frac{c_{S_c,n_f}^{(2)}}{2}
\right)
\right]\,.\nonumber
\end{align}
In this expression, $\zeta(\bar 3,\bar 1)$ is a multiple zeta value, where
\begin{equation}
\zeta(\bar 3,\bar 1) \simeq -0.117876\,.
\end{equation}
The coefficients of the two-loop anomalous dimension of the soft function $\gamma_S^{(1)}$ were calculated in \Refs{Frye:2016aiz,Bell:2018vaa,Bell:2018oqa} and are:
\begin{align}
\gamma_S^{(1)} &=  
2C_F \left(C_F \gamma_{S,C_F}^{(1)}+C_A \gamma_{S,C_A}^{(1)}+T_R n_f \gamma_{S,n_f}^{(1)}\right)\\
&\simeq2 C_F\left(17.0055\,C_F-20.4487\, C_A -10.9322\,T_R n_f\right)\,.\nonumber
\end{align}
The two-loop constants of the soft function $c_S^{(2)}$ were recently calculated by the \textsc{SoftServe} collaboration \cite{Bell:2018vaa,Bell:2018oqa,Bell:2020yzz}: 
\begin{align}
c_S^{(2)} &=  
C_F^2 c_{S,C_F}^{(2)}+C_F C_A c_{S,C_A}^{(2)}+C_FT_R n_f c_{S,n_f}^{(2)}\\
&\simeq74.523\,C_F^2-103.70\,C_F C_A +29.048\,C_FT_R n_f\,.\nonumber
\end{align}
The only unknowns in this expression are therefore the two-loop constants of the collinear-soft function.  As with the soft function's anomalous dimension and constants, we break up the collinear-soft constant into distinct color channel coefficients:
\begin{equation}
c_{S_c}^{(2)} =  C_F^2 c_{S_c,C_F}^{(2)}+C_F C_A c_{S_c,C_A}^{(2)}+C_FT_R n_f c_{S_c,n_f}^{(2)}\,.   
\end{equation}

To evaluate the right side of \Eq{eq:twoloop_ints}, we perform essentially the same procedure as at one-loop.  We generate about $10^{13}$ $e^+e^-\to$ hadrons events at ${\cal O}(\as^2)$ in {\tt EVENT2} and mMDT groom the final states with values of $\zcut$ ranging from $10^{-5}$ to $10^{-2}$ in factors of $\sqrt{10}$.  On these groomed events, we measure the heavy hemisphere mass $\rho$, and bin the resulting cross section over the range of $\log \rho \in[-20,0]$, in steps of $0.1$. For the second integral on the right side of \Eq{eq:twoloop_ints}, we also measure the ungroomed heavy hemisphere mass over the same range.  We then construct a smooth interpolating function for both integrands in \Eq{eq:twoloop_ints} which we then numerically integrate.  Because of the finite cutoffs inherent in {\tt EVENT2}, we cannot practically extend the first integral all the way down to 0.  Instead, we integrate over the reduced range of
\begin{equation}
\rho \in \left[
\frac{2\zcut}{400},2\zcut
\right]\,.
\end{equation} 
We have additionally verified that increasing or decreasing the lower bound by a factor of 2 only affects the results within quoted uncertainties.  With the integrals thus calculated, we then solve \Eq{eq:twoloop_ints} for each color channel's collinear-soft function two-loop constant coefficient.

\begin{figure}
    \centering
    \includegraphics[width=8cm]{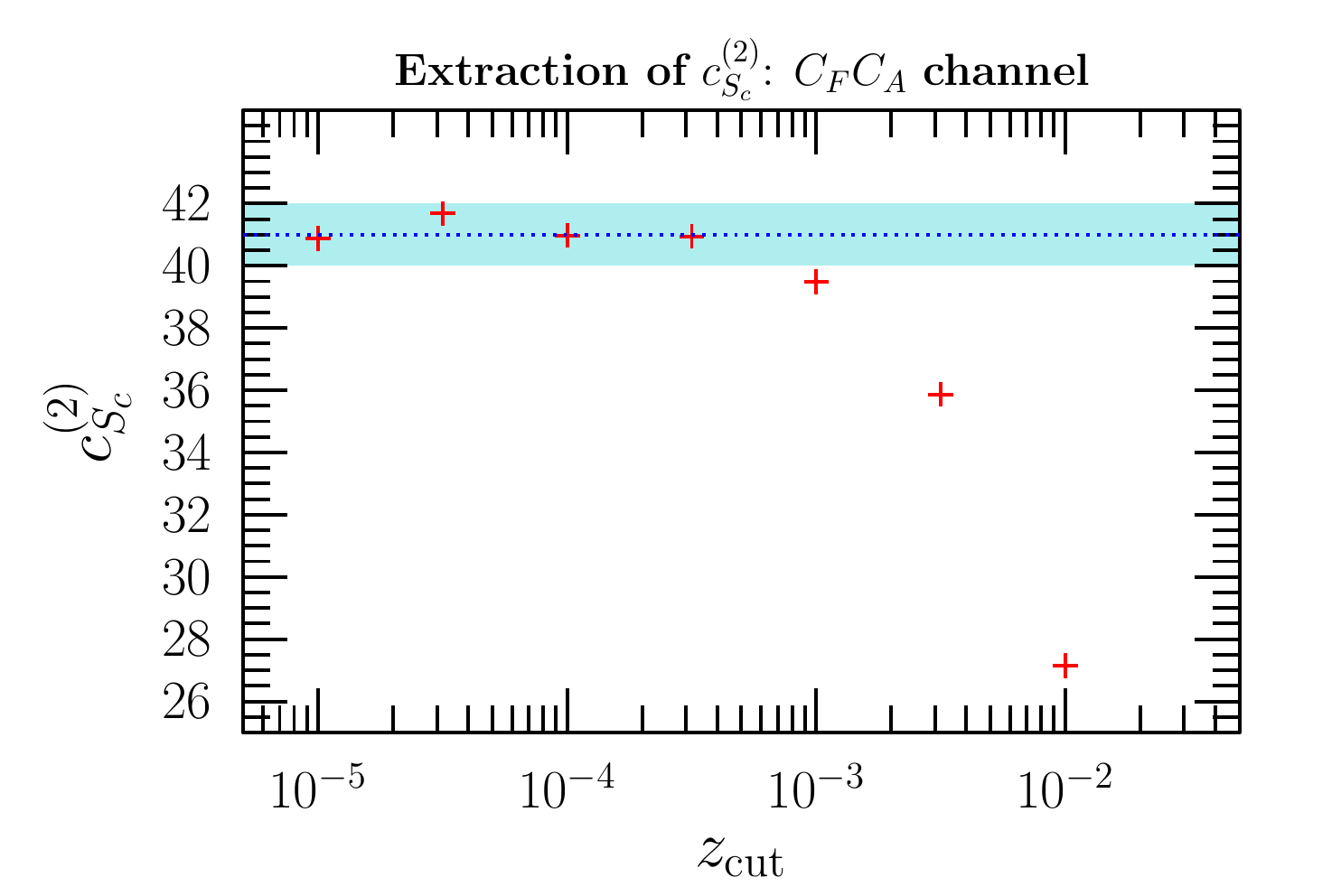}\hspace{0.2cm} 
    \includegraphics[width=8cm]{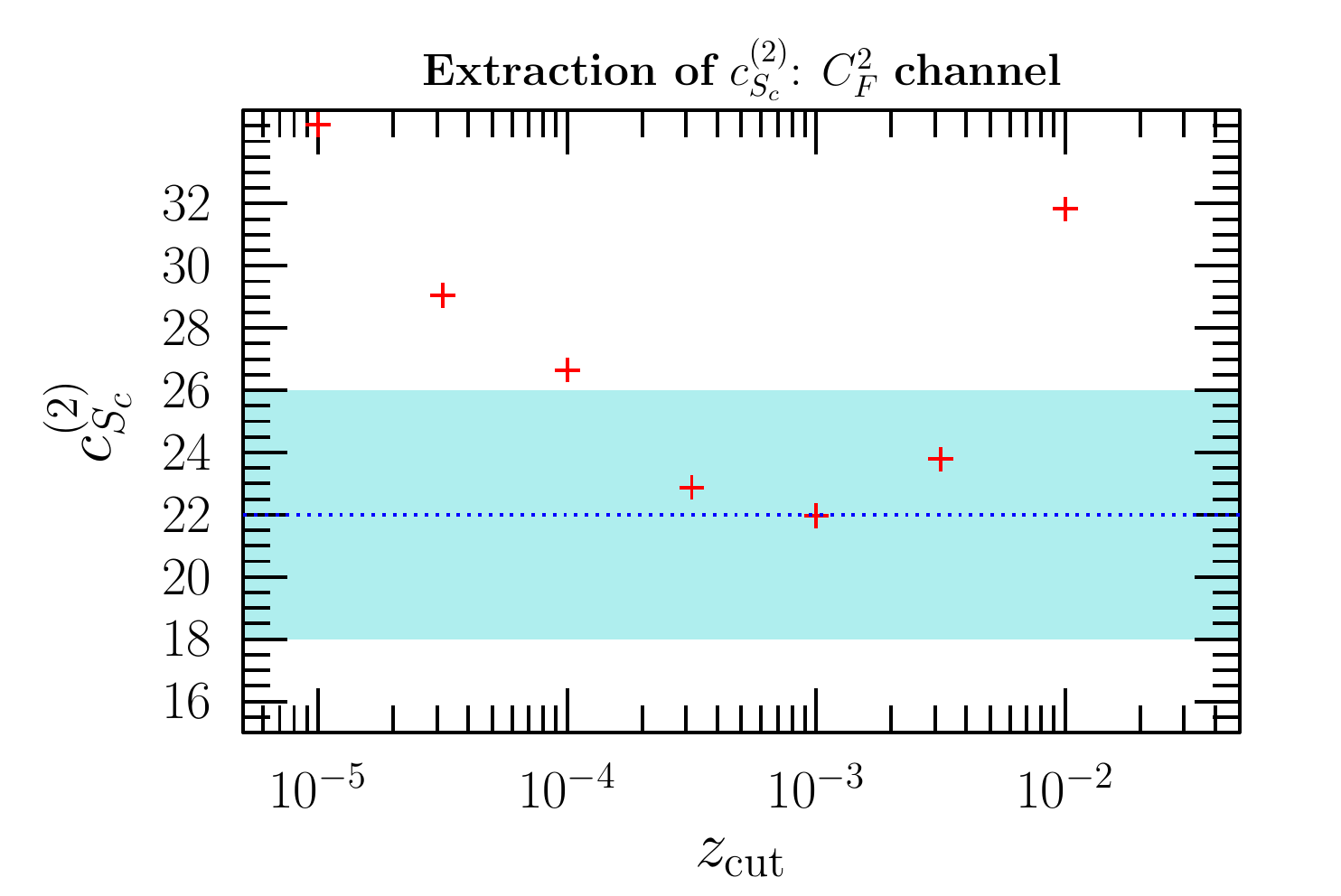}\\
    \vspace{0.2cm}
    \includegraphics[width=8cm]{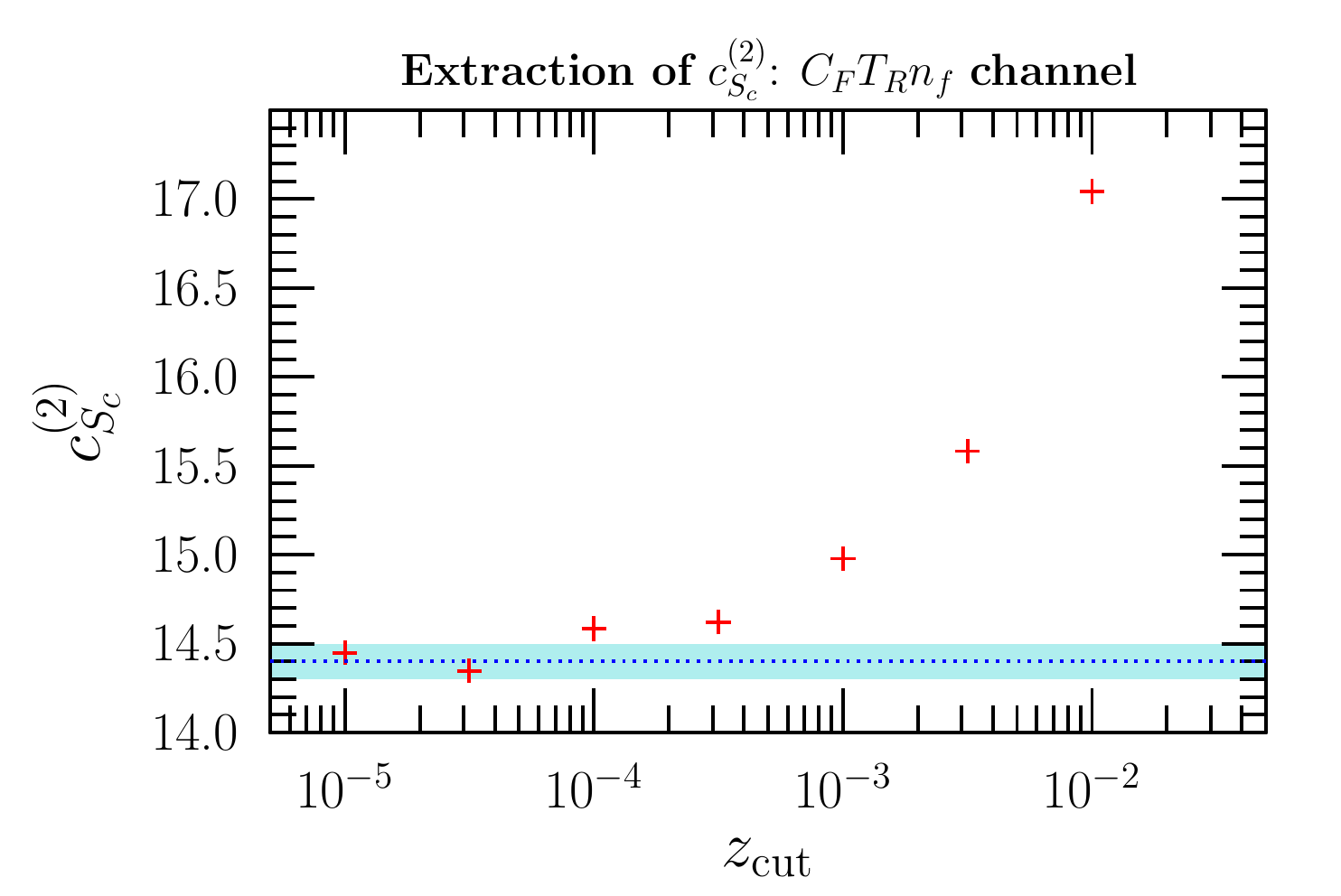}
    \caption{Plots of the values extracted from {\tt EVENT2} of the two-loop constants separated into distinct color channels of the collinear-soft function, as a function of $\zcut$.  Our claimed extracted value and its uncertainty is illustrated by the dotted line and shaded band, respectively.}
    \label{fig:twoloop_cs_const}
\end{figure}

The results of this procedure are plotted in \Fig{fig:twoloop_cs_const}. Here, we show the result of the two-loop constant extraction at each value of $\zcut$ considered.  We observe clear convergence as $\zcut\to 0$ for the $C_A$ and $T_R n_f$ channel coefficients, with the extracted value denoted by the dotted line, and the shaded band is our claimed uncertainty. For the $C_F$ channel, the convergence is less clear, with the extracted value of the two-loop constant appearing to diverge as $\zcut\to 0$.  In other contexts, it has been noted that the extraction of anomalous dimensions and constants in the $C_F$ channel with {\tt EVENT2} is problematic \cite{Chien:2010kc,Frye:2016aiz}, unless a significantly larger number of events are generated.  Thus, we take as our extracted value of the $C_F$ channel constant the lowest value determined through this procedure.  Further, if the approach to the true value were exponential in $\log \zcut$, as would be expected, then the difference between the extracted constants at different $\zcut$ values that differ by a factor should uniformly decrease.  Thus, we take our uncertainty to be twice the difference between the extracted value at $\zcut = 0.001$ and $\zcut = 0.00316$.\footnote{We have also attempted to use {\tt MCCSM} to determine the two-loop collinear-soft constant in the $C_F$ channel, given the results in the $C_A$ and $n_f T_R$ channels from {\tt EVENT2}.  However, we were never able to achieve the numerical precision of {\tt EVENT2} with comparable computational run times and as deep in the infrared, so could not perform a reliable extraction with {\tt MCCSM}.}

Finally, we obtain for the coefficients in the color decomposition of the two-loop constant of the collinear-soft function as
\begin{align}
c_{S_c}^{(2)} &= C_F^2 \left(22\pm 4\right)+C_F C_A\left(41\pm 1\right)+C_F T_R n_f \left(14.4\pm 0.1\right)\,.    
\end{align}

\section{\label{sec:threeloop}Extraction of Three-Loop Anomalous Dimensions}

We now turn to extraction of the three-loop non-cusp anomalous dimension of the mMDT global soft function, $\gamma_S^{(2)}$.  With the results quoted in the appendices and the extracted value of the two-loop constant of the collinear-soft function, we are able to solve the renormalization group equation through ${\cal O}(\as^3)$ order, and construct the differential cross section by inverse Laplace transformation.  For $\rho > 0$, the only unknown piece of the cross section is $\gamma_S^{(2)}$ and so its value can be extracted through comparison of the analytic prediction in the regime where $\rho \ll \zcut \ll 1$ to a next-to-next-to-leading fixed-order $e^+e^-\to $ hadrons event generator.  For some fixed values of $\zcut$, we can then fit the analytic cross section to match the predictions of the NNLO code for the value of the anomalous dimension and extrapolate to $\zcut \to 0$ to determine the $\gamma_S^{(2)}$.

The first step of this procedure is to solve the renormalization group equations to ${\cal O}(\as^3)$ to identify the cross section predicted by the factorization theorem. 
The ${\cal O}(\as^3)$ coefficient of the expansion is given in \eqn{eq:DC}
where we substituted explicitly the color factors of QCD ($C_F = 4/3$, $C_A = 3$, $T_R = 1/2$), and set the number of active quarks to $n_f = 5$.  We have also just used the central values of the collinear-soft constants, with no accounting for uncertainties.  The uncertainties that we will find for $\gamma_S^{(2)}$ will more than account for uncertainties in the collinear-soft function constants, so we can safely ignore them.  In this logarithmic form of the cross section, the anomalous dimension $\gamma_S^{(2)}$ appears as a constant offset of the cross section.  This suggests a procedure for its numerical extraction.

As the mMDT grooming removes double logarithms in $\rho$ to all orders, the singular term at ${\cal O}(\alpha_s^3)$, $D_{C,\rg}$, is only quadratic in $\log\rho$.  Thus, our task of determining $\gamma_S^{(2)}$ is reduced to finding the constant term of a parabola.  A parabola exhibits greatest sensitivity to the constant where its slope is smallest; that is, near its extremum, in our case a minimum.  Therefore, we would like to fit for $\gamma_S^{(2)}$ near the minimum of the logarithmic parabolic cross section.  However, the factorization theorem is only valid for $\rho \ll \zcut \ll 1$, so the fit is only sensible if the minimum is compatible with this singular limit.  These two competing requirements, fitting near the minimum and existing in the factorization regime, will greatly restrict the range of $\zcut$ that we can consider.

\begin{figure}
    \centering
    \includegraphics[width=8.5cm]{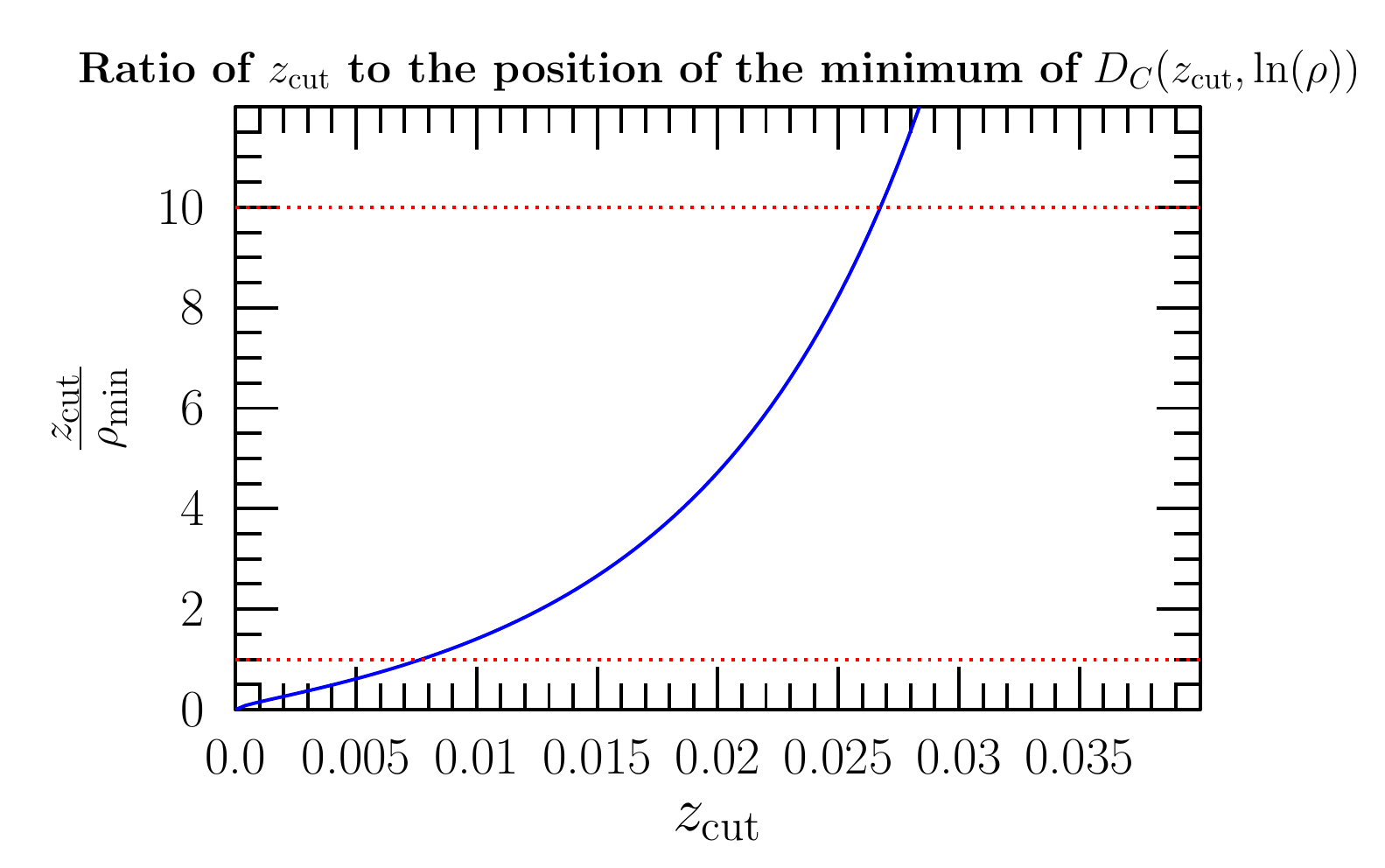}
    \caption{Plot of the ratio of $\zcut$ to the location of the minimum of the factorization theorem cross section at ${\cal O}(\as^3)$ logarithmic in $\rho$, $\rho_{\min}$.  Lines corresponding to ratios of 1 and 10 have been added to guide the eye.}
    \label{fig:min_range}
\end{figure}

Taking the derivative of the function $D_{C,\rg}$ given in \Eq{eq:DC} to determine the point $\rho_{\min}$ at which the cross section is minimized, we can then compare $\rho_{\min}$ to $\zcut$.  
We plot the result of this minimization and comparison to $\zcut$ in \Fig{fig:min_range} where the solid curve is the ratio $\zcut/\rho_{\min}$, as a function of $\zcut$.  The fixed ratios of 1 and 10 are also shown to guide the eye.  
We find that only for $\zcut \gtrsim 0.01$ is $\rho_{\min}$ smaller than $\zcut$, and only for $\zcut \gtrsim 0.03$ is there an order of magnitude between them.  As our factorization theorem requires a parametric separation between $\rho$ and $\zcut$, we will only consider values of $\zcut$ larger than $0.03$ in our fit for the anomalous dimension.  Further, demanding that $\zcut$ is relatively small, we will restrict to an upper bound on $\zcut$ of $0.1$.  While a limited range in $\zcut$, and relatively large values, we will see that it is sufficient for an extraction of the anomalous dimension.

For numerical generation of the cross section for the process $e^+e^-\to$ hadrons at ${\cal O}(\as^3)$ we use the {\tt MCCSM} (Monte Carlo for the CoLoRFulNNLO Subtraction Method) code \cite{Kardos:2016pic}. 
{\tt MCCSM} has previously been used to calculate distributions of standard event shapes \cite{DelDuca:2016csb,DelDuca:2016ily,Tulipant:2017ybb} as well as  mMDT and soft drop groomed event shapes \cite{Kardos:2018kth} in $e^+e^-$ collisions.  
To validate the implementation of the mMDT grooming we compared the resulting distributions at LO and NLO accuracy from the independent implementations in {\tt MCCSM} and {\tt EVENT2}.
We found perfect agreement between the predictions of the two codes.

\begin{figure}
    \centering
    \includegraphics[width=8.3cm]{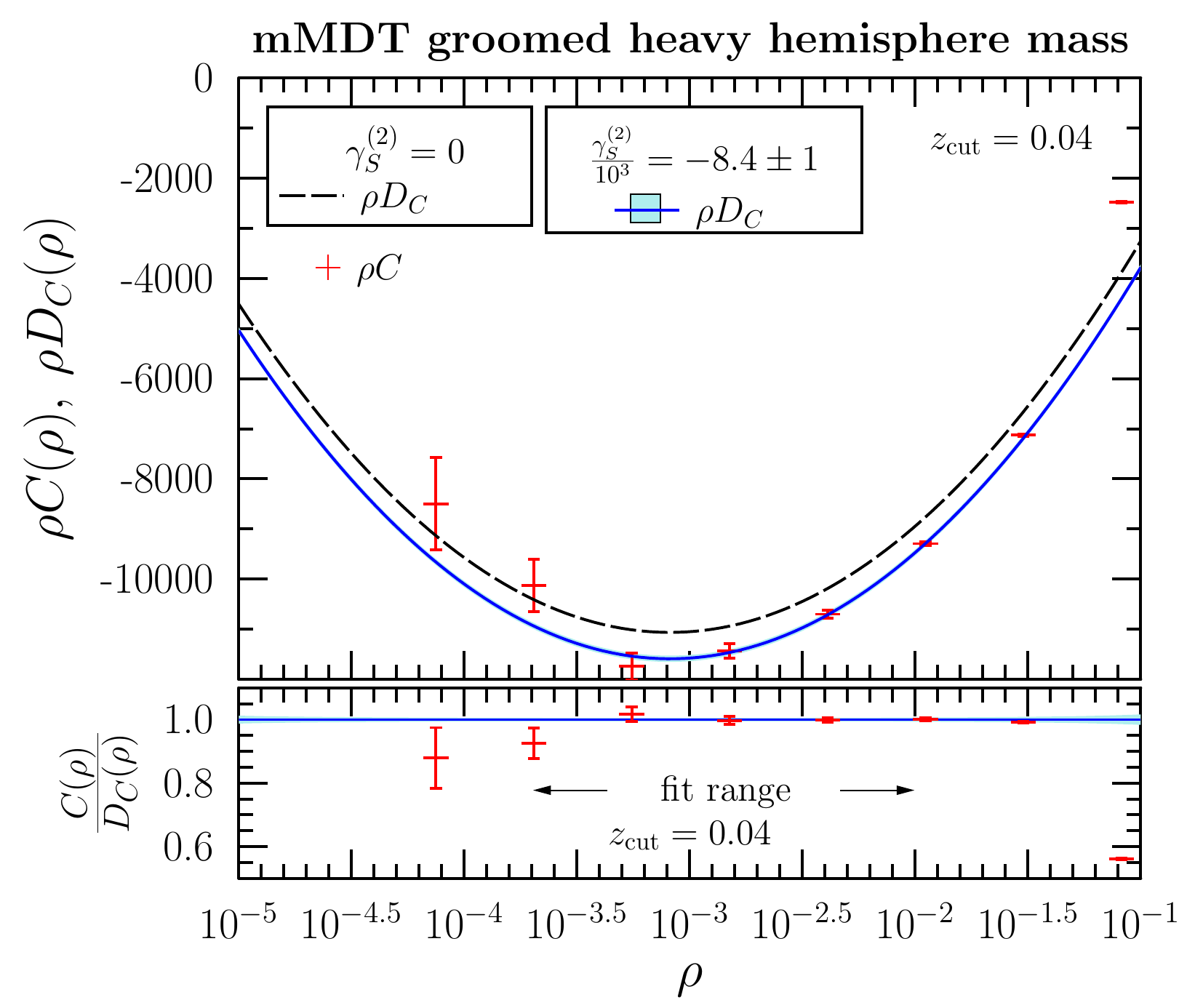}\ \ 
    \includegraphics[width=7.7cm]{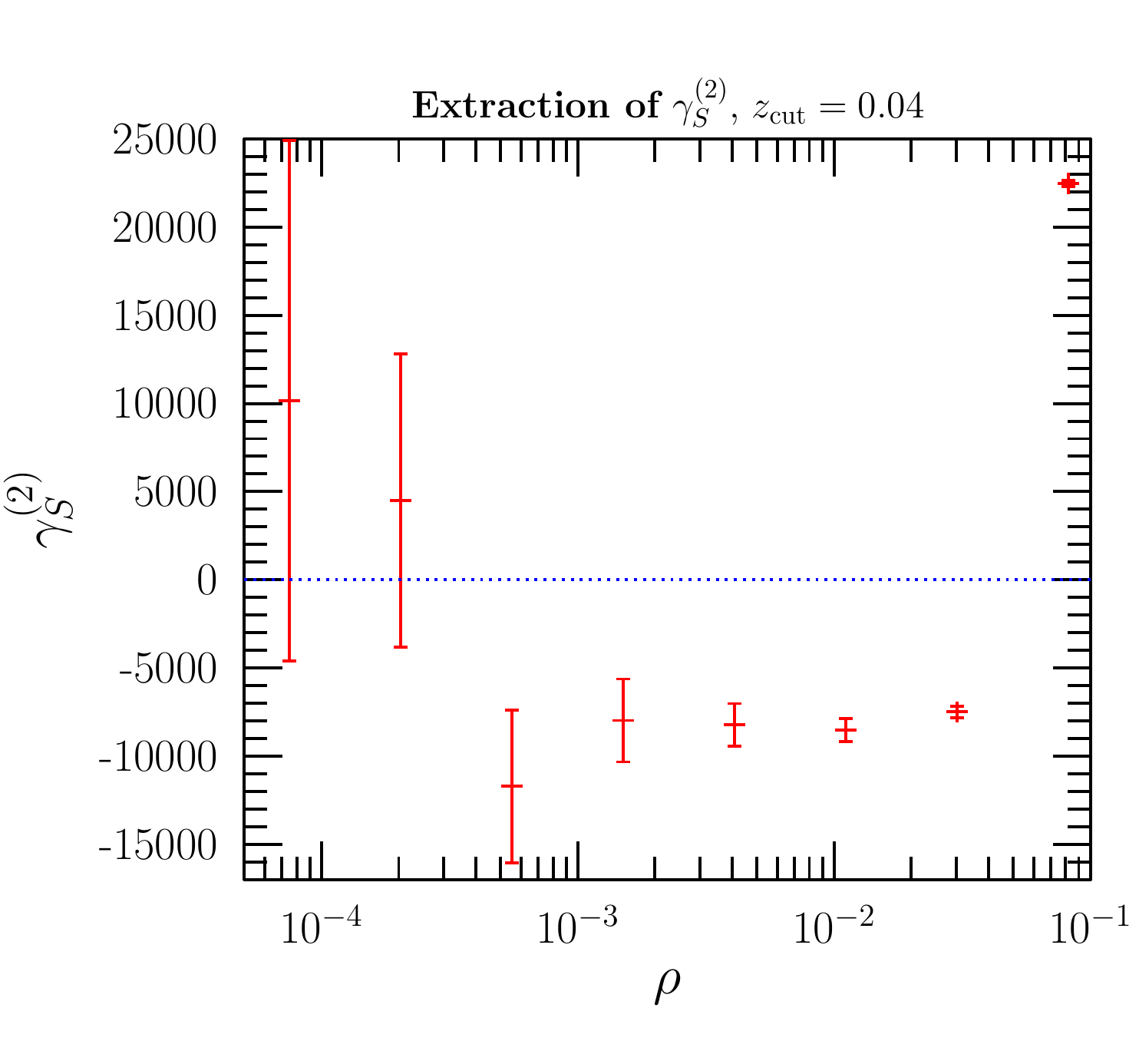}
    \caption{Left: Plot comparing the distribution of mMDT groomed heavy hemisphere mass at ${\cal O}(\as^3)$ from {\tt MCCSM} to the analytic prediction with $\gamma_S^{(2)}=0$ and $\gamma_S^{(2)}=-8400\pm 1000$.  Overall coupling dependence has been stripped off.  Right: Plot showing the extracted value of $\gamma_S^{(2)}$ from {\tt MCCSM} at each $\rho$ point.  For both plots, $\zcut = 0.04$.}
    \label{fig:cscomp_04}
\end{figure}

For extraction of the three-loop soft function anomalous dimension, we computed the distribution of the mMDT groomed heavy hemisphere mass for four values of $\zcut$ that satisfies the criteria established earlier: $\zcut = 0.04, 0.06,0.08,0.1$.  
Any Monte Carlo integrator must contain a lower cut \ymin\ on the phase space that any scaled two-particle invariant of the event must exceed.
Double-precision arithmetic limits the lowest value of \ymin\ to the range $10^{-8\pm 1}$. 
As for the extraction of $\gamma_S^{(2)}$ we need stable predictions for $\rho\ll\zcut$ we have to investigate carefully the lower limit on $\rho$ where our predictions are still reliable at a given \ymin. 
We found that we can use $\ymin=10^{-8}$ for computing the phase-space integrals of the real-virtual contributions and $\ymin=10^{-7}$ for the double-real contributions to obtain realiable distributions for $\rho \geq 5\cdot 10^{-5}$ \cite{Kardos:2019iwa}.
We used these values of the technical cut to generate 250 batches of events for the double-virtual and real-virtual contributions and 10 thousand batches for the most challenging double-real integrals, with 100 million events in each batch.
The events were binned logarithmically, with 10 equal bins in the range $-10<\log(\rho)<0$, i.e., we computed the coefficient of the distribution $\rho\frac{\rd\sigma_{\rg}}{\rd\rho}$ directly.
First we performed a flat average of the results of the batches of events.
The resulting distributions were fairly stable and smooth except for the first two bins with smallest values of $\rho$ due to occasional mis-binning of the squared matrix element and subtraction terms.
We filtered the results of the individual batches of events by selecting the ten batches with largest uncertainty in the second bin.
Removing the batches from the average one-by-one, we found compatible results for the distributions, but gradually decreasing uncertainty in the first two bins.
We produced our final distributions without removing any of the events, but employing a weighted (by the inverse of the uncertainties) average of the individual results of batches.
We have verified that the resulting distributions are consistent with the unfiltered and unweighted events, though with significantly smaller uncertainties in each histogram bin.  The resulting distributions are provided in \App{app:mccsm} for each value of $\zcut$ that was generated, and we also show it for $\zcut=0.04$ in \Fig{fig:cscomp_04} (left).

With these NNLO events generated, we can then compare {\tt MCCSM} to our analytic expression in \Eq{eq:DC}.  We first plot the two cross sections over one another to demonstrate sensitivity to the soft function anomalous dimension.  At left in \Fig{fig:cscomp_04}, we plot the {\tt MCCSM} cross section for the mMDT groomed heavy hemisphere mass with $\zcut = 0.04$, and provide two curves of the corresponding analytic prediction in which the soft function anomalous dimension takes one of two values: $\gamma_S^{(2)} = 0$ and $\gamma_S^{(2)}=-8400$, with the shaded band representing an uncertainty in its value of $\pm 1000$.   Clear sensitivity to the value of the anomalous dimension is observed, and the {\tt MCCSM} results prefer a negative value of $\gamma_S^{(2)}$.  This is illustrated more directly in the right panel of \Fig{fig:cscomp_04}.  Here, we have set the analytic prediction equal to the {\tt MCCSM} data and solved for $\gamma_S^{(2)}$ at each value of $\rho$.  Below about $\rho \simeq 0.01$, the value of the anomalous dimension at each data point is consistent within uncertainties of one another, demonstrating independence of $\rho$, as $\gamma_S^{(2)}$ must.

\begin{figure}
    \centering
    \includegraphics[width=9cm]{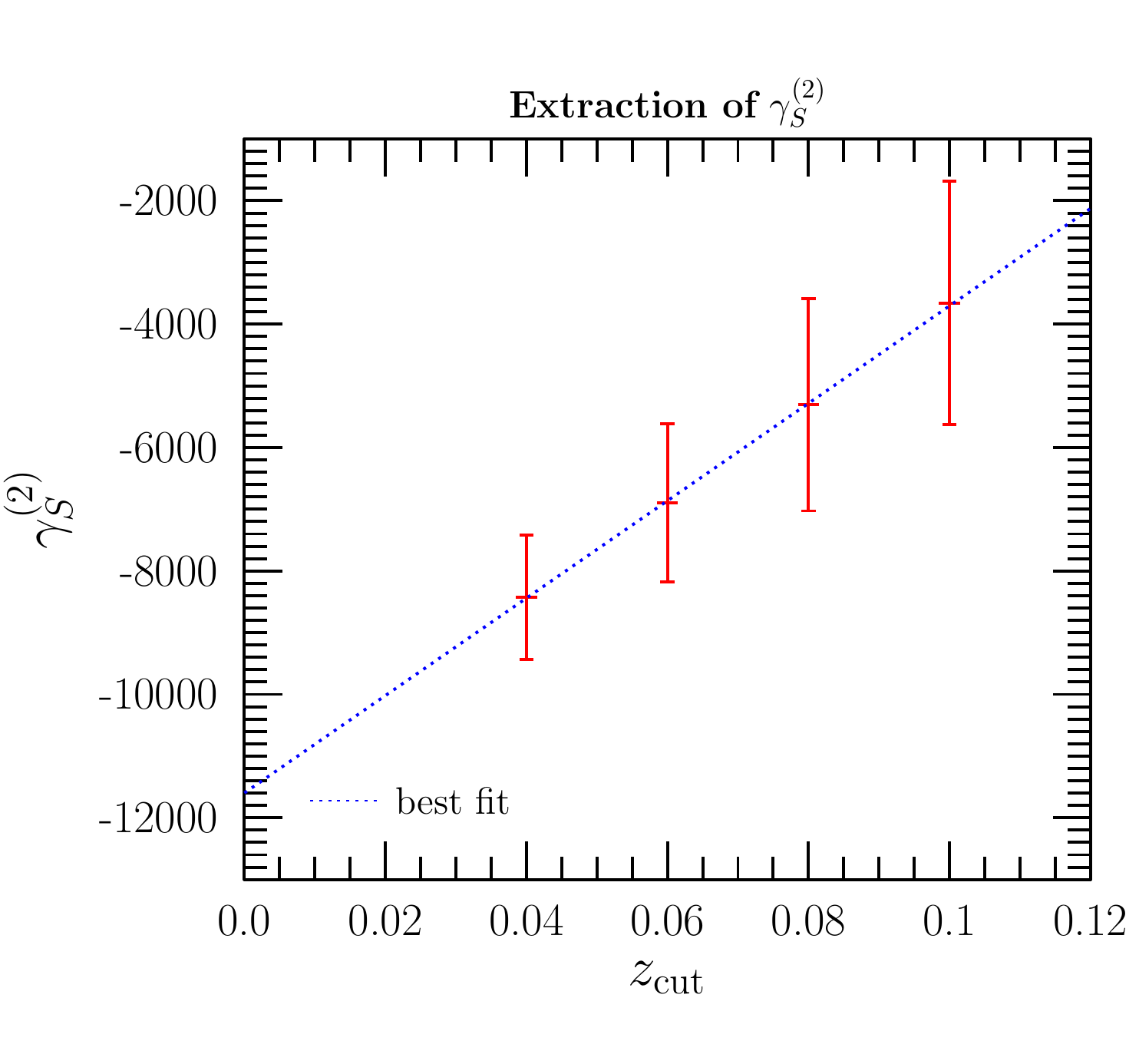}
    \caption{Plot of $\gamma_S^{(2)}$, with uncertainties, determined at each value of $\zcut$.  The best-fit line is dotted and its intercept at $\zcut = 0$ is the quoted extracted value of $\gamma_S^{(2)}$.}
    \label{fig:csfit}
\end{figure}

For each value of $\zcut$ that we generated, we determined the anomalous dimension through the following procedure.  Taking the plot on the right of \Fig{fig:cscomp_04} (and the corresponding plots for other values of $\zcut$), we fit for the anomalous dimension from data points in the range $\rho \in[0.0002,0.01]$.\footnote{We have verified that the extracted central value of $\gamma_S^{(2)}$ is consistent within uncertainties using a fit range of $\rho\in[0.000075,0.004]$, though with larger uncertainties.}  This range ensures that the largest value of $\rho$ is still significantly smaller than $\zcut$.  The contribution of a data point to the anomalous dimension and its uncertainty are weighted by the quoted variance of that data point.  So, the points at larger $\rho$ with smaller uncertainty dominate the fit.  The result of this fitting procedure is shown in \Fig{fig:csfit}.  As $\zcut$ decreases, the value of the extracted anomalous dimension also decreases, and the central values lie almost perfectly on a line.  Extrapolating the points in the $\zcut\to 0$ limit produces the desired three-loop anomalous dimension.  Performing a weighted fit of the points in \Fig{fig:csfit} to a line, we find the intercept, that is, $\gamma_S^{(2)}$, to be:
\begin{equation}
\gamma_S^{(2)} = -11600 \pm 2000 \qquad (n_f = 5)\,.
\end{equation}
We expect this claimed uncertainty to be a conservative overestimate due to the extremely linear behavior of the central values.\footnote{There are additional uncertainties in {\tt MCCSM} due to a numerical fit of part of the NNLO matrix element.  We find that the uncertainty stated here in the value of the anomalous dimension is representative of uncertainties that come from this fit in {\tt MCCSM}.}

\section{\label{sec:conclusions}Conclusions and Outlook}

Using numerical predictions at NLO and NNLO accuracies, we were able to extract two-loop constants and three-loop anomalous dimensions for the mMDT groomed mass factorization theorem in $e^+e^-\to$ hadrons events.  As a result, we are able to resum the groomed mass distribution to NNNLL accuracy, extending the predictions of \Ref{Frye:2016aiz} to one higher logarithmic order.  We present NNNLL+NNLO resummed and matched predictions for the heavy hemisphere mMDT groomed jet mass in a companion paper \cite{Kardos:2020gty}.

The results presented here also enable higher-precision calculations for groomed mass distributions for jets produced in hadron collisions.  The cross section for groomed mass $\rho$ of jets at a hadron collider can be expressed as
\begin{align}
\frac{\rd \sigma}{\rd \rho} = \sum_{i\in q,g} {\cal N}_i(p_T,y,\zcut,R)\, J_i(\rho)\otimes S_{c,i}(\rho,\zcut)\,.
\end{align}
At a hadron collider, both quark and gluon jets can be produced, so we have to sum over both possibilities.  In the factorization limit $\rho \ll \zcut \ll1$, all radiation that contributes to $\rho$ must be collinear, so the only dependence on the groomed jet mass lives in the universal jet and collinear-soft functions, appropriate for quarks or gluons.  The normalization factor ${\cal N}_i(p_T,y,\zcut,R)$ encodes the relative fraction of jets in the sample that are quark or gluon flavor, and can in principle depend on all other scales in the event, except for $\rho$: the jet's transverse momentum $p_T$, rapidity $y$, the value of $\zcut$, and the jet radius $R$.  The results presented here enable NNNLL resummation of all $\rho$ dependence at a hadron collider.

Extension of the mMDT groomed jet mass factorization theorem to three-loop order is one necessary step for complete perturbative control over the groomed jet mass distribution.  However, with multiple scales in a groomed jet, there are multiple regimes in which one may need to resum large logarithms.  No NGLs in the groomed mass $\rho$ are present in the regime where $\rho \ll \zcut$, by the collinearity of emissions that remain after grooming.  Yet, the NGLs are still there, they have just been pushed into the regime in which $\rho \simeq \zcut$.  Typical values of $\zcut$ of about $0.1$ may produce numerically-relevant NGLs that should be resummed to control them in the region of the cross section where grooming begins to dominate.  Conversely, our factorization theorem requires that $\zcut \ll 1$ to ensure that only soft emissions are groomed away.  However, the typical $\zcut\simeq 0.1$ may not be such a good approximation to $\zcut \ll 1$, and finite $\zcut$ contributions should be included.  This was studied in the original mMDT paper \cite{Dasgupta:2013ihk}, but a factorization theorem for this regime does not yet exist.

Additionally, our results in this paper are restricted to the mMDT groomer, or soft drop with $\beta = 0$.  To extract the two-loop collinear-soft constants with $\beta>0$, essentially the same procedure could be used as proposed in this paper, with appropriate modifications for the phase space boundaries when $\beta \neq 0$.  An analysis for general soft drop angular exponent $\beta$ is much more challenging at three-loops.  Our extraction of the three-loop anomalous dimension for mMDT grooming was simplified because mMDT removes double logarithms of the mass, but for $\beta > 0$, double logarithms generically exist.  What was a problem of finding the constant term of a parabola here turns into finding the constant term of a fifth-degree polynomial for general $\beta$.  A significant amount of data are known for general soft drop grooming at two-loop accuracy \cite{Bell:2018vaa,Bell:2018oqa,Bell:2020yzz}, but complete resummation to NNNLL accuracy will be challenging.  Soft drop grooming with non-zero $\beta$ can provide a handle on quark vs.~gluon fractions in a jet sample at a hadron collider \cite{Bendavid:2018nar} as well as provide just the right amount of grooming for boosted top quarks \cite{Hoang:2017kmk}.  Three-loop data for soft drop grooming is therefore highly desired.  We hope that the work presented in this paper inspires further  precision predictions for jet physics.

\begin{acknowledgments}
 We thank Guido Bell, Jim Talbert, and Rudi Rahn for providing their results for the two-loop constants of the global soft function.
 This work was facilitated in part by the Portland Institute for Computational Science and its resources acquired using NSF Grant DMS 1624776, by the grant K 125105 of the National Research, Development and
Innovation Fund in Hungary  and the Premium Postdoctoral Fellowship program of the Hungarian Academy of Sciences.
\end{acknowledgments}

\appendix

\section{\label{app:ingredients}Resummation Ingredients}

\subsection{$\beta$-function and Cusp Anomalous Dimension}

The QCD $\beta$-function is defined to be
\begin{equation}
\beta(\as)=\mu\frac{\partial\as}{\partial\mu} =-2\as\sum_{n=0}^\infty \beta_n \left(
\frac{\as}{4\pi}
\right)^{n+1}\,.
\end{equation}
For expansion of the cross section to three-loop order, we need the $\beta$-function to two-loop order \cite{Caswell:1974gg,Jones:1974mm,Egorian:1978zx}.  The first two coefficients are
\begin{align}
&
\hspace{-0.5cm}
\beta_0 =\frac{11}{3}C_A -\frac{4}{3}T_R n_f \,, \\
&
\hspace{-0.5cm}\beta_1 =\frac{34}{3}C_A^2-4T_R n_f\left(
C_F+\frac{5}{3}C_A
\right)  \nonumber\,.
\end{align}
Correspondingly, we need the cusp anomalous dimension
\begin{equation}\label{eq:cuspexp}
\Gamma_{\cusp} = \sum_{n=0}^\infty \Gamma_n \left(
\frac{\as}{4\pi}
\right)^{n+1}
\end{equation}
to three-loop order.  The first three coefficients of the cusp anomalous dimension are \cite{Korchemsky:1987wg,Vogt:2000ci,Berger:2002sv,Moch:2005tm}:
\begin{align}
\Gamma_0 &= 4 \,,\\
\Gamma_1 &=  4C_A\left(
\frac{67}{9}-\frac{\pi^2}{3}
\right) - \frac{80}{9}T_R n_f\,,\nonumber \\
\Gamma_2 &= 4C_A^2\left(
\frac{245}{6}-\frac{134\pi^2}{27}+\frac{11\pi^4}{45}+\frac{22}{3}\zeta_3
\right) +32 C_A T_R n_f\left(
-\frac{209}{108}+\frac{5\pi^2}{27}-\frac{7}{3}\zeta_3
\right)\nonumber \\
&
\hspace{1cm}
+4 C_F T_R n_f\left(
16 \zeta_3 - \frac{55}{3}
\right)-\frac{64}{27}T_R^2 n_f^2
\nonumber \,.
\end{align}

\subsection{Non-Cusp Anomalous Dimensions and Constants}

The non-cusp anomalous dimensions $\gamma$ and constants $c$ can be expanded in similar series in $\as$ as the $\beta$-function and cusp anomalous dimensions.  We have
\begin{align}
&\gamma = \sum_{n=0}^\infty \gamma^{(n)}\left(
\frac{\as}{4\pi}
\right)^{n+1}\,,
&c = \sum_{n=1}^\infty c^{(n)}\left(
\frac{\as}{4\pi}
\right)^{n}\,.
\end{align}
To ${\cal O}(\as^3)$ accuracy in terms with support away from 0, we need the non-cusp anomalous dimension to $\as^3$ order and the constants to $\as^2$ order.

\subsubsection{Hard Function}

The first three non-cusp anomalous dimensions of the hard function (quark form factor) are \cite{vanNeerven:1985xr,Matsuura:1988sm,Moch:2005id,Becher:2006mr}:
\begin{align}
\gamma_H^{(0)}&=-12 C_F\,,\\
\gamma_H^{(1)}&=\left(
-6+8\pi^2-96\zeta_3
\right)C_F^2+\left(
-\frac{1922}{27}-\frac{22}{3}\pi^2+104 \zeta_3
\right)C_F C_A+\left(
\frac{520}{27}+\frac{8}{3}\pi^2
\right) C_F n_f T_R\,,\nonumber\\
\gamma_H^{(2)} &=2C_F\left[C_F^2 \left( -29 - 6\pi^2 - \frac{16\pi^4}{5}
    - 136\zeta_3 + \frac{32\pi^2}{3}\,\zeta_3 + 480\zeta_5 \right) \right.\nonumber\\
   &
   \hspace{2cm}
   + C_F C_A \left( - \frac{151}{2} + \frac{410\pi^2}{9}
    + \frac{494\pi^4}{135} - \frac{1688}{3}\,\zeta_3
    - \frac{16\pi^2}{3}\,\zeta_3 - 240\zeta_5 \right) \nonumber\\
   &\hspace{2cm}+ C_A^2 \left( - \frac{139345}{1458} - \frac{7163\pi^2}{243}
    - \frac{83\pi^4}{45} + \frac{7052}{9}\,\zeta_3
    - \frac{88\pi^2}{9}\,\zeta_3 - 272\zeta_5 \right) \nonumber\\
   &\hspace{2cm}+ C_F T_F n_f \left( \frac{5906}{27} - \frac{52\pi^2}{9} 
    - \frac{56\pi^4}{27} + \frac{1024}{9}\,\zeta_3 \right) \nonumber\\
   &\hspace{2cm}+ C_A T_F n_f \left( - \frac{34636}{729}
    + \frac{5188\pi^2}{243} + \frac{44\pi^4}{45} - \frac{3856}{27}\,\zeta_3 
    \right) \nonumber\\
   &\hspace{2cm}\left.+ T_F^2 n_f^2 \left( \frac{19336}{729} - \frac{80\pi^2}{27} 
    - \frac{64}{27}\,\zeta_3 \right)\right]\,.\nonumber
\end{align}
The first two constants of the hard function are \cite{Moch:2005id,Gehrmann:2005pd}:
\begin{align}
c_H^{(1)}&=C_F\left(
-16+\frac{7\pi^2}{3}
\right)\,,\\
c_H^{(2)}&=
C_F^2 \left(
\frac{511}{4}-\frac{83}{3}\pi^2+\frac{67}{30}\pi^4-60\zeta_3
\right)+C_F C_A\left(
-\frac{51157}{324}+\frac{1061}{54}\pi^2-\frac{8}{45}\pi^4+\frac{626}{9}\zeta_3
\right)\nonumber\\
&
\hspace{1cm}+C_F T_R n_f\left(
\frac{4085}{81}-\frac{182}{27}\pi^2+\frac{8}{9}\zeta_3
\right)
\nonumber\,.
\end{align}

\subsubsection{Jet Function}

The first three non-cusp anomalous dimensions of the quark jet function are \cite{vanRitbergen:1997va,Moch:2005id,Becher:2006mr}:
\begin{align}
\gamma_J^{(0)}&=6 C_F\,,\\
\gamma_J^{(1)}&=C_F^2\left(
3-4\pi^2+48\zeta_3
\right)+C_F C_A\left(
\frac{1769}{27}+\frac{22}{9}\pi^2-80\zeta_3
\right)+C_F T_R n_f\left(
-\frac{484}{27}-\frac{8}{9}\pi^2
\right)\,,\nonumber\\
\gamma_J^{(2)}&=-2C_F\left[C_F^2 \left( - \frac{29}{2} - 3\pi^2 - \frac{8\pi^4}{5} - 68\zeta_3 
    + \frac{16\pi^2}{3}\,\zeta_3 + 240\zeta_5 \right) \right.\nonumber\\
   &\hspace{2cm}+ C_F C_A \left( - \frac{151}{4} + \frac{205\pi^2}{9}
    + \frac{247\pi^4}{135} - \frac{844}{3}\,\zeta_3
    - \frac{8\pi^2}{3}\,\zeta_3 - 120\zeta_5 \right) \nonumber\\
   &\hspace{2cm}+ C_A^2 \left( - \frac{412907}{2916} - \frac{419\pi^2}{243}
    - \frac{19\pi^4}{10} + \frac{5500}{9}\,\zeta_3
    - \frac{88\pi^2}{9}\,\zeta_3 - 232\zeta_5 \right) \nonumber\\
   &\hspace{2cm}+ C_F T_F n_f \left( \frac{4664}{27} - \frac{32\pi^2}{9}
    - \frac{164\pi^4}{135} + \frac{208}{9}\,\zeta_3 \right) \nonumber\\
   &\hspace{2cm}+ C_A T_F n_f \left( - \frac{5476}{729}
    + \frac{1180\pi^2}{243} + \frac{46\pi^4}{45}
    - \frac{2656}{27}\,\zeta_3 \right) \nonumber\\
   &\left.\hspace{2cm}+ T_F^2 n_f^2 \left( \frac{13828}{729} - \frac{80\pi^2}{81}
    - \frac{256}{27}\,\zeta_3 \right)\right]\,.\nonumber
\end{align}
The first two constants of the jet function are \cite{Becher:2008cf}:
\begin{align}
c_J^{(1)} &= C_F\left(
7-\frac{2\pi^2}{3}
\right)\,,\\
c_J^{(2)} &= C_F^2 \left(
\frac{205}{8}-\frac{97\pi^2}{12}+\frac{61\pi^4}{90}-6\zeta_3
\right)+C_F C_A\left(
\frac{53129}{648}-\frac{155\pi^2}{36}-\frac{37\pi^4}{180}-18\zeta_3
\right)\nonumber\\
&
\hspace{1cm}+C_F T_R n_f\left(
-\frac{4057}{162}+\frac{13\pi^2}{9}
\right)
\,.\nonumber
\end{align}

\subsubsection{mMDT Global Soft Function}

The first two coefficients of the non-cusp anomalous dimensions of the mMDT global soft function were calculated previously \cite{Frye:2016aiz,Bell:2018vaa} and the third coefficient is new in this work:
\begin{align}\label{eq:soft_anom}
\gamma_S^{(0)} &= 0\,,\\
\gamma_S^{(1)} &= 2C_F\left[
17.0055 C_F-20.4487 C_A-10.9322 T_R n_f
\right]\,,\nonumber\\
\gamma_S^{(2)} &= -11600 \pm 2000 \hspace{0.5cm}\text{($n_f=5$)}\,.\nonumber
\end{align}
In $\gamma_S^{(1)}$, the coefficient of the $C_F^2$ term is known analytically:
\begin{equation}
\frac{16\pi}{3}\text{Cl}_2\left(\frac{\pi}{3}\right)\simeq 17.0055\,,
\end{equation}
where Cl$_2(x)$ is the Clausen function:
\begin{equation}
\text{Cl}_2\left(\frac{\pi}{3}\right) = 1.01494160\dotsc\,.
\end{equation}
The $C_F C_A$ and $C_F T_R n_f$ coefficients in $\gamma_S^{(1)}$ are not known analytically, so we present their decimal expansion to 6 significant figures.  The value of $\gamma_S^{(2)}$ is evaluated with five active quarks, $n_f = 5$.

The first constant of the mMDT soft function was calculated in \cite{Frye:2016aiz} and the second constant was calculated by the \textsc{SoftServe} collaboration \cite{Bell:2018oqa,Bell:2018vaa,Bell:2020yzz}:
\begin{align}\label{eq:soft_const}
c_S^{(1)} &= -\pi^2 C_F\,,\\
c_S^{(2)} &= 74.523 C_F^2 - 103.70 C_F C_A + 29.048 C_FT_R n_f\,.\nonumber
\end{align}
Uncertainties in the values of these numerical expressions occur in the last digit, which we do not quote because they are vastly smaller than other relevant uncertainties.

\subsubsection{mMDT Collinear-Soft Function}

By renormalization group invariance, the anomalous dimension of the collinear-soft function can be found to all-orders through its relationship with the other anomalous dimensions in the factorization theorem:
\begin{equation}
\gamma_{S_c} = -\frac{\gamma_H + \gamma_S}{2} - \gamma_J\,.
\end{equation}
As such, we will not present the fixed-order expansion of the collinear-soft function's anomalous dimension here.  The first coefficient of the collinear-soft function's constant terms was calculated in \Ref{Frye:2016aiz} and the second coefficient is new to this work:
\begin{align}\label{eq:csoft_const}
c_{S_c}^{(1)} &= 0\,,\\
c_{S_c}^{(2)} &= C_F^2 \left(22\pm 4\right)+C_F C_A\left(41\pm 1\right)+C_F T_R n_f \left(14.4\pm 0.1\right)\,.\nonumber    
\end{align}

\section{\label{app:foexpansion}Fixed-Order Expansion}

The hemisphere mMDT groomed jet mass factorization theorem in $e^+e^-\to $ hadrons collisions is
\begin{equation}
\frac{1}{\sigma_0}\frac{\rd^2\sigma}{\rd\tau_\rL\, \rd\tau_\rR} = H(Q^2)S(\zcut) \left[
J(\tau_\rL)\otimes S_c(\tau_\rL,\zcut)
\right]  \left[
J(\tau_\rR)\otimes S_c(\tau_\rR,\zcut)
\right]\,,
\end{equation}
where $H(Q^2)$ is the hard function for quark--anti-quark production in $e^+e^-$ collisions, $S(\zcut)$ is the global soft function for mMDT grooming, $J(\tau_i)$ is the quark jet function for hemisphere mass $\tau_i$, and $S_c(\tau_i,\zcut)$ is the collinear-soft function for hemisphere mass $\tau_i$ with mMDT grooming.  The symbol $\otimes$ denotes convolution over the hemisphere mass $\tau_i$.  The convolution can be transformed into a simple product by Laplace transforming the jet and collinear-soft functions appropriately.  The Laplace transform of a function $F$ in the factorization theorem is
\begin{equation}
\tilde F(\nu) = \int_0^\infty\!\rd\tau\, e^{-\tau \nu}\,F(\tau)\,,
\end{equation}
and the Laplace-transformed factorization theorem is
\begin{equation}
\frac{\sigma(\nu_\rL, \nu_\rR)}{\sigma_0} = H(Q^2)S(\zcut)
\tilde J(\nu_\rL) \tilde S_c(\nu_\rL,\zcut)
\tilde J(\nu_\rR) \tilde S_c(\nu_\rR,\zcut)\,.
\end{equation}
All functions in this factorization theorem satisfy a simple renormalization group equation:
\begin{equation}
\mu \frac{\partial \tilde F}{\partial \mu} = \left(
d_F \Gamma_{\cusp}\log\frac{\mu^2}{\mu_F^2}+\gamma_F
\right)\tilde F\,,
\quad
\tilde F = H,\:S,\:\tilde J,\:\tilde S_c
\end{equation}
where $\mu$ is the renormalization scale.  The expression in parentheses is $\tilde F$'s anomalous dimension where $\Gamma_{\cusp}$ is the cusp anomalous dimension, $d_F$ is a coefficient particular to function $F$, $\mu_F$ is the canonical scale of $F$, and $\gamma_F$ is the non-cusp anomalous dimension of $\tilde F$.

The anomalous dimension equation can be solved iteratively in powers of $\as$.  Through ${\cal O}(\as^3)$, the solution is
\begin{align}
\tilde F &= 1+\frac{\as}{4\pi}\left[
\frac{d_F \Gamma_0}{4}\log^2\frac{\mu^2}{\mu_F^2}+\frac{\gamma_F^{(0)}}{2}\log\frac{\mu^2}{\mu_F^2}+c_F^{(1)}
\right]+\left(
\frac{\as}{4\pi}
\right)^2\left[
\frac{d_F^2 \Gamma_0^2}{32}\log^4\frac{\mu^2}{\mu_F^2}\right.\\
&
\hspace{0.5cm}+\frac{d_F \Gamma_0}{4}\left(
\frac{\beta_0}{3}+\frac{\gamma_F^{(0)}}{2}
\right)\log^3\frac{\mu^2}{\mu_F^2}+\left(
\frac{\beta_0 \gamma_F^{(0)}}{4}+\frac{(\gamma_F^{(0)})^2}{8}+\frac{d_F \Gamma_0 c_F^{(1)}}{4}+\frac{d_F \Gamma_1}{4}
\right)\log^2\frac{\mu^2}{\mu^2_F}\nonumber\\
&\hspace{0.5cm}\left.+\left(
\beta_0 c_F^{(1)}+\frac{\gamma_F^{(0)}c_F^{(1)}}{2}+\frac{\gamma_F^{(1)}}{2}
\right)\log\frac{\mu^2}{\mu^2_F}+c_F^{(2)}
\right]+\left(
\frac{\as}{4\pi}
\right)^3\left[
\frac{d_F^3\Gamma_0^3}{384}\log^6\frac{\mu^2}{\mu_F^2}\right.\nonumber\\
&\hspace{0.5cm}
+\frac{d_F^2 \Gamma_0^2}{16}\left(\frac{\beta_0}{3}+\frac{\gamma_F^{(0)}}{4}\right)\log^5\frac{\mu^2}{\mu_F^2}+d_F \Gamma_0\left(\frac{\beta_0^2}{24}+\frac{5}{48} \beta_0\gamma_F^{(0)}+\frac{(\gamma_F^{(0)})^2}{32}+\frac{d_F\Gamma_0 c_F^{(1)}}{32}+\frac{d_F \Gamma_1}{16}\right)\log^4\frac{\mu^2}{\mu_F^2}\nonumber\\
&\hspace{0.5cm}+\left(\frac{\beta_0^2\gamma_F^{(0)}}{6}+\frac{\beta_0(\gamma_F^{(0)})^2}{8}+\frac{(\gamma_F^{(0)})^3}{48}+\frac{d_F\Gamma_0\beta_0 c_F^{(1)}}{3}+\frac{d_F\Gamma_0\beta_1}{12}+\frac{d_F \Gamma_0 \gamma_F^{(0)}c_F^{(1)}}{8}+\frac{d_F \Gamma_0 \gamma_F^{(1)}}{8}\right.\nonumber\\
&\left.
\hspace{1cm}+\frac{d_F\Gamma_1 \beta_0 }{6}+\frac{d_F\Gamma_1\gamma_F^{(0)}}{8}\right)\log^3\frac{\mu^2}{\mu_F^2}+\left(\beta_0^2 c_F^{(1)}+\frac{3}{4}\beta_0 \gamma_F^{(0)}c_F^{(1)}+\frac{\beta_1 \gamma_F^{(0)}}{4}+\frac{(\gamma_F^{(0)})^2 c_F^{(1)}}{8}\right.\nonumber\\
&\left.
\hspace{1cm}
+\frac{d_F \Gamma_0 c_F^{(2)}}{4}+\frac{\beta_0 \gamma_F^{(1)}}{2}+\frac{\gamma_F^{(0)}\gamma_F^{(1)}}{4}+\frac{d_F\Gamma_1 c_F^{(1)}}{4}+\frac{d_F\Gamma_2}{4}\right)\log^2\frac{\mu^2}{\mu_F^2}\nonumber\\
&\left.
\hspace{0.5cm}+\left(2\beta_0 c_F^{(2)}+\beta_1c_F^{(1)}+\frac{\gamma_F^{(0)}c_F^{(2)}}{2}+\frac{\gamma_F^{(1)}c_F^{(1)}}{2}+\frac{\gamma_F^{(2)}}{2}\right)\log\frac{\mu^2}{\mu_F^2}+c_F^{(3)}
\right]+{\cal O}(\as^4)\,.
\nonumber
\end{align}

For the functions in the factorization theorem, the $d_F$ coefficients are:
\begin{align}
&d_H = -2 C_F\,, &d_J = 2 C_F\,,\\
&d_S = 2C_F\,, &d_{S_c} = -2C_F\,.\nonumber
\end{align}
The canonical scales $\mu_F^2$ of the Laplace-space functions are:
\begin{align}
&\mu_H^2 = Q^2\,, &\mu_J^2 = \frac{Q^2}{4\nu}\,,\\
&\mu_S^2 = \zcut^2 Q^2\,, &\mu_{S_c}^2 = \frac{\zcut Q^2}{4\nu}\,.\nonumber
\end{align}

Then, to find the cross section in real space through ${\cal O}(\as^3)$, we multiply the Laplace-space functions together, expand in $\as$, and then inverse Laplace transform.  Because mMDT grooming removes double logarithms, the inverse Laplace transformation is relatively easy, and one only needs three transformations:
\begin{align}
{\cal L}^{-1}(\log \nu) &= -\left(\frac{1}{\tau}\right)_+\,,\\
{\cal L}^{-1}(\log^2 \nu) &= 2\left(\frac{\log \tau}{\tau}\right)_+-\frac{\pi^2}{6}\delta(\tau)\,,\nonumber\\
{\cal L}^{-1}(\log^3 \nu) &= -3\left(\frac{\log^2 \tau}{\tau}\right)_++\frac{\pi^2}{2}\left(\frac{1}{\tau}\right)_+-2\zeta_3\,\delta(\tau)\,.\nonumber
\end{align}
The $+$-functions are defined to integrate to 0 on $\tau\in[0,1]$.

\section{\label{app:mmdt_res}mMDT Groomed Heavy Hemisphere Mass Results}

Through ${\cal O}(\as^3)$, the singular cross section for the heavy hemisphere mMDT groomed mass can be written as
\begin{equation}
\frac{\rd\sigma_{\rg,\LP}}{\rd\rho} = \delta(\rho)D_{\delta,\rg}+\frac{\as}{2\pi}(D_{A,\rg}(\rho))_++\left(\frac{\as}{2\pi}\right)^2(D_{B,\rg}(\rho))_++\left(\frac{\as}{2\pi}\right)^3(D_{C,\rg}(\rho))_+\,.
\end{equation}
The coefficient functions are
\begin{align}
D_{\delta,\rg}&=1+\frac{\as}{4\pi}\left[C_F
\left(
-2+12 \log 2+16\log 2\log\zcut+4\log^2\zcut
\right)
\right]\\
&
\hspace{0.5cm}
+\left(\frac{\as}{4\pi}\right)^2\left[
C_F^2\left(
4+\frac{4}{3}\pi^2-\frac{7}{90}\pi^4-72\zeta_3-18\log 2+72\log^2 2+96 \zeta_3\log2\right.\right.\nonumber\\
&\left.\left.
\hspace{2cm}
+c_{S,C_F}^{(2)}+2c_{S_c,C_F}^{(2)}-4\pi^2\log(4\zcut)-\gamma_{S,C_F}^{(1)}\log(4\zcut)-8\log\zcut\log(16\zcut)\right.\right.\nonumber\\
&\left.\left.
\hspace{2cm}
-\frac{8}{3}\pi^2\log^2\zcut+192\log^2 2\log\zcut+48\log2\log^2\zcut+128\log^2 2\log^2\zcut\right.\right.\nonumber\\
&
\hspace{2cm}
+64\log2\log^3\zcut+8\log^4\zcut
\biggr)\nonumber\\
&
\hspace{1.5cm}
+C_F C_A\left(
\frac{493}{81}+\frac{497}{54}\pi^2-\frac{53}{90}\pi^4+\frac{4694}{27}\log 2-\frac{22}{9}\pi^2\log2+44\log^2 2+\frac{302}{9}\zeta_3\right.\nonumber\\
&\left.
\hspace{2cm}
-104\zeta_3\log 2+c_{S,C_A}^{(2)}+2c_{S_c,C_A}^{(2)}-\gamma_{S,C_A}^{(1)}\log(4\zcut)+\frac{808}{27}\log\zcut\right.\nonumber\\
&\left.
\hspace{2cm}
+\frac{1072}{9}\log2\log\zcut+\frac{11}{3}\pi^2\log\zcut-28\zeta_3\log\zcut -\frac{4}{3}\pi^2\log\zcut\log(16\zcut)\right.\nonumber\\
&\left.
\hspace{2cm}
+\frac{268}{9}\log^2\zcut+\frac{176}{3}\log^2 2\log\zcut-\frac{44}{3}\log^2\zcut\log(4\zcut)
\right)\nonumber\\
&\left.
\hspace{1.5cm}
+C_F n_f T_R\left(
\frac{28}{81}-\frac{86}{27}\pi^2-\frac{1528}{27}\log 2+\frac{8}{9}\pi^2\log 2-16\log^2 2+\frac{8}{9}\zeta_3\right.\right.\nonumber\\
&\left.\left.
\hspace{2cm}
+c_{S,n_f}^{(2)}+2c_{S_c,n_f}^{(2)}-\gamma_{S,n_f}^{(1)}\log(4\zcut)-\frac{224}{27}\log\zcut-\frac{4}{3}\pi^2\log\zcut\right.\right.\nonumber\\
&\left.\left.
\hspace{2cm}
-\frac{64}{3}\log^2 2\log\zcut-\frac{80}{9}\log\zcut\log(16\zcut)+\frac{16}{3}\log^2\zcut\log(4\zcut)
\right)
\right]\nonumber\,.
\end{align}
In this expression, we have left the two-loop soft and collinear-soft function constants implicit, as well as the two-loop non-cusp anomalous dimension of the soft function.  In the subscripts of these terms, we have identified the corresponding color channel, which is just the coefficient of that term in the equations \ref{eq:soft_anom}, \ref{eq:soft_const}, and \ref{eq:csoft_const}.

The terms with support away from $\rho = 0$ are:
\begin{align}
D_{A,\rg}(\rho) &= \frac{1}{\rho}\left[C_F\left(-3-4\log\zcut\right)\right]\,,\\
D_{B,\rg}(\rho) &= \frac{1}{\rho}\left[C_F^2\bigg(\left(3+4\log\zcut\right)^2\log\rho-18\log2+4\log\zcut-48\log 2\log\zcut\right.\nonumber\\
&\left.
\hspace{2cm}-6\log^2\zcut-32\log2\log^2\zcut-8\log^3\zcut+\frac{9}{4}+\pi^2-12\zeta_3+\frac{\gamma_{S,C_F}^{(1)}}{4}\right)\nonumber\\
&
\hspace{1cm}
+C_F C_A\left(
\frac{11}{6}(3+4\log\zcut)\log\rho-11\log 2+\left(
-\frac{134}{9}+\frac{2}{3}\pi^2
\right)\log\zcut\right.\nonumber\\
&
\left.\left.
\hspace{2cm}+\frac{11}{3}\log\frac{\zcut}{16}\log\zcut-\frac{2347}{108}+\frac{11}{36}\pi^2
+13\zeta_3+\frac{\gamma_{S,C_A}^{(1)}}{4}\right)\right.\nonumber\\
&\left.
\hspace{1cm}
+C_F n_f T_R\left(
-\frac{2}{3}(3+4\log\zcut)
\log\rho+4\log2+\frac{40}{9}\log\zcut+\frac{16}{3}\log2\log\zcut\right.\right.\nonumber\\
&\left.\left.
\hspace{2cm}-\frac{4}{3}\log^2\zcut+\frac{191}{27}-\frac{\pi^2}{9}+\frac{\gamma_{S,n_F}^{(1)}}{4}\right)
\right]\,.\nonumber
\end{align}
The $D_{C,\rg}$ coefficient is extremely unieldy, so we only quote here the approximate numerical function:
\begin{align}\label{eq:DC}
D_{C,\rg} &\simeq \frac{1}{\rho}\bigg[\left(
-75.85\log^3\zcut-334.22\log^2\zcut-451.70 \log\zcut-182.78
\right)\log^2\rho\\
&
\hspace{0cm}
+\left(75.85\log^4\zcut+269.56\log^3\zcut+1008.64\log^2\zcut+1762.95\log\zcut+877.52
\right)\log\rho\nonumber\\
&
\hspace{0cm}
-18.96\log^5\zcut-37.59\log^4\zcut-230.06\log^3\zcut-724.49\log^2\zcut-1641.62\log\zcut\nonumber\\
&
\hspace{0cm}-1944.97+\frac{\gamma_S^{(2)}}{16}\bigg]\,.
\nonumber
\end{align}
Here we substituted explicitly the color factors of QCD ($C_F = 4/3$, $C_A = 3$, $T_R = 1/2$), and set the number of active quarks to $n_f = 5$.  We have also just used the central values of the collinear-soft constants, with no accounting for their uncertainties.

\section{\label{app:mccsm}MCCSM Data at ${\cal O}(\as^3)$}

Generating the data from {\tt MCCSM} for the extraction of the three-loop non-cusp anomalous dimension of the soft function $\gamma_S^{(2)}$ took about a century of CPU time.  For faster evaluation later, we provide the mMDT groomed heavy hemisphere distributions here, with $\zcut = 0.04,0.06,0.08,0.1$.  For the differential cross section of the mMDT groomed heavy hemisphere mass, {\tt MCCSM} calculates the $\rho$-dependent factors $A$, $B$, and $C$ at leading, next-to-leading, and next-to-next-to-leading fixed order, respectively.  These can then be combined to evaluate the cross section through ${\cal O}(\as^3)$ as:
\begin{align}\label{eq:mccsm_xsec_gen}
\frac{\rho}{\sigma_0}\frac{\rd\sigma_{\text{FO}}}{\rd\rho} &= \frac{\as}{2\pi}A+\left(
\frac{\as}{2\pi}
\right)^2\left[
B+A\beta_0 \log\frac{\mu}{Q}
\right]\\
&
\hspace{2cm}+\left(
\frac{\as}{2\pi}
\right)^3\left[
C+2B\beta_0 \log\frac{\mu}{Q}+A\left(\frac{\beta_1}{2} \log\frac{\mu}{Q}+\beta_0^2\log^2\frac{\mu}{Q}
\right)
\right]\,.\nonumber
\end{align}
Here, $\as = \as(\mu)$, the strong coupling evaluated at the renormalization scale $\mu$, $\beta_0$ and $\beta_1$ are the QCD $\beta$-function coefficients, and $Q$ is the center-of-mass collision energy.  When $\mu=Q$, the terms proportional to the $\beta$-function vanish, and the expression simplifies.  To extract $\gamma_S^{(2)}$, we need the distribution of $C$ as a function of $\rho$.  The value of $C$ for $\log \rho\in[-9.5,-0.5]$ in steps of 1 and all $\zcut$ values we consider are listed in \Tab{tab:mccsmtab}.  The entries in this table were determined using the filtering and weighting procedure described in \Sec{sec:threeloop}.

\begin{table}
\begin{center}
\begin{tabular}{c| c c c c c}
 $\log\rho$\ \ & \ \ $\zcut = 0.04$\ \  & \ \ $\zcut = 0.06$ \ \ &\ \  $\zcut = 0.08$ \ \ & \ \ $\zcut = 0.1$ \\
 \hline
$-9.5$\ \ &\ \  $-8495\pm922$\ \  &\ \  $-9361\pm831$\ \   & \ \ $-7872\pm804$\ \  & \ \ $-6543\pm762$  \\
$-8.5$\ \ &\ \  $-10127\pm519$\ \  &\ \  $-9529\pm489$\ \   & \ \ $-7835\pm456$\ \  & \ \ $-5973\pm444$  \\
$-7.5$\ \ &\ \  $-11742\pm271$\ \  &\ \  $-10185\pm253$\ \   & \ \ $-8245\pm235$\ \  & \ \ $-6553\pm218$  \\
$-6.5$\ \ &\ \  $-11435\pm147$\ \  &\ \  $-9183\pm134$\ \   & \ \ $-7256\pm120$\ \  & \ \ $-5653\pm112$  \\
$-5.5$\ \ &\ \  $-10701\pm 76$\ \  &\ \  $-8040\pm70$\ \   & \ \ $-6089\pm60$\ \  & \ \ $-4560\pm56$  \\
$-4.5$\ \ &\ \  $-9293\pm41$\ \  &\ \  $-6615\pm35$\ \   & \ \ $-4817\pm 32$\ \  & \ \ $-3506\pm29$  \\
$-3.5$\ \ &\ \  $-7127\pm21$\ \  &\ \  $-4892\pm18$\ \   & \ \ $-3429\pm 16$\ \  & \ \ $-2390\pm15$  \\
$-2.5$\ \ &\ \  $-2475\pm 12$\ \  &\ \  $-2247\pm 10$\ \   & \ \ $-1642\pm 9$\ \  & \ \ $-1175\pm 8$  \\
$-1.5$\ \ &\ \  $2803\pm5$\ \  &\ \  $2334\pm4$\ \   & \ \ $1781\pm 4$\ \  & \ \ $1442\pm 4$  \\
$-0.5$\ \ &\ \  $1608.0\pm 1.2$\ \  &\ \  $1593.8\pm 1.1$\ \   & \ \ $1574.6\pm 1.1$\ \  & \ \ $1548.7\pm 1.1$  \\
\end{tabular}
\end{center}
\caption{Table of values with uncertainties of the $\as^3$ factor $C$ in the form of the differential cross section of \Eq{eq:mccsm_xsec_gen} for the mMDT groomed heavy hemisphere mass and the quoted value of $\zcut$.  These values correspond to five active quarks, $n_f = 5$.}\label{tab:mccsmtab}
\end{table}

\bibliography{SoftDropNNNLL}

\providecommand{\noopsort}[1]{}\providecommand{\singleletter}[1]{#1}%
\begin{thebibliography}{97}%
\makeatletter
\providecommand \@ifxundefined [1]{%
 \@ifx{#1\undefined}
}%
\providecommand \@ifnum [1]{%
 \ifnum #1\expandafter \@firstoftwo
 \else \expandafter \@secondoftwo
 \fi
}%
\providecommand \@ifx [1]{%
 \ifx #1\expandafter \@firstoftwo
 \else \expandafter \@secondoftwo
 \fi
}%
\providecommand \natexlab [1]{#1}%
\providecommand \enquote  [1]{``#1''}%
\providecommand \bibnamefont  [1]{#1}%
\providecommand \bibfnamefont [1]{#1}%
\providecommand \citenamefont [1]{#1}%
\providecommand \href@noop [0]{\@secondoftwo}%
\providecommand \href [0]{\begingroup \@sanitize@url \@href}%
\providecommand \@href[1]{\@@startlink{#1}\@@href}%
\providecommand \@@href[1]{\endgroup#1\@@endlink}%
\providecommand \@sanitize@url [0]{\catcode `\\12\catcode `\$12\catcode
  `\&12\catcode `\#12\catcode `\^12\catcode `\_12\catcode `\%12\relax}%
\providecommand \@@startlink[1]{}%
\providecommand \@@endlink[0]{}%
\providecommand \url  [0]{\begingroup\@sanitize@url \@url }%
\providecommand \@url [1]{\endgroup\@href {#1}{\urlprefix }}%
\providecommand \urlprefix  [0]{URL }%
\providecommand \Eprint [0]{\href }%
\providecommand \doibase [0]{http://dx.doi.org/}%
\providecommand \selectlanguage [0]{\@gobble}%
\providecommand \bibinfo  [0]{\@secondoftwo}%
\providecommand \bibfield  [0]{\@secondoftwo}%
\providecommand \translation [1]{[#1]}%
\providecommand \BibitemOpen [0]{}%
\providecommand \bibitemStop [0]{}%
\providecommand \bibitemNoStop [0]{.\EOS\space}%
\providecommand \EOS [0]{\spacefactor3000\relax}%
\providecommand \BibitemShut  [1]{\csname bibitem#1\endcsname}%
\let\auto@bib@innerbib\@empty
\bibitem [{\citenamefont {Larkoski}\ \emph
  {et~al.}(2017{\natexlab{a}})\citenamefont {Larkoski}, \citenamefont {Moult},\
  and\ \citenamefont {Nachman}}]{Larkoski:2017jix}%
  \BibitemOpen
  \bibfield  {author} {\bibinfo {author} {\bibfnamefont {A.~J.}\ \bibnamefont
  {Larkoski}}, \bibinfo {author} {\bibfnamefont {I.}~\bibnamefont {Moult}}, \
  and\ \bibinfo {author} {\bibfnamefont {B.}~\bibnamefont {Nachman}},\ }\href
  {\doibase 10.1016/j.physrep.2019.11.001} {\  (\bibinfo {year}
  {2017}{\natexlab{a}}),\ 10.1016/j.physrep.2019.11.001},\ \Eprint
  {http://arxiv.org/abs/1709.04464} {arXiv:1709.04464 [hep-ph]} \BibitemShut
  {NoStop}%
\bibitem [{\citenamefont {Kogler}\ \emph {et~al.}(2019)\citenamefont {Kogler}
  \emph {et~al.}}]{Asquith:2018igt}%
  \BibitemOpen
  \bibfield  {author} {\bibinfo {author} {\bibfnamefont {R.}~\bibnamefont
  {Kogler}} \emph {et~al.},\ }\href {\doibase 10.1103/RevModPhys.91.045003}
  {\bibfield  {journal} {\bibinfo  {journal} {Rev. Mod. Phys.}\ }\textbf
  {\bibinfo {volume} {91}},\ \bibinfo {pages} {045003} (\bibinfo {year}
  {2019})},\ \Eprint {http://arxiv.org/abs/1803.06991} {arXiv:1803.06991
  [hep-ex]} \BibitemShut {NoStop}%
\bibitem [{\citenamefont {Dasgupta}\ \emph
  {et~al.}(2013{\natexlab{a}})\citenamefont {Dasgupta}, \citenamefont
  {Fregoso}, \citenamefont {Marzani},\ and\ \citenamefont
  {Salam}}]{Dasgupta:2013ihk}%
  \BibitemOpen
  \bibfield  {author} {\bibinfo {author} {\bibfnamefont {M.}~\bibnamefont
  {Dasgupta}}, \bibinfo {author} {\bibfnamefont {A.}~\bibnamefont {Fregoso}},
  \bibinfo {author} {\bibfnamefont {S.}~\bibnamefont {Marzani}}, \ and\
  \bibinfo {author} {\bibfnamefont {G.~P.}\ \bibnamefont {Salam}},\ }\href
  {\doibase 10.1007/JHEP09(2013)029} {\bibfield  {journal} {\bibinfo  {journal}
  {JHEP}\ }\textbf {\bibinfo {volume} {09}},\ \bibinfo {pages} {029} (\bibinfo
  {year} {2013}{\natexlab{a}})},\ \Eprint {http://arxiv.org/abs/1307.0007}
  {arXiv:1307.0007 [hep-ph]} \BibitemShut {NoStop}%
\bibitem [{\citenamefont {Dasgupta}\ \emph
  {et~al.}(2013{\natexlab{b}})\citenamefont {Dasgupta}, \citenamefont
  {Fregoso}, \citenamefont {Marzani},\ and\ \citenamefont
  {Powling}}]{Dasgupta:2013via}%
  \BibitemOpen
  \bibfield  {author} {\bibinfo {author} {\bibfnamefont {M.}~\bibnamefont
  {Dasgupta}}, \bibinfo {author} {\bibfnamefont {A.}~\bibnamefont {Fregoso}},
  \bibinfo {author} {\bibfnamefont {S.}~\bibnamefont {Marzani}}, \ and\
  \bibinfo {author} {\bibfnamefont {A.}~\bibnamefont {Powling}},\ }\href
  {\doibase 10.1140/epjc/s10052-013-2623-3} {\bibfield  {journal} {\bibinfo
  {journal} {Eur. Phys. J.}\ }\textbf {\bibinfo {volume} {C73}},\ \bibinfo
  {pages} {2623} (\bibinfo {year} {2013}{\natexlab{b}})},\ \Eprint
  {http://arxiv.org/abs/1307.0013} {arXiv:1307.0013 [hep-ph]} \BibitemShut
  {NoStop}%
\bibitem [{\citenamefont {Larkoski}\ \emph
  {et~al.}(2014{\natexlab{a}})\citenamefont {Larkoski}, \citenamefont
  {Marzani}, \citenamefont {Soyez},\ and\ \citenamefont
  {Thaler}}]{Larkoski:2014wba}%
  \BibitemOpen
  \bibfield  {author} {\bibinfo {author} {\bibfnamefont {A.~J.}\ \bibnamefont
  {Larkoski}}, \bibinfo {author} {\bibfnamefont {S.}~\bibnamefont {Marzani}},
  \bibinfo {author} {\bibfnamefont {G.}~\bibnamefont {Soyez}}, \ and\ \bibinfo
  {author} {\bibfnamefont {J.}~\bibnamefont {Thaler}},\ }\href {\doibase
  10.1007/JHEP05(2014)146} {\bibfield  {journal} {\bibinfo  {journal} {JHEP}\
  }\textbf {\bibinfo {volume} {05}},\ \bibinfo {pages} {146} (\bibinfo {year}
  {2014}{\natexlab{a}})},\ \Eprint {http://arxiv.org/abs/1402.2657}
  {arXiv:1402.2657 [hep-ph]} \BibitemShut {NoStop}%
\bibitem [{\citenamefont {Dasgupta}\ and\ \citenamefont
  {Salam}(2001)}]{Dasgupta:2001sh}%
  \BibitemOpen
  \bibfield  {author} {\bibinfo {author} {\bibfnamefont {M.}~\bibnamefont
  {Dasgupta}}\ and\ \bibinfo {author} {\bibfnamefont {G.~P.}\ \bibnamefont
  {Salam}},\ }\href {\doibase 10.1016/S0370-2693(01)00725-0} {\bibfield
  {journal} {\bibinfo  {journal} {Phys. Lett.}\ }\textbf {\bibinfo {volume}
  {B512}},\ \bibinfo {pages} {323} (\bibinfo {year} {2001})},\ \Eprint
  {http://arxiv.org/abs/hep-ph/0104277} {arXiv:hep-ph/0104277 [hep-ph]}
  \BibitemShut {NoStop}%
\bibitem [{\citenamefont {Frye}\ \emph
  {et~al.}(2016{\natexlab{a}})\citenamefont {Frye}, \citenamefont {Larkoski},
  \citenamefont {Schwartz},\ and\ \citenamefont {Yan}}]{Frye:2016okc}%
  \BibitemOpen
  \bibfield  {author} {\bibinfo {author} {\bibfnamefont {C.}~\bibnamefont
  {Frye}}, \bibinfo {author} {\bibfnamefont {A.~J.}\ \bibnamefont {Larkoski}},
  \bibinfo {author} {\bibfnamefont {M.~D.}\ \bibnamefont {Schwartz}}, \ and\
  \bibinfo {author} {\bibfnamefont {K.}~\bibnamefont {Yan}},\ }\href@noop {} {\
   (\bibinfo {year} {2016}{\natexlab{a}})},\ \Eprint
  {http://arxiv.org/abs/1603.06375} {arXiv:1603.06375 [hep-ph]} \BibitemShut
  {NoStop}%
\bibitem [{\citenamefont {Frye}\ \emph
  {et~al.}(2016{\natexlab{b}})\citenamefont {Frye}, \citenamefont {Larkoski},
  \citenamefont {Schwartz},\ and\ \citenamefont {Yan}}]{Frye:2016aiz}%
  \BibitemOpen
  \bibfield  {author} {\bibinfo {author} {\bibfnamefont {C.}~\bibnamefont
  {Frye}}, \bibinfo {author} {\bibfnamefont {A.~J.}\ \bibnamefont {Larkoski}},
  \bibinfo {author} {\bibfnamefont {M.~D.}\ \bibnamefont {Schwartz}}, \ and\
  \bibinfo {author} {\bibfnamefont {K.}~\bibnamefont {Yan}},\ }\href {\doibase
  10.1007/JHEP07(2016)064} {\bibfield  {journal} {\bibinfo  {journal} {JHEP}\
  }\textbf {\bibinfo {volume} {07}},\ \bibinfo {pages} {064} (\bibinfo {year}
  {2016}{\natexlab{b}})},\ \Eprint {http://arxiv.org/abs/1603.09338}
  {arXiv:1603.09338 [hep-ph]} \BibitemShut {NoStop}%
\bibitem [{\citenamefont {Marzani}\ \emph {et~al.}(2017)\citenamefont
  {Marzani}, \citenamefont {Schunk},\ and\ \citenamefont
  {Soyez}}]{Marzani:2017mva}%
  \BibitemOpen
  \bibfield  {author} {\bibinfo {author} {\bibfnamefont {S.}~\bibnamefont
  {Marzani}}, \bibinfo {author} {\bibfnamefont {L.}~\bibnamefont {Schunk}}, \
  and\ \bibinfo {author} {\bibfnamefont {G.}~\bibnamefont {Soyez}},\ }\href
  {\doibase 10.1007/JHEP07(2017)132} {\bibfield  {journal} {\bibinfo  {journal}
  {JHEP}\ }\textbf {\bibinfo {volume} {07}},\ \bibinfo {pages} {132} (\bibinfo
  {year} {2017})},\ \Eprint {http://arxiv.org/abs/1704.02210} {arXiv:1704.02210
  [hep-ph]} \BibitemShut {NoStop}%
\bibitem [{\citenamefont {Marzani}\ \emph {et~al.}(2018)\citenamefont
  {Marzani}, \citenamefont {Schunk},\ and\ \citenamefont
  {Soyez}}]{Marzani:2017kqd}%
  \BibitemOpen
  \bibfield  {author} {\bibinfo {author} {\bibfnamefont {S.}~\bibnamefont
  {Marzani}}, \bibinfo {author} {\bibfnamefont {L.}~\bibnamefont {Schunk}}, \
  and\ \bibinfo {author} {\bibfnamefont {G.}~\bibnamefont {Soyez}},\ }\href
  {\doibase 10.1140/epjc/s10052-018-5579-5} {\bibfield  {journal} {\bibinfo
  {journal} {Eur. Phys. J.}\ }\textbf {\bibinfo {volume} {C78}},\ \bibinfo
  {pages} {96} (\bibinfo {year} {2018})},\ \Eprint
  {http://arxiv.org/abs/1712.05105} {arXiv:1712.05105 [hep-ph]} \BibitemShut
  {NoStop}%
\bibitem [{\citenamefont {Makris}\ \emph {et~al.}(2018)\citenamefont {Makris},
  \citenamefont {Neill},\ and\ \citenamefont {Vaidya}}]{Makris:2017arq}%
  \BibitemOpen
  \bibfield  {author} {\bibinfo {author} {\bibfnamefont {Y.}~\bibnamefont
  {Makris}}, \bibinfo {author} {\bibfnamefont {D.}~\bibnamefont {Neill}}, \
  and\ \bibinfo {author} {\bibfnamefont {V.}~\bibnamefont {Vaidya}},\ }\href
  {\doibase 10.1007/JHEP07(2018)167} {\bibfield  {journal} {\bibinfo  {journal}
  {JHEP}\ }\textbf {\bibinfo {volume} {07}},\ \bibinfo {pages} {167} (\bibinfo
  {year} {2018})},\ \Eprint {http://arxiv.org/abs/1712.07653} {arXiv:1712.07653
  [hep-ph]} \BibitemShut {NoStop}%
\bibitem [{\citenamefont {Kang}\ \emph {et~al.}(2018)\citenamefont {Kang},
  \citenamefont {Lee}, \citenamefont {Liu},\ and\ \citenamefont
  {Ringer}}]{Kang:2018jwa}%
  \BibitemOpen
  \bibfield  {author} {\bibinfo {author} {\bibfnamefont {Z.-B.}\ \bibnamefont
  {Kang}}, \bibinfo {author} {\bibfnamefont {K.}~\bibnamefont {Lee}}, \bibinfo
  {author} {\bibfnamefont {X.}~\bibnamefont {Liu}}, \ and\ \bibinfo {author}
  {\bibfnamefont {F.}~\bibnamefont {Ringer}},\ }\href {\doibase
  10.1007/JHEP10(2018)137} {\bibfield  {journal} {\bibinfo  {journal} {JHEP}\
  }\textbf {\bibinfo {volume} {10}},\ \bibinfo {pages} {137} (\bibinfo {year}
  {2018})},\ \Eprint {http://arxiv.org/abs/1803.03645} {arXiv:1803.03645
  [hep-ph]} \BibitemShut {NoStop}%
\bibitem [{\citenamefont {Chay}\ and\ \citenamefont
  {Kim}(2019)}]{Chay:2018pvp}%
  \BibitemOpen
  \bibfield  {author} {\bibinfo {author} {\bibfnamefont {J.}~\bibnamefont
  {Chay}}\ and\ \bibinfo {author} {\bibfnamefont {C.}~\bibnamefont {Kim}},\
  }\href {\doibase 10.3938/jkps.74.439} {\bibfield  {journal} {\bibinfo
  {journal} {J. Korean Phys. Soc.}\ }\textbf {\bibinfo {volume} {74}},\
  \bibinfo {pages} {439} (\bibinfo {year} {2019})},\ \Eprint
  {http://arxiv.org/abs/1806.01712} {arXiv:1806.01712 [hep-ph]} \BibitemShut
  {NoStop}%
\bibitem [{\citenamefont {Baron}\ \emph {et~al.}(2018)\citenamefont {Baron},
  \citenamefont {Marzani},\ and\ \citenamefont {Theeuwes}}]{Baron:2018nfz}%
  \BibitemOpen
  \bibfield  {author} {\bibinfo {author} {\bibfnamefont {J.}~\bibnamefont
  {Baron}}, \bibinfo {author} {\bibfnamefont {S.}~\bibnamefont {Marzani}}, \
  and\ \bibinfo {author} {\bibfnamefont {V.}~\bibnamefont {Theeuwes}},\ }\href
  {\doibase 10.1007/JHEP08(2018)105, 10.1007/JHEP05(2019)056} {\bibfield
  {journal} {\bibinfo  {journal} {JHEP}\ }\textbf {\bibinfo {volume} {08}},\
  \bibinfo {pages} {105} (\bibinfo {year} {2018})},\ \bibinfo {note} {[erratum:
  JHEP05,056(2019)]},\ \Eprint {http://arxiv.org/abs/1803.04719}
  {arXiv:1803.04719 [hep-ph]} \BibitemShut {NoStop}%
\bibitem [{\citenamefont {Makris}\ and\ \citenamefont
  {Vaidya}(2018)}]{Makris:2018npl}%
  \BibitemOpen
  \bibfield  {author} {\bibinfo {author} {\bibfnamefont {Y.}~\bibnamefont
  {Makris}}\ and\ \bibinfo {author} {\bibfnamefont {V.}~\bibnamefont
  {Vaidya}},\ }\href {\doibase 10.1007/JHEP10(2018)019} {\bibfield  {journal}
  {\bibinfo  {journal} {JHEP}\ }\textbf {\bibinfo {volume} {10}},\ \bibinfo
  {pages} {019} (\bibinfo {year} {2018})},\ \Eprint
  {http://arxiv.org/abs/1807.09805} {arXiv:1807.09805 [hep-ph]} \BibitemShut
  {NoStop}%
\bibitem [{\citenamefont {Kang}\ \emph
  {et~al.}(2019{\natexlab{a}})\citenamefont {Kang}, \citenamefont {Lee},
  \citenamefont {Liu},\ and\ \citenamefont {Ringer}}]{Kang:2018vgn}%
  \BibitemOpen
  \bibfield  {author} {\bibinfo {author} {\bibfnamefont {Z.-B.}\ \bibnamefont
  {Kang}}, \bibinfo {author} {\bibfnamefont {K.}~\bibnamefont {Lee}}, \bibinfo
  {author} {\bibfnamefont {X.}~\bibnamefont {Liu}}, \ and\ \bibinfo {author}
  {\bibfnamefont {F.}~\bibnamefont {Ringer}},\ }\href {\doibase
  10.1016/j.physletb.2019.04.018} {\bibfield  {journal} {\bibinfo  {journal}
  {Phys. Lett.}\ }\textbf {\bibinfo {volume} {B793}},\ \bibinfo {pages} {41}
  (\bibinfo {year} {2019}{\natexlab{a}})},\ \Eprint
  {http://arxiv.org/abs/1811.06983} {arXiv:1811.06983 [hep-ph]} \BibitemShut
  {NoStop}%
\bibitem [{\citenamefont {Lee}\ \emph {et~al.}(2019{\natexlab{a}})\citenamefont
  {Lee}, \citenamefont {Shrivastava},\ and\ \citenamefont
  {Vaidya}}]{Lee:2019lge}%
  \BibitemOpen
  \bibfield  {author} {\bibinfo {author} {\bibfnamefont {C.}~\bibnamefont
  {Lee}}, \bibinfo {author} {\bibfnamefont {P.}~\bibnamefont {Shrivastava}}, \
  and\ \bibinfo {author} {\bibfnamefont {V.}~\bibnamefont {Vaidya}},\ }\href
  {\doibase 10.1007/JHEP09(2019)045} {\bibfield  {journal} {\bibinfo  {journal}
  {JHEP}\ }\textbf {\bibinfo {volume} {09}},\ \bibinfo {pages} {045} (\bibinfo
  {year} {2019}{\natexlab{a}})},\ \Eprint {http://arxiv.org/abs/1901.09095}
  {arXiv:1901.09095 [hep-ph]} \BibitemShut {NoStop}%
\bibitem [{\citenamefont {Marzani}\ \emph {et~al.}(2019)\citenamefont
  {Marzani}, \citenamefont {Reichelt}, \citenamefont {Schumann}, \citenamefont
  {Soyez},\ and\ \citenamefont {Theeuwes}}]{Marzani:2019evv}%
  \BibitemOpen
  \bibfield  {author} {\bibinfo {author} {\bibfnamefont {S.}~\bibnamefont
  {Marzani}}, \bibinfo {author} {\bibfnamefont {D.}~\bibnamefont {Reichelt}},
  \bibinfo {author} {\bibfnamefont {S.}~\bibnamefont {Schumann}}, \bibinfo
  {author} {\bibfnamefont {G.}~\bibnamefont {Soyez}}, \ and\ \bibinfo {author}
  {\bibfnamefont {V.}~\bibnamefont {Theeuwes}},\ }\href {\doibase
  10.1007/JHEP11(2019)179} {\bibfield  {journal} {\bibinfo  {journal} {JHEP}\
  }\textbf {\bibinfo {volume} {11}},\ \bibinfo {pages} {179} (\bibinfo {year}
  {2019})},\ \Eprint {http://arxiv.org/abs/1906.10504} {arXiv:1906.10504
  [hep-ph]} \BibitemShut {NoStop}%
\bibitem [{\citenamefont {Gutierrez-Reyes}\ \emph {et~al.}(2019)\citenamefont
  {Gutierrez-Reyes}, \citenamefont {Makris}, \citenamefont {Vaidya},
  \citenamefont {Scimemi},\ and\ \citenamefont
  {Zoppi}}]{Gutierrez-Reyes:2019msa}%
  \BibitemOpen
  \bibfield  {author} {\bibinfo {author} {\bibfnamefont {D.}~\bibnamefont
  {Gutierrez-Reyes}}, \bibinfo {author} {\bibfnamefont {Y.}~\bibnamefont
  {Makris}}, \bibinfo {author} {\bibfnamefont {V.}~\bibnamefont {Vaidya}},
  \bibinfo {author} {\bibfnamefont {I.}~\bibnamefont {Scimemi}}, \ and\
  \bibinfo {author} {\bibfnamefont {L.}~\bibnamefont {Zoppi}},\ }\href
  {\doibase 10.1007/JHEP08(2019)161} {\bibfield  {journal} {\bibinfo  {journal}
  {JHEP}\ }\textbf {\bibinfo {volume} {08}},\ \bibinfo {pages} {161} (\bibinfo
  {year} {2019})},\ \Eprint {http://arxiv.org/abs/1907.05896} {arXiv:1907.05896
  [hep-ph]} \BibitemShut {NoStop}%
\bibitem [{\citenamefont {Cal}\ \emph {et~al.}(2019)\citenamefont {Cal},
  \citenamefont {Neill}, \citenamefont {Ringer},\ and\ \citenamefont
  {Waalewijn}}]{Cal:2019gxa}%
  \BibitemOpen
  \bibfield  {author} {\bibinfo {author} {\bibfnamefont {P.}~\bibnamefont
  {Cal}}, \bibinfo {author} {\bibfnamefont {D.}~\bibnamefont {Neill}}, \bibinfo
  {author} {\bibfnamefont {F.}~\bibnamefont {Ringer}}, \ and\ \bibinfo {author}
  {\bibfnamefont {W.~J.}\ \bibnamefont {Waalewijn}},\ }\href@noop {} {\
  (\bibinfo {year} {2019})},\ \Eprint {http://arxiv.org/abs/1911.06840}
  {arXiv:1911.06840 [hep-ph]} \BibitemShut {NoStop}%
\bibitem [{\citenamefont {Kang}\ \emph
  {et~al.}(2019{\natexlab{b}})\citenamefont {Kang}, \citenamefont {Lee},
  \citenamefont {Liu}, \citenamefont {Neill},\ and\ \citenamefont
  {Ringer}}]{Kang:2019prh}%
  \BibitemOpen
  \bibfield  {author} {\bibinfo {author} {\bibfnamefont {Z.-B.}\ \bibnamefont
  {Kang}}, \bibinfo {author} {\bibfnamefont {K.}~\bibnamefont {Lee}}, \bibinfo
  {author} {\bibfnamefont {X.}~\bibnamefont {Liu}}, \bibinfo {author}
  {\bibfnamefont {D.}~\bibnamefont {Neill}}, \ and\ \bibinfo {author}
  {\bibfnamefont {F.}~\bibnamefont {Ringer}},\ }\href@noop {} {\  (\bibinfo
  {year} {2019}{\natexlab{b}})},\ \Eprint {http://arxiv.org/abs/1908.01783}
  {arXiv:1908.01783 [hep-ph]} \BibitemShut {NoStop}%
\bibitem [{\citenamefont {Aaboud}\ \emph {et~al.}(2018)\citenamefont {Aaboud}
  \emph {et~al.}}]{Aaboud:2017qwh}%
  \BibitemOpen
  \bibfield  {author} {\bibinfo {author} {\bibfnamefont {M.}~\bibnamefont
  {Aaboud}} \emph {et~al.} (\bibinfo {collaboration} {ATLAS}),\ }\href
  {\doibase 10.1103/PhysRevLett.121.092001} {\bibfield  {journal} {\bibinfo
  {journal} {Phys. Rev. Lett.}\ }\textbf {\bibinfo {volume} {121}},\ \bibinfo
  {pages} {092001} (\bibinfo {year} {2018})},\ \Eprint
  {http://arxiv.org/abs/1711.08341} {arXiv:1711.08341 [hep-ex]} \BibitemShut
  {NoStop}%
\bibitem [{\citenamefont {Sirunyan}\ \emph {et~al.}(2018)\citenamefont
  {Sirunyan} \emph {et~al.}}]{Sirunyan:2018xdh}%
  \BibitemOpen
  \bibfield  {author} {\bibinfo {author} {\bibfnamefont {A.~M.}\ \bibnamefont
  {Sirunyan}} \emph {et~al.} (\bibinfo {collaboration} {CMS}),\ }\href
  {\doibase 10.1007/JHEP11(2018)113} {\bibfield  {journal} {\bibinfo  {journal}
  {JHEP}\ }\textbf {\bibinfo {volume} {11}},\ \bibinfo {pages} {113} (\bibinfo
  {year} {2018})},\ \Eprint {http://arxiv.org/abs/1807.05974} {arXiv:1807.05974
  [hep-ex]} \BibitemShut {NoStop}%
\bibitem [{\citenamefont {Aad}\ \emph {et~al.}(2019)\citenamefont {Aad} \emph
  {et~al.}}]{Aad:2019vyi}%
  \BibitemOpen
  \bibfield  {author} {\bibinfo {author} {\bibfnamefont {G.}~\bibnamefont
  {Aad}} \emph {et~al.} (\bibinfo {collaboration} {ATLAS}),\ }\href@noop {} {\
  (\bibinfo {year} {2019})},\ \Eprint {http://arxiv.org/abs/1912.09837}
  {arXiv:1912.09837 [hep-ex]} \BibitemShut {NoStop}%
\bibitem [{\citenamefont {Salam}\ \emph {et~al.}(2017)\citenamefont {Salam},
  \citenamefont {Schunk},\ and\ \citenamefont {Soyez}}]{Salam:2016yht}%
  \BibitemOpen
  \bibfield  {author} {\bibinfo {author} {\bibfnamefont {G.~P.}\ \bibnamefont
  {Salam}}, \bibinfo {author} {\bibfnamefont {L.}~\bibnamefont {Schunk}}, \
  and\ \bibinfo {author} {\bibfnamefont {G.}~\bibnamefont {Soyez}},\ }\href
  {\doibase 10.1007/JHEP03(2017)022} {\bibfield  {journal} {\bibinfo  {journal}
  {JHEP}\ }\textbf {\bibinfo {volume} {03}},\ \bibinfo {pages} {022} (\bibinfo
  {year} {2017})},\ \Eprint {http://arxiv.org/abs/1612.03917} {arXiv:1612.03917
  [hep-ph]} \BibitemShut {NoStop}%
\bibitem [{\citenamefont {Larkoski}\ \emph
  {et~al.}(2017{\natexlab{b}})\citenamefont {Larkoski}, \citenamefont {Moult},\
  and\ \citenamefont {Neill}}]{Larkoski:2017iuy}%
  \BibitemOpen
  \bibfield  {author} {\bibinfo {author} {\bibfnamefont {A.~J.}\ \bibnamefont
  {Larkoski}}, \bibinfo {author} {\bibfnamefont {I.}~\bibnamefont {Moult}}, \
  and\ \bibinfo {author} {\bibfnamefont {D.}~\bibnamefont {Neill}},\
  }\href@noop {} {\  (\bibinfo {year} {2017}{\natexlab{b}})},\ \Eprint
  {http://arxiv.org/abs/1708.06760} {arXiv:1708.06760 [hep-ph]} \BibitemShut
  {NoStop}%
\bibitem [{\citenamefont {Larkoski}\ \emph {et~al.}(2018)\citenamefont
  {Larkoski}, \citenamefont {Moult},\ and\ \citenamefont
  {Neill}}]{Larkoski:2017cqq}%
  \BibitemOpen
  \bibfield  {author} {\bibinfo {author} {\bibfnamefont {A.~J.}\ \bibnamefont
  {Larkoski}}, \bibinfo {author} {\bibfnamefont {I.}~\bibnamefont {Moult}}, \
  and\ \bibinfo {author} {\bibfnamefont {D.}~\bibnamefont {Neill}},\ }\href
  {\doibase 10.1007/JHEP02(2018)144} {\bibfield  {journal} {\bibinfo  {journal}
  {JHEP}\ }\textbf {\bibinfo {volume} {02}},\ \bibinfo {pages} {144} (\bibinfo
  {year} {2018})},\ \Eprint {http://arxiv.org/abs/1710.00014} {arXiv:1710.00014
  [hep-ph]} \BibitemShut {NoStop}%
\bibitem [{\citenamefont {Napoletano}\ and\ \citenamefont
  {Soyez}(2018)}]{Napoletano:2018ohv}%
  \BibitemOpen
  \bibfield  {author} {\bibinfo {author} {\bibfnamefont {D.}~\bibnamefont
  {Napoletano}}\ and\ \bibinfo {author} {\bibfnamefont {G.}~\bibnamefont
  {Soyez}},\ }\href {\doibase 10.1007/JHEP12(2018)031} {\bibfield  {journal}
  {\bibinfo  {journal} {JHEP}\ }\textbf {\bibinfo {volume} {12}},\ \bibinfo
  {pages} {031} (\bibinfo {year} {2018})},\ \Eprint
  {http://arxiv.org/abs/1809.04602} {arXiv:1809.04602 [hep-ph]} \BibitemShut
  {NoStop}%
\bibitem [{\citenamefont {Chien}\ and\ \citenamefont
  {Stewart}(2019)}]{Chien:2019osu}%
  \BibitemOpen
  \bibfield  {author} {\bibinfo {author} {\bibfnamefont {Y.-T.}\ \bibnamefont
  {Chien}}\ and\ \bibinfo {author} {\bibfnamefont {I.~W.}\ \bibnamefont
  {Stewart}},\ }\href@noop {} {\  (\bibinfo {year} {2019})},\ \Eprint
  {http://arxiv.org/abs/1907.11107} {arXiv:1907.11107 [hep-ph]} \BibitemShut
  {NoStop}%
\bibitem [{Ben(2018)}]{Bendavid:2018nar}%
  \BibitemOpen
  \href {http://lss.fnal.gov/archive/2018/conf/fermilab-conf-18-122-cd-t.pdf}
  {\emph {\bibinfo {title} {{Les Houches 2017: Physics at TeV Colliders
  Standard Model Working Group Report}}}}\ (\bibinfo {year} {2018})\ \Eprint
  {http://arxiv.org/abs/1803.07977} {arXiv:1803.07977 [hep-ph]} \BibitemShut
  {NoStop}%
\bibitem [{\citenamefont {Hoang}\ \emph
  {et~al.}(2019{\natexlab{a}})\citenamefont {Hoang}, \citenamefont {Mantry},
  \citenamefont {Pathak},\ and\ \citenamefont {Stewart}}]{Hoang:2017kmk}%
  \BibitemOpen
  \bibfield  {author} {\bibinfo {author} {\bibfnamefont {A.~H.}\ \bibnamefont
  {Hoang}}, \bibinfo {author} {\bibfnamefont {S.}~\bibnamefont {Mantry}},
  \bibinfo {author} {\bibfnamefont {A.}~\bibnamefont {Pathak}}, \ and\ \bibinfo
  {author} {\bibfnamefont {I.~W.}\ \bibnamefont {Stewart}},\ }\href {\doibase
  10.1103/PhysRevD.100.074021} {\bibfield  {journal} {\bibinfo  {journal}
  {Phys. Rev.}\ }\textbf {\bibinfo {volume} {D100}},\ \bibinfo {pages} {074021}
  (\bibinfo {year} {2019}{\natexlab{a}})},\ \Eprint
  {http://arxiv.org/abs/1708.02586} {arXiv:1708.02586 [hep-ph]} \BibitemShut
  {NoStop}%
\bibitem [{\citenamefont {Kardos}\ \emph {et~al.}(2016)\citenamefont {Kardos},
  \citenamefont {Somogyi},\ and\ \citenamefont
  {Tr\'ocs\'anyi}}]{Kardos:2016pic}%
  \BibitemOpen
  \bibfield  {author} {\bibinfo {author} {\bibfnamefont {A.}~\bibnamefont
  {Kardos}}, \bibinfo {author} {\bibfnamefont {G.}~\bibnamefont {Somogyi}}, \
  and\ \bibinfo {author} {\bibfnamefont {Z.}~\bibnamefont {Tr\'ocs\'anyi}},\
  }\bibfield  {booktitle} {\emph {\bibinfo {booktitle} {{Proceedings, 13th DESY
  Workshop on Elementary Particle Physics: Loops and Legs in Quantum Field
  Theory (LL2016): Leipzig, Germany, April 24-29, 2016}}},\ }\href {\doibase
  10.22323/1.260.0021} {\bibfield  {journal} {\bibinfo  {journal} {PoS}\
  }\textbf {\bibinfo {volume} {LL2016}},\ \bibinfo {pages} {021} (\bibinfo
  {year} {2016})}\BibitemShut {NoStop}%
\bibitem [{\citenamefont {Kardos}\ \emph {et~al.}(2018)\citenamefont {Kardos},
  \citenamefont {Somogyi},\ and\ \citenamefont
  {Tr\'ocs\'anyi}}]{Kardos:2018kth}%
  \BibitemOpen
  \bibfield  {author} {\bibinfo {author} {\bibfnamefont {A.}~\bibnamefont
  {Kardos}}, \bibinfo {author} {\bibfnamefont {G.}~\bibnamefont {Somogyi}}, \
  and\ \bibinfo {author} {\bibfnamefont {Z.}~\bibnamefont {Tr\'ocs\'anyi}},\
  }\href {\doibase 10.1016/j.physletb.2018.10.014} {\bibfield  {journal}
  {\bibinfo  {journal} {Phys. Lett.}\ }\textbf {\bibinfo {volume} {B786}},\
  \bibinfo {pages} {313} (\bibinfo {year} {2018})},\ \Eprint
  {http://arxiv.org/abs/1807.11472} {arXiv:1807.11472 [hep-ph]} \BibitemShut
  {NoStop}%
\bibitem [{\citenamefont {Hoang}\ \emph
  {et~al.}(2019{\natexlab{b}})\citenamefont {Hoang}, \citenamefont {Mantry},
  \citenamefont {Pathak},\ and\ \citenamefont {Stewart}}]{Hoang:2019ceu}%
  \BibitemOpen
  \bibfield  {author} {\bibinfo {author} {\bibfnamefont {A.~H.}\ \bibnamefont
  {Hoang}}, \bibinfo {author} {\bibfnamefont {S.}~\bibnamefont {Mantry}},
  \bibinfo {author} {\bibfnamefont {A.}~\bibnamefont {Pathak}}, \ and\ \bibinfo
  {author} {\bibfnamefont {I.~W.}\ \bibnamefont {Stewart}},\ }\href {\doibase
  10.1007/JHEP12(2019)002} {\bibfield  {journal} {\bibinfo  {journal} {JHEP}\
  }\textbf {\bibinfo {volume} {12}},\ \bibinfo {pages} {002} (\bibinfo {year}
  {2019}{\natexlab{b}})},\ \Eprint {http://arxiv.org/abs/1906.11843}
  {arXiv:1906.11843 [hep-ph]} \BibitemShut {NoStop}%
\bibitem [{\citenamefont {Badger}\ \emph {et~al.}(2018)\citenamefont {Badger},
  \citenamefont {Brønnum-Hansen}, \citenamefont {Hartanto},\ and\
  \citenamefont {Peraro}}]{Badger:2017jhb}%
  \BibitemOpen
  \bibfield  {author} {\bibinfo {author} {\bibfnamefont {S.}~\bibnamefont
  {Badger}}, \bibinfo {author} {\bibfnamefont {C.}~\bibnamefont
  {Brønnum-Hansen}}, \bibinfo {author} {\bibfnamefont {H.~B.}\ \bibnamefont
  {Hartanto}}, \ and\ \bibinfo {author} {\bibfnamefont {T.}~\bibnamefont
  {Peraro}},\ }\href {\doibase 10.1103/PhysRevLett.120.092001} {\bibfield
  {journal} {\bibinfo  {journal} {Phys. Rev. Lett.}\ }\textbf {\bibinfo
  {volume} {120}},\ \bibinfo {pages} {092001} (\bibinfo {year} {2018})},\
  \Eprint {http://arxiv.org/abs/1712.02229} {arXiv:1712.02229 [hep-ph]}
  \BibitemShut {NoStop}%
\bibitem [{\citenamefont {Abreu}\ \emph
  {et~al.}(2018{\natexlab{a}})\citenamefont {Abreu}, \citenamefont
  {Febres~Cordero}, \citenamefont {Ita}, \citenamefont {Page},\ and\
  \citenamefont {Zeng}}]{Abreu:2017hqn}%
  \BibitemOpen
  \bibfield  {author} {\bibinfo {author} {\bibfnamefont {S.}~\bibnamefont
  {Abreu}}, \bibinfo {author} {\bibfnamefont {F.}~\bibnamefont
  {Febres~Cordero}}, \bibinfo {author} {\bibfnamefont {H.}~\bibnamefont {Ita}},
  \bibinfo {author} {\bibfnamefont {B.}~\bibnamefont {Page}}, \ and\ \bibinfo
  {author} {\bibfnamefont {M.}~\bibnamefont {Zeng}},\ }\href {\doibase
  10.1103/PhysRevD.97.116014} {\bibfield  {journal} {\bibinfo  {journal} {Phys.
  Rev.}\ }\textbf {\bibinfo {volume} {D97}},\ \bibinfo {pages} {116014}
  (\bibinfo {year} {2018}{\natexlab{a}})},\ \Eprint
  {http://arxiv.org/abs/1712.03946} {arXiv:1712.03946 [hep-ph]} \BibitemShut
  {NoStop}%
\bibitem [{\citenamefont {Chawdhry}\ \emph {et~al.}(2019)\citenamefont
  {Chawdhry}, \citenamefont {Lim},\ and\ \citenamefont
  {Mitov}}]{Chawdhry:2018awn}%
  \BibitemOpen
  \bibfield  {author} {\bibinfo {author} {\bibfnamefont {H.~A.}\ \bibnamefont
  {Chawdhry}}, \bibinfo {author} {\bibfnamefont {M.~A.}\ \bibnamefont {Lim}}, \
  and\ \bibinfo {author} {\bibfnamefont {A.}~\bibnamefont {Mitov}},\ }\href
  {\doibase 10.1103/PhysRevD.99.076011} {\bibfield  {journal} {\bibinfo
  {journal} {Phys. Rev.}\ }\textbf {\bibinfo {volume} {D99}},\ \bibinfo {pages}
  {076011} (\bibinfo {year} {2019})},\ \Eprint
  {http://arxiv.org/abs/1805.09182} {arXiv:1805.09182 [hep-ph]} \BibitemShut
  {NoStop}%
\bibitem [{\citenamefont {Abreu}\ \emph
  {et~al.}(2018{\natexlab{b}})\citenamefont {Abreu}, \citenamefont
  {Febres~Cordero}, \citenamefont {Ita}, \citenamefont {Page},\ and\
  \citenamefont {Sotnikov}}]{Abreu:2018jgq}%
  \BibitemOpen
  \bibfield  {author} {\bibinfo {author} {\bibfnamefont {S.}~\bibnamefont
  {Abreu}}, \bibinfo {author} {\bibfnamefont {F.}~\bibnamefont
  {Febres~Cordero}}, \bibinfo {author} {\bibfnamefont {H.}~\bibnamefont {Ita}},
  \bibinfo {author} {\bibfnamefont {B.}~\bibnamefont {Page}}, \ and\ \bibinfo
  {author} {\bibfnamefont {V.}~\bibnamefont {Sotnikov}},\ }\href {\doibase
  10.1007/JHEP11(2018)116} {\bibfield  {journal} {\bibinfo  {journal} {JHEP}\
  }\textbf {\bibinfo {volume} {11}},\ \bibinfo {pages} {116} (\bibinfo {year}
  {2018}{\natexlab{b}})},\ \Eprint {http://arxiv.org/abs/1809.09067}
  {arXiv:1809.09067 [hep-ph]} \BibitemShut {NoStop}%
\bibitem [{\citenamefont {Badger}\ \emph
  {et~al.}(2019{\natexlab{a}})\citenamefont {Badger}, \citenamefont
  {Brønnum-Hansen}, \citenamefont {Hartanto},\ and\ \citenamefont
  {Peraro}}]{Badger:2018enw}%
  \BibitemOpen
  \bibfield  {author} {\bibinfo {author} {\bibfnamefont {S.}~\bibnamefont
  {Badger}}, \bibinfo {author} {\bibfnamefont {C.}~\bibnamefont
  {Brønnum-Hansen}}, \bibinfo {author} {\bibfnamefont {H.~B.}\ \bibnamefont
  {Hartanto}}, \ and\ \bibinfo {author} {\bibfnamefont {T.}~\bibnamefont
  {Peraro}},\ }\href {\doibase 10.1007/JHEP01(2019)186} {\bibfield  {journal}
  {\bibinfo  {journal} {JHEP}\ }\textbf {\bibinfo {volume} {01}},\ \bibinfo
  {pages} {186} (\bibinfo {year} {2019}{\natexlab{a}})},\ \Eprint
  {http://arxiv.org/abs/1811.11699} {arXiv:1811.11699 [hep-ph]} \BibitemShut
  {NoStop}%
\bibitem [{\citenamefont {Abreu}\ \emph
  {et~al.}(2019{\natexlab{a}})\citenamefont {Abreu}, \citenamefont {Dormans},
  \citenamefont {Febres~Cordero}, \citenamefont {Ita},\ and\ \citenamefont
  {Page}}]{Abreu:2018zmy}%
  \BibitemOpen
  \bibfield  {author} {\bibinfo {author} {\bibfnamefont {S.}~\bibnamefont
  {Abreu}}, \bibinfo {author} {\bibfnamefont {J.}~\bibnamefont {Dormans}},
  \bibinfo {author} {\bibfnamefont {F.}~\bibnamefont {Febres~Cordero}},
  \bibinfo {author} {\bibfnamefont {H.}~\bibnamefont {Ita}}, \ and\ \bibinfo
  {author} {\bibfnamefont {B.}~\bibnamefont {Page}},\ }\href {\doibase
  10.1103/PhysRevLett.122.082002} {\bibfield  {journal} {\bibinfo  {journal}
  {Phys. Rev. Lett.}\ }\textbf {\bibinfo {volume} {122}},\ \bibinfo {pages}
  {082002} (\bibinfo {year} {2019}{\natexlab{a}})},\ \Eprint
  {http://arxiv.org/abs/1812.04586} {arXiv:1812.04586 [hep-ph]} \BibitemShut
  {NoStop}%
\bibitem [{\citenamefont {Chicherin}\ \emph {et~al.}(2019)\citenamefont
  {Chicherin}, \citenamefont {Gehrmann}, \citenamefont {Henn}, \citenamefont
  {Wasser}, \citenamefont {Zhang},\ and\ \citenamefont
  {Zoia}}]{Chicherin:2018old}%
  \BibitemOpen
  \bibfield  {author} {\bibinfo {author} {\bibfnamefont {D.}~\bibnamefont
  {Chicherin}}, \bibinfo {author} {\bibfnamefont {T.}~\bibnamefont {Gehrmann}},
  \bibinfo {author} {\bibfnamefont {J.~M.}\ \bibnamefont {Henn}}, \bibinfo
  {author} {\bibfnamefont {P.}~\bibnamefont {Wasser}}, \bibinfo {author}
  {\bibfnamefont {Y.}~\bibnamefont {Zhang}}, \ and\ \bibinfo {author}
  {\bibfnamefont {S.}~\bibnamefont {Zoia}},\ }\href {\doibase
  10.1103/PhysRevLett.123.041603} {\bibfield  {journal} {\bibinfo  {journal}
  {Phys. Rev. Lett.}\ }\textbf {\bibinfo {volume} {123}},\ \bibinfo {pages}
  {041603} (\bibinfo {year} {2019})},\ \Eprint
  {http://arxiv.org/abs/1812.11160} {arXiv:1812.11160 [hep-ph]} \BibitemShut
  {NoStop}%
\bibitem [{\citenamefont {Abreu}\ \emph
  {et~al.}(2019{\natexlab{b}})\citenamefont {Abreu}, \citenamefont {Dormans},
  \citenamefont {Febres~Cordero}, \citenamefont {Ita}, \citenamefont {Page},\
  and\ \citenamefont {Sotnikov}}]{Abreu:2019odu}%
  \BibitemOpen
  \bibfield  {author} {\bibinfo {author} {\bibfnamefont {S.}~\bibnamefont
  {Abreu}}, \bibinfo {author} {\bibfnamefont {J.}~\bibnamefont {Dormans}},
  \bibinfo {author} {\bibfnamefont {F.}~\bibnamefont {Febres~Cordero}},
  \bibinfo {author} {\bibfnamefont {H.}~\bibnamefont {Ita}}, \bibinfo {author}
  {\bibfnamefont {B.}~\bibnamefont {Page}}, \ and\ \bibinfo {author}
  {\bibfnamefont {V.}~\bibnamefont {Sotnikov}},\ }\href {\doibase
  10.1007/JHEP05(2019)084} {\bibfield  {journal} {\bibinfo  {journal} {JHEP}\
  }\textbf {\bibinfo {volume} {05}},\ \bibinfo {pages} {084} (\bibinfo {year}
  {2019}{\natexlab{b}})},\ \Eprint {http://arxiv.org/abs/1904.00945}
  {arXiv:1904.00945 [hep-ph]} \BibitemShut {NoStop}%
\bibitem [{\citenamefont {Badger}\ \emph
  {et~al.}(2019{\natexlab{b}})\citenamefont {Badger}, \citenamefont
  {Chicherin}, \citenamefont {Gehrmann}, \citenamefont {Heinrich},
  \citenamefont {Henn}, \citenamefont {Peraro}, \citenamefont {Wasser},
  \citenamefont {Zhang},\ and\ \citenamefont {Zoia}}]{Badger:2019djh}%
  \BibitemOpen
  \bibfield  {author} {\bibinfo {author} {\bibfnamefont {S.}~\bibnamefont
  {Badger}}, \bibinfo {author} {\bibfnamefont {D.}~\bibnamefont {Chicherin}},
  \bibinfo {author} {\bibfnamefont {T.}~\bibnamefont {Gehrmann}}, \bibinfo
  {author} {\bibfnamefont {G.}~\bibnamefont {Heinrich}}, \bibinfo {author}
  {\bibfnamefont {J.~M.}\ \bibnamefont {Henn}}, \bibinfo {author}
  {\bibfnamefont {T.}~\bibnamefont {Peraro}}, \bibinfo {author} {\bibfnamefont
  {P.}~\bibnamefont {Wasser}}, \bibinfo {author} {\bibfnamefont
  {Y.}~\bibnamefont {Zhang}}, \ and\ \bibinfo {author} {\bibfnamefont
  {S.}~\bibnamefont {Zoia}},\ }\href {\doibase 10.1103/PhysRevLett.123.071601}
  {\bibfield  {journal} {\bibinfo  {journal} {Phys. Rev. Lett.}\ }\textbf
  {\bibinfo {volume} {123}},\ \bibinfo {pages} {071601} (\bibinfo {year}
  {2019}{\natexlab{b}})},\ \Eprint {http://arxiv.org/abs/1905.03733}
  {arXiv:1905.03733 [hep-ph]} \BibitemShut {NoStop}%
\bibitem [{\citenamefont {Hartanto}\ \emph {et~al.}(2019)\citenamefont
  {Hartanto}, \citenamefont {Badger}, \citenamefont {Brønnum-Hansen},\ and\
  \citenamefont {Peraro}}]{Hartanto:2019uvl}%
  \BibitemOpen
  \bibfield  {author} {\bibinfo {author} {\bibfnamefont {H.~B.}\ \bibnamefont
  {Hartanto}}, \bibinfo {author} {\bibfnamefont {S.}~\bibnamefont {Badger}},
  \bibinfo {author} {\bibfnamefont {C.}~\bibnamefont {Brønnum-Hansen}}, \ and\
  \bibinfo {author} {\bibfnamefont {T.}~\bibnamefont {Peraro}},\ }\href
  {\doibase 10.1007/JHEP09(2019)119} {\bibfield  {journal} {\bibinfo  {journal}
  {JHEP}\ }\textbf {\bibinfo {volume} {09}},\ \bibinfo {pages} {119} (\bibinfo
  {year} {2019})},\ \Eprint {http://arxiv.org/abs/1906.11862} {arXiv:1906.11862
  [hep-ph]} \BibitemShut {NoStop}%
\bibitem [{\citenamefont {Becher}\ and\ \citenamefont
  {Schwartz}(2008)}]{Becher:2008cf}%
  \BibitemOpen
  \bibfield  {author} {\bibinfo {author} {\bibfnamefont {T.}~\bibnamefont
  {Becher}}\ and\ \bibinfo {author} {\bibfnamefont {M.~D.}\ \bibnamefont
  {Schwartz}},\ }\href {\doibase 10.1088/1126-6708/2008/07/034} {\bibfield
  {journal} {\bibinfo  {journal} {JHEP}\ }\textbf {\bibinfo {volume} {07}},\
  \bibinfo {pages} {034} (\bibinfo {year} {2008})},\ \Eprint
  {http://arxiv.org/abs/0803.0342} {arXiv:0803.0342 [hep-ph]} \BibitemShut
  {NoStop}%
\bibitem [{\citenamefont {Abbate}\ \emph {et~al.}(2011)\citenamefont {Abbate},
  \citenamefont {Fickinger}, \citenamefont {Hoang}, \citenamefont {Mateu},\
  and\ \citenamefont {Stewart}}]{Abbate:2010xh}%
  \BibitemOpen
  \bibfield  {author} {\bibinfo {author} {\bibfnamefont {R.}~\bibnamefont
  {Abbate}}, \bibinfo {author} {\bibfnamefont {M.}~\bibnamefont {Fickinger}},
  \bibinfo {author} {\bibfnamefont {A.~H.}\ \bibnamefont {Hoang}}, \bibinfo
  {author} {\bibfnamefont {V.}~\bibnamefont {Mateu}}, \ and\ \bibinfo {author}
  {\bibfnamefont {I.~W.}\ \bibnamefont {Stewart}},\ }\href {\doibase
  10.1103/PhysRevD.83.074021} {\bibfield  {journal} {\bibinfo  {journal} {Phys.
  Rev.}\ }\textbf {\bibinfo {volume} {D83}},\ \bibinfo {pages} {074021}
  (\bibinfo {year} {2011})},\ \Eprint {http://arxiv.org/abs/1006.3080}
  {arXiv:1006.3080 [hep-ph]} \BibitemShut {NoStop}%
\bibitem [{\citenamefont {Hoang}\ \emph
  {et~al.}(2015{\natexlab{a}})\citenamefont {Hoang}, \citenamefont
  {Kolodrubetz}, \citenamefont {Mateu},\ and\ \citenamefont
  {Stewart}}]{Hoang:2014wka}%
  \BibitemOpen
  \bibfield  {author} {\bibinfo {author} {\bibfnamefont {A.~H.}\ \bibnamefont
  {Hoang}}, \bibinfo {author} {\bibfnamefont {D.~W.}\ \bibnamefont
  {Kolodrubetz}}, \bibinfo {author} {\bibfnamefont {V.}~\bibnamefont {Mateu}},
  \ and\ \bibinfo {author} {\bibfnamefont {I.~W.}\ \bibnamefont {Stewart}},\
  }\href {\doibase 10.1103/PhysRevD.91.094017} {\bibfield  {journal} {\bibinfo
  {journal} {Phys. Rev.}\ }\textbf {\bibinfo {volume} {D91}},\ \bibinfo {pages}
  {094017} (\bibinfo {year} {2015}{\natexlab{a}})},\ \Eprint
  {http://arxiv.org/abs/1411.6633} {arXiv:1411.6633 [hep-ph]} \BibitemShut
  {NoStop}%
\bibitem [{\citenamefont {Hoang}\ \emph
  {et~al.}(2015{\natexlab{b}})\citenamefont {Hoang}, \citenamefont
  {Kolodrubetz}, \citenamefont {Mateu},\ and\ \citenamefont
  {Stewart}}]{Hoang:2015hka}%
  \BibitemOpen
  \bibfield  {author} {\bibinfo {author} {\bibfnamefont {A.~H.}\ \bibnamefont
  {Hoang}}, \bibinfo {author} {\bibfnamefont {D.~W.}\ \bibnamefont
  {Kolodrubetz}}, \bibinfo {author} {\bibfnamefont {V.}~\bibnamefont {Mateu}},
  \ and\ \bibinfo {author} {\bibfnamefont {I.~W.}\ \bibnamefont {Stewart}},\
  }\href {\doibase 10.1103/PhysRevD.91.094018} {\bibfield  {journal} {\bibinfo
  {journal} {Phys. Rev.}\ }\textbf {\bibinfo {volume} {D91}},\ \bibinfo {pages}
  {094018} (\bibinfo {year} {2015}{\natexlab{b}})},\ \Eprint
  {http://arxiv.org/abs/1501.04111} {arXiv:1501.04111 [hep-ph]} \BibitemShut
  {NoStop}%
\bibitem [{\citenamefont {Bell}\ \emph {et~al.}(2019)\citenamefont {Bell},
  \citenamefont {Rahn},\ and\ \citenamefont {Talbert}}]{Bell:2018oqa}%
  \BibitemOpen
  \bibfield  {author} {\bibinfo {author} {\bibfnamefont {G.}~\bibnamefont
  {Bell}}, \bibinfo {author} {\bibfnamefont {R.}~\bibnamefont {Rahn}}, \ and\
  \bibinfo {author} {\bibfnamefont {J.}~\bibnamefont {Talbert}},\ }\href
  {\doibase 10.1007/JHEP07(2019)101} {\bibfield  {journal} {\bibinfo  {journal}
  {JHEP}\ }\textbf {\bibinfo {volume} {07}},\ \bibinfo {pages} {101} (\bibinfo
  {year} {2019})},\ \Eprint {http://arxiv.org/abs/1812.08690} {arXiv:1812.08690
  [hep-ph]} \BibitemShut {NoStop}%
\bibitem [{\citenamefont {Bell}\ \emph {et~al.}(2018)\citenamefont {Bell},
  \citenamefont {Rahn},\ and\ \citenamefont {Talbert}}]{Bell:2018vaa}%
  \BibitemOpen
  \bibfield  {author} {\bibinfo {author} {\bibfnamefont {G.}~\bibnamefont
  {Bell}}, \bibinfo {author} {\bibfnamefont {R.}~\bibnamefont {Rahn}}, \ and\
  \bibinfo {author} {\bibfnamefont {J.}~\bibnamefont {Talbert}},\ }\href
  {\doibase 10.1016/j.nuclphysb.2018.09.026} {\bibfield  {journal} {\bibinfo
  {journal} {Nucl. Phys.}\ }\textbf {\bibinfo {volume} {B936}},\ \bibinfo
  {pages} {520} (\bibinfo {year} {2018})},\ \Eprint
  {http://arxiv.org/abs/1805.12414} {arXiv:1805.12414 [hep-ph]} \BibitemShut
  {NoStop}%
\bibitem [{\citenamefont {Bell}\ \emph {et~al.}(2020)\citenamefont {Bell},
  \citenamefont {Rahn},\ and\ \citenamefont {Talbert}}]{Bell:2020yzz}%
  \BibitemOpen
  \bibfield  {author} {\bibinfo {author} {\bibfnamefont {G.}~\bibnamefont
  {Bell}}, \bibinfo {author} {\bibfnamefont {R.}~\bibnamefont {Rahn}}, \ and\
  \bibinfo {author} {\bibfnamefont {J.}~\bibnamefont {Talbert}},\ }\href@noop
  {} {\  (\bibinfo {year} {2020})},\ \Eprint {http://arxiv.org/abs/2004.08396}
  {arXiv:2004.08396 [hep-ph]} \BibitemShut {NoStop}%
\bibitem [{\citenamefont {Catani}\ and\ \citenamefont
  {Seymour}(1997)}]{Catani:1996vz}%
  \BibitemOpen
  \bibfield  {author} {\bibinfo {author} {\bibfnamefont {S.}~\bibnamefont
  {Catani}}\ and\ \bibinfo {author} {\bibfnamefont {M.~H.}\ \bibnamefont
  {Seymour}},\ }\href {\doibase 10.1016/S0550-3213(96)00589-5,
  10.1016/S0550-3213(98)81022-5} {\bibfield  {journal} {\bibinfo  {journal}
  {Nucl. Phys.}\ }\textbf {\bibinfo {volume} {B485}},\ \bibinfo {pages} {291}
  (\bibinfo {year} {1997})},\ \bibinfo {note} {[Erratum: Nucl.
  Phys.B510,503(1998)]},\ \Eprint {http://arxiv.org/abs/hep-ph/9605323}
  {arXiv:hep-ph/9605323 [hep-ph]} \BibitemShut {NoStop}%
\bibitem [{\citenamefont {Kardos}\ \emph {et~al.}(2020)\citenamefont {Kardos},
  \citenamefont {Larkoski},\ and\ \citenamefont
  {Tr\'ocs\'anyi}}]{Kardos:2020gty}%
  \BibitemOpen
  \bibfield  {author} {\bibinfo {author} {\bibfnamefont {A.}~\bibnamefont
  {Kardos}}, \bibinfo {author} {\bibfnamefont {A.~J.}\ \bibnamefont
  {Larkoski}}, \ and\ \bibinfo {author} {\bibfnamefont {Z.}~\bibnamefont
  {Tr\'ocs\'anyi}},\ }\href@noop {} {\  (\bibinfo {year} {2020})},\ \Eprint
  {http://arxiv.org/abs/2002.00942} {arXiv:2002.00942 [hep-ph]} \BibitemShut
  {NoStop}%
\bibitem [{\citenamefont {Brandt}\ \emph {et~al.}(1964)\citenamefont {Brandt},
  \citenamefont {Peyrou}, \citenamefont {Sosnowski},\ and\ \citenamefont
  {Wroblewski}}]{Brandt:1964sa}%
  \BibitemOpen
  \bibfield  {author} {\bibinfo {author} {\bibfnamefont {S.}~\bibnamefont
  {Brandt}}, \bibinfo {author} {\bibfnamefont {C.}~\bibnamefont {Peyrou}},
  \bibinfo {author} {\bibfnamefont {R.}~\bibnamefont {Sosnowski}}, \ and\
  \bibinfo {author} {\bibfnamefont {A.}~\bibnamefont {Wroblewski}},\ }\href
  {\doibase 10.1016/0031-9163(64)91176-X} {\bibfield  {journal} {\bibinfo
  {journal} {Phys. Lett.}\ }\textbf {\bibinfo {volume} {12}},\ \bibinfo {pages}
  {57} (\bibinfo {year} {1964})}\BibitemShut {NoStop}%
\bibitem [{\citenamefont {Farhi}(1977)}]{Farhi:1977sg}%
  \BibitemOpen
  \bibfield  {author} {\bibinfo {author} {\bibfnamefont {E.}~\bibnamefont
  {Farhi}},\ }\href {\doibase 10.1103/PhysRevLett.39.1587} {\bibfield
  {journal} {\bibinfo  {journal} {Phys. Rev. Lett.}\ }\textbf {\bibinfo
  {volume} {39}},\ \bibinfo {pages} {1587} (\bibinfo {year}
  {1977})}\BibitemShut {NoStop}%
\bibitem [{\citenamefont {Georgi}\ and\ \citenamefont
  {Machacek}(1977)}]{Georgi:1977sf}%
  \BibitemOpen
  \bibfield  {author} {\bibinfo {author} {\bibfnamefont {H.}~\bibnamefont
  {Georgi}}\ and\ \bibinfo {author} {\bibfnamefont {M.}~\bibnamefont
  {Machacek}},\ }\href {\doibase 10.1103/PhysRevLett.39.1237} {\bibfield
  {journal} {\bibinfo  {journal} {Phys. Rev. Lett.}\ }\textbf {\bibinfo
  {volume} {39}},\ \bibinfo {pages} {1237} (\bibinfo {year}
  {1977})}\BibitemShut {NoStop}%
\bibitem [{\citenamefont {Larkoski}\ \emph
  {et~al.}(2014{\natexlab{b}})\citenamefont {Larkoski}, \citenamefont {Neill},\
  and\ \citenamefont {Thaler}}]{Larkoski:2014uqa}%
  \BibitemOpen
  \bibfield  {author} {\bibinfo {author} {\bibfnamefont {A.~J.}\ \bibnamefont
  {Larkoski}}, \bibinfo {author} {\bibfnamefont {D.}~\bibnamefont {Neill}}, \
  and\ \bibinfo {author} {\bibfnamefont {J.}~\bibnamefont {Thaler}},\ }\href
  {\doibase 10.1007/JHEP04(2014)017} {\bibfield  {journal} {\bibinfo  {journal}
  {JHEP}\ }\textbf {\bibinfo {volume} {04}},\ \bibinfo {pages} {017} (\bibinfo
  {year} {2014}{\natexlab{b}})},\ \Eprint {http://arxiv.org/abs/1401.2158}
  {arXiv:1401.2158 [hep-ph]} \BibitemShut {NoStop}%
\bibitem [{\citenamefont {Dokshitzer}\ \emph {et~al.}(1997)\citenamefont
  {Dokshitzer}, \citenamefont {Leder}, \citenamefont {Moretti},\ and\
  \citenamefont {Webber}}]{Dokshitzer:1997in}%
  \BibitemOpen
  \bibfield  {author} {\bibinfo {author} {\bibfnamefont {Y.~L.}\ \bibnamefont
  {Dokshitzer}}, \bibinfo {author} {\bibfnamefont {G.~D.}\ \bibnamefont
  {Leder}}, \bibinfo {author} {\bibfnamefont {S.}~\bibnamefont {Moretti}}, \
  and\ \bibinfo {author} {\bibfnamefont {B.~R.}\ \bibnamefont {Webber}},\
  }\href {\doibase 10.1088/1126-6708/1997/08/001} {\bibfield  {journal}
  {\bibinfo  {journal} {JHEP}\ }\textbf {\bibinfo {volume} {08}},\ \bibinfo
  {pages} {001} (\bibinfo {year} {1997})},\ \Eprint
  {http://arxiv.org/abs/hep-ph/9707323} {arXiv:hep-ph/9707323 [hep-ph]}
  \BibitemShut {NoStop}%
\bibitem [{\citenamefont {Wobisch}\ and\ \citenamefont
  {Wengler}(1998)}]{Wobisch:1998wt}%
  \BibitemOpen
  \bibfield  {author} {\bibinfo {author} {\bibfnamefont {M.}~\bibnamefont
  {Wobisch}}\ and\ \bibinfo {author} {\bibfnamefont {T.}~\bibnamefont
  {Wengler}},\ }in\ \href@noop {} {\emph {\bibinfo {booktitle} {{Monte Carlo
  generators for HERA physics. Proceedings, Workshop, Hamburg, Germany,
  1998-1999}}}}\ (\bibinfo {year} {1998})\ pp.\ \bibinfo {pages} {270--279},\
  \Eprint {http://arxiv.org/abs/hep-ph/9907280} {arXiv:hep-ph/9907280 [hep-ph]}
  \BibitemShut {NoStop}%
\bibitem [{\citenamefont {Catani}\ \emph {et~al.}(1993)\citenamefont {Catani},
  \citenamefont {Trentadue}, \citenamefont {Turnock},\ and\ \citenamefont
  {Webber}}]{Catani:1992ua}%
  \BibitemOpen
  \bibfield  {author} {\bibinfo {author} {\bibfnamefont {S.}~\bibnamefont
  {Catani}}, \bibinfo {author} {\bibfnamefont {L.}~\bibnamefont {Trentadue}},
  \bibinfo {author} {\bibfnamefont {G.}~\bibnamefont {Turnock}}, \ and\
  \bibinfo {author} {\bibfnamefont {B.~R.}\ \bibnamefont {Webber}},\ }\href
  {\doibase 10.1016/0550-3213(93)90271-P} {\bibfield  {journal} {\bibinfo
  {journal} {Nucl. Phys.}\ }\textbf {\bibinfo {volume} {B407}},\ \bibinfo
  {pages} {3} (\bibinfo {year} {1993})}\BibitemShut {NoStop}%
\bibitem [{\citenamefont {Gracey}(1994)}]{Gracey:1994nn}%
  \BibitemOpen
  \bibfield  {author} {\bibinfo {author} {\bibfnamefont {J.~A.}\ \bibnamefont
  {Gracey}},\ }\href {\doibase 10.1016/0370-2693(94)90502-9} {\bibfield
  {journal} {\bibinfo  {journal} {Phys. Lett.}\ }\textbf {\bibinfo {volume}
  {B322}},\ \bibinfo {pages} {141} (\bibinfo {year} {1994})},\ \Eprint
  {http://arxiv.org/abs/hep-ph/9401214} {arXiv:hep-ph/9401214 [hep-ph]}
  \BibitemShut {NoStop}%
\bibitem [{\citenamefont {Beneke}\ and\ \citenamefont
  {Braun}(1995)}]{Beneke:1995pq}%
  \BibitemOpen
  \bibfield  {author} {\bibinfo {author} {\bibfnamefont {M.}~\bibnamefont
  {Beneke}}\ and\ \bibinfo {author} {\bibfnamefont {V.~M.}\ \bibnamefont
  {Braun}},\ }\href {\doibase 10.1016/0550-3213(95)00439-Y} {\bibfield
  {journal} {\bibinfo  {journal} {Nucl. Phys.}\ }\textbf {\bibinfo {volume}
  {B454}},\ \bibinfo {pages} {253} (\bibinfo {year} {1995})},\ \Eprint
  {http://arxiv.org/abs/hep-ph/9506452} {arXiv:hep-ph/9506452 [hep-ph]}
  \BibitemShut {NoStop}%
\bibitem [{\citenamefont {Davies}\ \emph {et~al.}(2017)\citenamefont {Davies},
  \citenamefont {Vogt}, \citenamefont {Ruijl}, \citenamefont {Ueda},\ and\
  \citenamefont {Vermaseren}}]{Davies:2016jie}%
  \BibitemOpen
  \bibfield  {author} {\bibinfo {author} {\bibfnamefont {J.}~\bibnamefont
  {Davies}}, \bibinfo {author} {\bibfnamefont {A.}~\bibnamefont {Vogt}},
  \bibinfo {author} {\bibfnamefont {B.}~\bibnamefont {Ruijl}}, \bibinfo
  {author} {\bibfnamefont {T.}~\bibnamefont {Ueda}}, \ and\ \bibinfo {author}
  {\bibfnamefont {J.~A.~M.}\ \bibnamefont {Vermaseren}},\ }\href {\doibase
  10.1016/j.nuclphysb.2016.12.012} {\bibfield  {journal} {\bibinfo  {journal}
  {Nucl. Phys.}\ }\textbf {\bibinfo {volume} {B915}},\ \bibinfo {pages} {335}
  (\bibinfo {year} {2017})},\ \Eprint {http://arxiv.org/abs/1610.07477}
  {arXiv:1610.07477 [hep-ph]} \BibitemShut {NoStop}%
\bibitem [{\citenamefont {Henn}\ \emph {et~al.}(2017)\citenamefont {Henn},
  \citenamefont {Smirnov}, \citenamefont {Smirnov}, \citenamefont
  {Steinhauser},\ and\ \citenamefont {Lee}}]{Lee:2016ixa}%
  \BibitemOpen
  \bibfield  {author} {\bibinfo {author} {\bibfnamefont {J.}~\bibnamefont
  {Henn}}, \bibinfo {author} {\bibfnamefont {A.~V.}\ \bibnamefont {Smirnov}},
  \bibinfo {author} {\bibfnamefont {V.~A.}\ \bibnamefont {Smirnov}}, \bibinfo
  {author} {\bibfnamefont {M.}~\bibnamefont {Steinhauser}}, \ and\ \bibinfo
  {author} {\bibfnamefont {R.~N.}\ \bibnamefont {Lee}},\ }\href {\doibase
  10.1007/JHEP03(2017)139} {\bibfield  {journal} {\bibinfo  {journal} {JHEP}\
  }\textbf {\bibinfo {volume} {03}},\ \bibinfo {pages} {139} (\bibinfo {year}
  {2017})},\ \Eprint {http://arxiv.org/abs/1612.04389} {arXiv:1612.04389
  [hep-ph]} \BibitemShut {NoStop}%
\bibitem [{\citenamefont {Henn}\ \emph {et~al.}(2016)\citenamefont {Henn},
  \citenamefont {Smirnov}, \citenamefont {Smirnov},\ and\ \citenamefont
  {Steinhauser}}]{Henn:2016men}%
  \BibitemOpen
  \bibfield  {author} {\bibinfo {author} {\bibfnamefont {J.~M.}\ \bibnamefont
  {Henn}}, \bibinfo {author} {\bibfnamefont {A.~V.}\ \bibnamefont {Smirnov}},
  \bibinfo {author} {\bibfnamefont {V.~A.}\ \bibnamefont {Smirnov}}, \ and\
  \bibinfo {author} {\bibfnamefont {M.}~\bibnamefont {Steinhauser}},\ }\href
  {\doibase 10.1007/JHEP05(2016)066} {\bibfield  {journal} {\bibinfo  {journal}
  {JHEP}\ }\textbf {\bibinfo {volume} {05}},\ \bibinfo {pages} {066} (\bibinfo
  {year} {2016})},\ \Eprint {http://arxiv.org/abs/1604.03126} {arXiv:1604.03126
  [hep-ph]} \BibitemShut {NoStop}%
\bibitem [{\citenamefont {Moch}\ \emph {et~al.}(2017)\citenamefont {Moch},
  \citenamefont {Ruijl}, \citenamefont {Ueda}, \citenamefont {Vermaseren},\
  and\ \citenamefont {Vogt}}]{Moch:2017uml}%
  \BibitemOpen
  \bibfield  {author} {\bibinfo {author} {\bibfnamefont {S.}~\bibnamefont
  {Moch}}, \bibinfo {author} {\bibfnamefont {B.}~\bibnamefont {Ruijl}},
  \bibinfo {author} {\bibfnamefont {T.}~\bibnamefont {Ueda}}, \bibinfo {author}
  {\bibfnamefont {J.~A.~M.}\ \bibnamefont {Vermaseren}}, \ and\ \bibinfo
  {author} {\bibfnamefont {A.}~\bibnamefont {Vogt}},\ }\href {\doibase
  10.1007/JHEP10(2017)041} {\bibfield  {journal} {\bibinfo  {journal} {JHEP}\
  }\textbf {\bibinfo {volume} {10}},\ \bibinfo {pages} {041} (\bibinfo {year}
  {2017})},\ \Eprint {http://arxiv.org/abs/1707.08315} {arXiv:1707.08315
  [hep-ph]} \BibitemShut {NoStop}%
\bibitem [{\citenamefont {Grozin}(2018)}]{Grozin:2018vdn}%
  \BibitemOpen
  \bibfield  {author} {\bibinfo {author} {\bibfnamefont {A.}~\bibnamefont
  {Grozin}},\ }\href {\doibase 10.1007/JHEP06(2018)073,
  10.1007/JHEP01(2019)134} {\bibfield  {journal} {\bibinfo  {journal} {JHEP}\
  }\textbf {\bibinfo {volume} {06}},\ \bibinfo {pages} {073} (\bibinfo {year}
  {2018})},\ \bibinfo {note} {[Addendum: JHEP01,134(2019)]},\ \Eprint
  {http://arxiv.org/abs/1805.05050} {arXiv:1805.05050 [hep-ph]} \BibitemShut
  {NoStop}%
\bibitem [{\citenamefont {Moch}\ \emph {et~al.}(2018)\citenamefont {Moch},
  \citenamefont {Ruijl}, \citenamefont {Ueda}, \citenamefont {Vermaseren},\
  and\ \citenamefont {Vogt}}]{Moch:2018wjh}%
  \BibitemOpen
  \bibfield  {author} {\bibinfo {author} {\bibfnamefont {S.}~\bibnamefont
  {Moch}}, \bibinfo {author} {\bibfnamefont {B.}~\bibnamefont {Ruijl}},
  \bibinfo {author} {\bibfnamefont {T.}~\bibnamefont {Ueda}}, \bibinfo {author}
  {\bibfnamefont {J.~A.~M.}\ \bibnamefont {Vermaseren}}, \ and\ \bibinfo
  {author} {\bibfnamefont {A.}~\bibnamefont {Vogt}},\ }\href {\doibase
  10.1016/j.physletb.2018.06.017} {\bibfield  {journal} {\bibinfo  {journal}
  {Phys. Lett.}\ }\textbf {\bibinfo {volume} {B782}},\ \bibinfo {pages} {627}
  (\bibinfo {year} {2018})},\ \Eprint {http://arxiv.org/abs/1805.09638}
  {arXiv:1805.09638 [hep-ph]} \BibitemShut {NoStop}%
\bibitem [{\citenamefont {Brüser}\ \emph {et~al.}(2019)\citenamefont
  {Brüser}, \citenamefont {Grozin}, \citenamefont {Henn},\ and\ \citenamefont
  {Stahlhofen}}]{Bruser:2019auj}%
  \BibitemOpen
  \bibfield  {author} {\bibinfo {author} {\bibfnamefont {R.}~\bibnamefont
  {Brüser}}, \bibinfo {author} {\bibfnamefont {A.}~\bibnamefont {Grozin}},
  \bibinfo {author} {\bibfnamefont {J.~M.}\ \bibnamefont {Henn}}, \ and\
  \bibinfo {author} {\bibfnamefont {M.}~\bibnamefont {Stahlhofen}},\ }\href
  {\doibase 10.1007/JHEP05(2019)186} {\bibfield  {journal} {\bibinfo  {journal}
  {JHEP}\ }\textbf {\bibinfo {volume} {05}},\ \bibinfo {pages} {186} (\bibinfo
  {year} {2019})},\ \Eprint {http://arxiv.org/abs/1902.05076} {arXiv:1902.05076
  [hep-ph]} \BibitemShut {NoStop}%
\bibitem [{\citenamefont {Henn}\ \emph
  {et~al.}(2019{\natexlab{a}})\citenamefont {Henn}, \citenamefont {Peraro},
  \citenamefont {Stahlhofen},\ and\ \citenamefont {Wasser}}]{Henn:2019rmi}%
  \BibitemOpen
  \bibfield  {author} {\bibinfo {author} {\bibfnamefont {J.~M.}\ \bibnamefont
  {Henn}}, \bibinfo {author} {\bibfnamefont {T.}~\bibnamefont {Peraro}},
  \bibinfo {author} {\bibfnamefont {M.}~\bibnamefont {Stahlhofen}}, \ and\
  \bibinfo {author} {\bibfnamefont {P.}~\bibnamefont {Wasser}},\ }\href
  {\doibase 10.1103/PhysRevLett.122.201602} {\bibfield  {journal} {\bibinfo
  {journal} {Phys. Rev. Lett.}\ }\textbf {\bibinfo {volume} {122}},\ \bibinfo
  {pages} {201602} (\bibinfo {year} {2019}{\natexlab{a}})},\ \Eprint
  {http://arxiv.org/abs/1901.03693} {arXiv:1901.03693 [hep-ph]} \BibitemShut
  {NoStop}%
\bibitem [{\citenamefont {von Manteuffel}\ and\ \citenamefont
  {Schabinger}(2019)}]{vonManteuffel:2019wbj}%
  \BibitemOpen
  \bibfield  {author} {\bibinfo {author} {\bibfnamefont {A.}~\bibnamefont {von
  Manteuffel}}\ and\ \bibinfo {author} {\bibfnamefont {R.~M.}\ \bibnamefont
  {Schabinger}},\ }\href {\doibase 10.1103/PhysRevD.99.094014} {\bibfield
  {journal} {\bibinfo  {journal} {Phys. Rev.}\ }\textbf {\bibinfo {volume}
  {D99}},\ \bibinfo {pages} {094014} (\bibinfo {year} {2019})},\ \Eprint
  {http://arxiv.org/abs/1902.08208} {arXiv:1902.08208 [hep-ph]} \BibitemShut
  {NoStop}%
\bibitem [{\citenamefont {Lee}\ \emph {et~al.}(2019{\natexlab{b}})\citenamefont
  {Lee}, \citenamefont {Smirnov}, \citenamefont {Smirnov},\ and\ \citenamefont
  {Steinhauser}}]{Lee:2019zop}%
  \BibitemOpen
  \bibfield  {author} {\bibinfo {author} {\bibfnamefont {R.~N.}\ \bibnamefont
  {Lee}}, \bibinfo {author} {\bibfnamefont {A.~V.}\ \bibnamefont {Smirnov}},
  \bibinfo {author} {\bibfnamefont {V.~A.}\ \bibnamefont {Smirnov}}, \ and\
  \bibinfo {author} {\bibfnamefont {M.}~\bibnamefont {Steinhauser}},\ }\href
  {\doibase 10.1007/JHEP02(2019)172} {\bibfield  {journal} {\bibinfo  {journal}
  {JHEP}\ }\textbf {\bibinfo {volume} {02}},\ \bibinfo {pages} {172} (\bibinfo
  {year} {2019}{\natexlab{b}})},\ \Eprint {http://arxiv.org/abs/1901.02898}
  {arXiv:1901.02898 [hep-ph]} \BibitemShut {NoStop}%
\bibitem [{\citenamefont {Henn}\ \emph
  {et~al.}(2019{\natexlab{b}})\citenamefont {Henn}, \citenamefont
  {Korchemsky},\ and\ \citenamefont {Mistlberger}}]{Henn:2019swt}%
  \BibitemOpen
  \bibfield  {author} {\bibinfo {author} {\bibfnamefont {J.~M.}\ \bibnamefont
  {Henn}}, \bibinfo {author} {\bibfnamefont {G.~P.}\ \bibnamefont
  {Korchemsky}}, \ and\ \bibinfo {author} {\bibfnamefont {B.}~\bibnamefont
  {Mistlberger}},\ }\href@noop {} {\  (\bibinfo {year} {2019}{\natexlab{b}})},\
  \Eprint {http://arxiv.org/abs/1911.10174} {arXiv:1911.10174 [hep-th]}
  \BibitemShut {NoStop}%
\bibitem [{\citenamefont {von Manteuffel}\ \emph {et~al.}(2020)\citenamefont
  {von Manteuffel}, \citenamefont {Panzer},\ and\ \citenamefont
  {Schabinger}}]{vonManteuffel:2020vjv}%
  \BibitemOpen
  \bibfield  {author} {\bibinfo {author} {\bibfnamefont {A.}~\bibnamefont {von
  Manteuffel}}, \bibinfo {author} {\bibfnamefont {E.}~\bibnamefont {Panzer}}, \
  and\ \bibinfo {author} {\bibfnamefont {R.~M.}\ \bibnamefont {Schabinger}},\
  }\href@noop {} {\  (\bibinfo {year} {2020})},\ \Eprint
  {http://arxiv.org/abs/2002.04617} {arXiv:2002.04617 [hep-ph]} \BibitemShut
  {NoStop}%
\bibitem [{\citenamefont {van Ritbergen}\ \emph {et~al.}(1997)\citenamefont
  {van Ritbergen}, \citenamefont {Vermaseren},\ and\ \citenamefont
  {Larin}}]{vanRitbergen:1997va}%
  \BibitemOpen
  \bibfield  {author} {\bibinfo {author} {\bibfnamefont {T.}~\bibnamefont {van
  Ritbergen}}, \bibinfo {author} {\bibfnamefont {J.~A.~M.}\ \bibnamefont
  {Vermaseren}}, \ and\ \bibinfo {author} {\bibfnamefont {S.~A.}\ \bibnamefont
  {Larin}},\ }\href {\doibase 10.1016/S0370-2693(97)00370-5} {\bibfield
  {journal} {\bibinfo  {journal} {Phys. Lett.}\ }\textbf {\bibinfo {volume}
  {B400}},\ \bibinfo {pages} {379} (\bibinfo {year} {1997})},\ \Eprint
  {http://arxiv.org/abs/hep-ph/9701390} {arXiv:hep-ph/9701390 [hep-ph]}
  \BibitemShut {NoStop}%
\bibitem [{\citenamefont {Moch}\ \emph
  {et~al.}(2005{\natexlab{a}})\citenamefont {Moch}, \citenamefont
  {Vermaseren},\ and\ \citenamefont {Vogt}}]{Moch:2005id}%
  \BibitemOpen
  \bibfield  {author} {\bibinfo {author} {\bibfnamefont {S.}~\bibnamefont
  {Moch}}, \bibinfo {author} {\bibfnamefont {J.~A.~M.}\ \bibnamefont
  {Vermaseren}}, \ and\ \bibinfo {author} {\bibfnamefont {A.}~\bibnamefont
  {Vogt}},\ }\href {\doibase 10.1088/1126-6708/2005/08/049} {\bibfield
  {journal} {\bibinfo  {journal} {JHEP}\ }\textbf {\bibinfo {volume} {08}},\
  \bibinfo {pages} {049} (\bibinfo {year} {2005}{\natexlab{a}})},\ \Eprint
  {http://arxiv.org/abs/hep-ph/0507039} {arXiv:hep-ph/0507039 [hep-ph]}
  \BibitemShut {NoStop}%
\bibitem [{\citenamefont {Gehrmann}\ \emph {et~al.}(2005)\citenamefont
  {Gehrmann}, \citenamefont {Huber},\ and\ \citenamefont
  {Maitre}}]{Gehrmann:2005pd}%
  \BibitemOpen
  \bibfield  {author} {\bibinfo {author} {\bibfnamefont {T.}~\bibnamefont
  {Gehrmann}}, \bibinfo {author} {\bibfnamefont {T.}~\bibnamefont {Huber}}, \
  and\ \bibinfo {author} {\bibfnamefont {D.}~\bibnamefont {Maitre}},\ }\href
  {\doibase 10.1016/j.physletb.2005.07.019} {\bibfield  {journal} {\bibinfo
  {journal} {Phys. Lett.}\ }\textbf {\bibinfo {volume} {B622}},\ \bibinfo
  {pages} {295} (\bibinfo {year} {2005})},\ \Eprint
  {http://arxiv.org/abs/hep-ph/0507061} {arXiv:hep-ph/0507061 [hep-ph]}
  \BibitemShut {NoStop}%
\bibitem [{\citenamefont {Becher}\ \emph {et~al.}(2007)\citenamefont {Becher},
  \citenamefont {Neubert},\ and\ \citenamefont {Pecjak}}]{Becher:2006mr}%
  \BibitemOpen
  \bibfield  {author} {\bibinfo {author} {\bibfnamefont {T.}~\bibnamefont
  {Becher}}, \bibinfo {author} {\bibfnamefont {M.}~\bibnamefont {Neubert}}, \
  and\ \bibinfo {author} {\bibfnamefont {B.~D.}\ \bibnamefont {Pecjak}},\
  }\href {\doibase 10.1088/1126-6708/2007/01/076} {\bibfield  {journal}
  {\bibinfo  {journal} {JHEP}\ }\textbf {\bibinfo {volume} {01}},\ \bibinfo
  {pages} {076} (\bibinfo {year} {2007})},\ \Eprint
  {http://arxiv.org/abs/hep-ph/0607228} {arXiv:hep-ph/0607228 [hep-ph]}
  \BibitemShut {NoStop}%
\bibitem [{\citenamefont {Baikov}\ \emph {et~al.}(2009)\citenamefont {Baikov},
  \citenamefont {Chetyrkin}, \citenamefont {Smirnov}, \citenamefont {Smirnov},\
  and\ \citenamefont {Steinhauser}}]{Baikov:2009bg}%
  \BibitemOpen
  \bibfield  {author} {\bibinfo {author} {\bibfnamefont {P.~A.}\ \bibnamefont
  {Baikov}}, \bibinfo {author} {\bibfnamefont {K.~G.}\ \bibnamefont
  {Chetyrkin}}, \bibinfo {author} {\bibfnamefont {A.~V.}\ \bibnamefont
  {Smirnov}}, \bibinfo {author} {\bibfnamefont {V.~A.}\ \bibnamefont
  {Smirnov}}, \ and\ \bibinfo {author} {\bibfnamefont {M.}~\bibnamefont
  {Steinhauser}},\ }\href {\doibase 10.1103/PhysRevLett.102.212002} {\bibfield
  {journal} {\bibinfo  {journal} {Phys. Rev. Lett.}\ }\textbf {\bibinfo
  {volume} {102}},\ \bibinfo {pages} {212002} (\bibinfo {year} {2009})},\
  \Eprint {http://arxiv.org/abs/0902.3519} {arXiv:0902.3519 [hep-ph]}
  \BibitemShut {NoStop}%
\bibitem [{\citenamefont {Lee}\ \emph {et~al.}(2010)\citenamefont {Lee},
  \citenamefont {Smirnov},\ and\ \citenamefont {Smirnov}}]{Lee:2010cga}%
  \BibitemOpen
  \bibfield  {author} {\bibinfo {author} {\bibfnamefont {R.~N.}\ \bibnamefont
  {Lee}}, \bibinfo {author} {\bibfnamefont {A.~V.}\ \bibnamefont {Smirnov}}, \
  and\ \bibinfo {author} {\bibfnamefont {V.~A.}\ \bibnamefont {Smirnov}},\
  }\href {\doibase 10.1007/JHEP04(2010)020} {\bibfield  {journal} {\bibinfo
  {journal} {JHEP}\ }\textbf {\bibinfo {volume} {04}},\ \bibinfo {pages} {020}
  (\bibinfo {year} {2010})},\ \Eprint {http://arxiv.org/abs/1001.2887}
  {arXiv:1001.2887 [hep-ph]} \BibitemShut {NoStop}%
\bibitem [{\citenamefont {Gehrmann}\ \emph {et~al.}(2010)\citenamefont
  {Gehrmann}, \citenamefont {Glover}, \citenamefont {Huber}, \citenamefont
  {Ikizlerli},\ and\ \citenamefont {Studerus}}]{Gehrmann:2010ue}%
  \BibitemOpen
  \bibfield  {author} {\bibinfo {author} {\bibfnamefont {T.}~\bibnamefont
  {Gehrmann}}, \bibinfo {author} {\bibfnamefont {E.~W.~N.}\ \bibnamefont
  {Glover}}, \bibinfo {author} {\bibfnamefont {T.}~\bibnamefont {Huber}},
  \bibinfo {author} {\bibfnamefont {N.}~\bibnamefont {Ikizlerli}}, \ and\
  \bibinfo {author} {\bibfnamefont {C.}~\bibnamefont {Studerus}},\ }\href
  {\doibase 10.1007/JHEP06(2010)094} {\bibfield  {journal} {\bibinfo  {journal}
  {JHEP}\ }\textbf {\bibinfo {volume} {06}},\ \bibinfo {pages} {094} (\bibinfo
  {year} {2010})},\ \Eprint {http://arxiv.org/abs/1004.3653} {arXiv:1004.3653
  [hep-ph]} \BibitemShut {NoStop}%
\bibitem [{\citenamefont {Chien}\ and\ \citenamefont
  {Schwartz}(2010)}]{Chien:2010kc}%
  \BibitemOpen
  \bibfield  {author} {\bibinfo {author} {\bibfnamefont {Y.-T.}\ \bibnamefont
  {Chien}}\ and\ \bibinfo {author} {\bibfnamefont {M.~D.}\ \bibnamefont
  {Schwartz}},\ }\href {\doibase 10.1007/JHEP08(2010)058} {\bibfield  {journal}
  {\bibinfo  {journal} {JHEP}\ }\textbf {\bibinfo {volume} {08}},\ \bibinfo
  {pages} {058} (\bibinfo {year} {2010})},\ \Eprint
  {http://arxiv.org/abs/1005.1644} {arXiv:1005.1644 [hep-ph]} \BibitemShut
  {NoStop}%
\bibitem [{\citenamefont {Kelley}\ \emph {et~al.}(2011)\citenamefont {Kelley},
  \citenamefont {Schwartz}, \citenamefont {Schabinger},\ and\ \citenamefont
  {Zhu}}]{Kelley:2011ng}%
  \BibitemOpen
  \bibfield  {author} {\bibinfo {author} {\bibfnamefont {R.}~\bibnamefont
  {Kelley}}, \bibinfo {author} {\bibfnamefont {M.~D.}\ \bibnamefont
  {Schwartz}}, \bibinfo {author} {\bibfnamefont {R.~M.}\ \bibnamefont
  {Schabinger}}, \ and\ \bibinfo {author} {\bibfnamefont {H.~X.}\ \bibnamefont
  {Zhu}},\ }\href {\doibase 10.1103/PhysRevD.84.045022} {\bibfield  {journal}
  {\bibinfo  {journal} {Phys. Rev.}\ }\textbf {\bibinfo {volume} {D84}},\
  \bibinfo {pages} {045022} (\bibinfo {year} {2011})},\ \Eprint
  {http://arxiv.org/abs/1105.3676} {arXiv:1105.3676 [hep-ph]} \BibitemShut
  {NoStop}%
\bibitem [{\citenamefont {Monni}\ \emph {et~al.}(2011)\citenamefont {Monni},
  \citenamefont {Gehrmann},\ and\ \citenamefont {Luisoni}}]{Monni:2011gb}%
  \BibitemOpen
  \bibfield  {author} {\bibinfo {author} {\bibfnamefont {P.~F.}\ \bibnamefont
  {Monni}}, \bibinfo {author} {\bibfnamefont {T.}~\bibnamefont {Gehrmann}}, \
  and\ \bibinfo {author} {\bibfnamefont {G.}~\bibnamefont {Luisoni}},\ }\href
  {\doibase 10.1007/JHEP08(2011)010} {\bibfield  {journal} {\bibinfo  {journal}
  {JHEP}\ }\textbf {\bibinfo {volume} {08}},\ \bibinfo {pages} {010} (\bibinfo
  {year} {2011})},\ \Eprint {http://arxiv.org/abs/1105.4560} {arXiv:1105.4560
  [hep-ph]} \BibitemShut {NoStop}%
\bibitem [{\citenamefont {Del~Duca}\ \emph
  {et~al.}(2016{\natexlab{a}})\citenamefont {Del~Duca}, \citenamefont {Duhr},
  \citenamefont {Kardos}, \citenamefont {Somogyi},\ and\ \citenamefont
  {Tr\'ocs\'anyi}}]{DelDuca:2016csb}%
  \BibitemOpen
  \bibfield  {author} {\bibinfo {author} {\bibfnamefont {V.}~\bibnamefont
  {Del~Duca}}, \bibinfo {author} {\bibfnamefont {C.}~\bibnamefont {Duhr}},
  \bibinfo {author} {\bibfnamefont {A.}~\bibnamefont {Kardos}}, \bibinfo
  {author} {\bibfnamefont {G.}~\bibnamefont {Somogyi}}, \ and\ \bibinfo
  {author} {\bibfnamefont {Z.}~\bibnamefont {Tr\'ocs\'anyi}},\ }\href {\doibase
  10.1103/PhysRevLett.117.152004} {\bibfield  {journal} {\bibinfo  {journal}
  {Phys. Rev. Lett.}\ }\textbf {\bibinfo {volume} {117}},\ \bibinfo {pages}
  {152004} (\bibinfo {year} {2016}{\natexlab{a}})},\ \Eprint
  {http://arxiv.org/abs/1603.08927} {arXiv:1603.08927 [hep-ph]} \BibitemShut
  {NoStop}%
\bibitem [{\citenamefont {Del~Duca}\ \emph
  {et~al.}(2016{\natexlab{b}})\citenamefont {Del~Duca}, \citenamefont {Duhr},
  \citenamefont {Kardos}, \citenamefont {Somogyi}, \citenamefont {Sz\H{o}r},
  \citenamefont {Tr\'ocs\'anyi},\ and\ \citenamefont
  {Tulip\'ant}}]{DelDuca:2016ily}%
  \BibitemOpen
  \bibfield  {author} {\bibinfo {author} {\bibfnamefont {V.}~\bibnamefont
  {Del~Duca}}, \bibinfo {author} {\bibfnamefont {C.}~\bibnamefont {Duhr}},
  \bibinfo {author} {\bibfnamefont {A.}~\bibnamefont {Kardos}}, \bibinfo
  {author} {\bibfnamefont {G.}~\bibnamefont {Somogyi}}, \bibinfo {author}
  {\bibfnamefont {Z.}~\bibnamefont {Sz\H{o}r}}, \bibinfo {author}
  {\bibfnamefont {Z.}~\bibnamefont {Tr\'ocs\'anyi}}, \ and\ \bibinfo {author}
  {\bibfnamefont {Z.}~\bibnamefont {Tulip\'ant}},\ }\href {\doibase
  10.1103/PhysRevD.94.074019} {\bibfield  {journal} {\bibinfo  {journal} {Phys.
  Rev.}\ }\textbf {\bibinfo {volume} {D94}},\ \bibinfo {pages} {074019}
  (\bibinfo {year} {2016}{\natexlab{b}})},\ \Eprint
  {http://arxiv.org/abs/1606.03453} {arXiv:1606.03453 [hep-ph]} \BibitemShut
  {NoStop}%
\bibitem [{\citenamefont {Tulip\'ant}\ \emph {et~al.}(2017)\citenamefont
  {Tulip\'ant}, \citenamefont {Kardos},\ and\ \citenamefont
  {Somogyi}}]{Tulipant:2017ybb}%
  \BibitemOpen
  \bibfield  {author} {\bibinfo {author} {\bibfnamefont {Z.}~\bibnamefont
  {Tulip\'ant}}, \bibinfo {author} {\bibfnamefont {A.}~\bibnamefont {Kardos}},
  \ and\ \bibinfo {author} {\bibfnamefont {G.}~\bibnamefont {Somogyi}},\ }\href
  {\doibase 10.1140/epjc/s10052-017-5320-9} {\bibfield  {journal} {\bibinfo
  {journal} {Eur. Phys. J.}\ }\textbf {\bibinfo {volume} {C77}},\ \bibinfo
  {pages} {749} (\bibinfo {year} {2017})},\ \Eprint
  {http://arxiv.org/abs/1708.04093} {arXiv:1708.04093 [hep-ph]} \BibitemShut
  {NoStop}%
\bibitem [{\citenamefont {Kardos}\ \emph {et~al.}(2019)\citenamefont {Kardos},
  \citenamefont {Larkoski},\ and\ \citenamefont
  {Tr\'ocs\'anyi}}]{Kardos:2019iwa}%
  \BibitemOpen
  \bibfield  {author} {\bibinfo {author} {\bibfnamefont {A.}~\bibnamefont
  {Kardos}}, \bibinfo {author} {\bibfnamefont {A.}~\bibnamefont {Larkoski}}, \
  and\ \bibinfo {author} {\bibfnamefont {Z.}~\bibnamefont {Tr\'ocs\'anyi}},\
  }\bibfield  {booktitle} {\emph {\bibinfo {booktitle} {{Proceedings, 43rd
  International Conference of Theoretical Physics: Matter to the Deepest,
  Recent Developments In Physics Of Fundamental Interactions (MTTD2019):
  Chorzów/Katowice, Katowice, Poland, September 1-6, 2019}}},\ }\href
  {\doibase 10.5506/APhysPolB.50.1891} {\bibfield  {journal} {\bibinfo
  {journal} {Acta Phys. Polon.}\ }\textbf {\bibinfo {volume} {B50}},\ \bibinfo
  {pages} {1891} (\bibinfo {year} {2019})}\BibitemShut {NoStop}%
\bibitem [{\citenamefont {Caswell}(1974)}]{Caswell:1974gg}%
  \BibitemOpen
  \bibfield  {author} {\bibinfo {author} {\bibfnamefont {W.~E.}\ \bibnamefont
  {Caswell}},\ }\href {\doibase 10.1103/PhysRevLett.33.244} {\bibfield
  {journal} {\bibinfo  {journal} {Phys. Rev. Lett.}\ }\textbf {\bibinfo
  {volume} {33}},\ \bibinfo {pages} {244} (\bibinfo {year} {1974})}\BibitemShut
  {NoStop}%
\bibitem [{\citenamefont {Jones}(1974)}]{Jones:1974mm}%
  \BibitemOpen
  \bibfield  {author} {\bibinfo {author} {\bibfnamefont {D.~R.~T.}\
  \bibnamefont {Jones}},\ }\href {\doibase 10.1016/0550-3213(74)90093-5}
  {\bibfield  {journal} {\bibinfo  {journal} {Nucl. Phys.}\ }\textbf {\bibinfo
  {volume} {B75}},\ \bibinfo {pages} {531} (\bibinfo {year}
  {1974})}\BibitemShut {NoStop}%
\bibitem [{\citenamefont {Egorian}\ and\ \citenamefont
  {Tarasov}(1979)}]{Egorian:1978zx}%
  \BibitemOpen
  \bibfield  {author} {\bibinfo {author} {\bibfnamefont {E.}~\bibnamefont
  {Egorian}}\ and\ \bibinfo {author} {\bibfnamefont {O.~V.}\ \bibnamefont
  {Tarasov}},\ }\href@noop {} {\bibfield  {journal} {\bibinfo  {journal} {Teor.
  Mat. Fiz.}\ }\textbf {\bibinfo {volume} {41}},\ \bibinfo {pages} {26}
  (\bibinfo {year} {1979})},\ \bibinfo {note} {[Theor. Math.
  Phys.41,863(1979)]}\BibitemShut {NoStop}%
\bibitem [{\citenamefont {Korchemsky}\ and\ \citenamefont
  {Radyushkin}(1987)}]{Korchemsky:1987wg}%
  \BibitemOpen
  \bibfield  {author} {\bibinfo {author} {\bibfnamefont {G.~P.}\ \bibnamefont
  {Korchemsky}}\ and\ \bibinfo {author} {\bibfnamefont {A.~V.}\ \bibnamefont
  {Radyushkin}},\ }\href {\doibase 10.1016/0550-3213(87)90277-X} {\bibfield
  {journal} {\bibinfo  {journal} {Nucl. Phys.}\ }\textbf {\bibinfo {volume}
  {B283}},\ \bibinfo {pages} {342} (\bibinfo {year} {1987})}\BibitemShut
  {NoStop}%
\bibitem [{\citenamefont {Vogt}(2001)}]{Vogt:2000ci}%
  \BibitemOpen
  \bibfield  {author} {\bibinfo {author} {\bibfnamefont {A.}~\bibnamefont
  {Vogt}},\ }\href {\doibase 10.1016/S0370-2693(00)01344-7} {\bibfield
  {journal} {\bibinfo  {journal} {Phys. Lett.}\ }\textbf {\bibinfo {volume}
  {B497}},\ \bibinfo {pages} {228} (\bibinfo {year} {2001})},\ \Eprint
  {http://arxiv.org/abs/hep-ph/0010146} {arXiv:hep-ph/0010146 [hep-ph]}
  \BibitemShut {NoStop}%
\bibitem [{\citenamefont {Berger}(2002)}]{Berger:2002sv}%
  \BibitemOpen
  \bibfield  {author} {\bibinfo {author} {\bibfnamefont {C.~F.}\ \bibnamefont
  {Berger}},\ }\href {\doibase 10.1103/PhysRevD.66.116002} {\bibfield
  {journal} {\bibinfo  {journal} {Phys. Rev.}\ }\textbf {\bibinfo {volume}
  {D66}},\ \bibinfo {pages} {116002} (\bibinfo {year} {2002})},\ \Eprint
  {http://arxiv.org/abs/hep-ph/0209107} {arXiv:hep-ph/0209107 [hep-ph]}
  \BibitemShut {NoStop}%
\bibitem [{\citenamefont {Moch}\ \emph
  {et~al.}(2005{\natexlab{b}})\citenamefont {Moch}, \citenamefont
  {Vermaseren},\ and\ \citenamefont {Vogt}}]{Moch:2005tm}%
  \BibitemOpen
  \bibfield  {author} {\bibinfo {author} {\bibfnamefont {S.}~\bibnamefont
  {Moch}}, \bibinfo {author} {\bibfnamefont {J.~A.~M.}\ \bibnamefont
  {Vermaseren}}, \ and\ \bibinfo {author} {\bibfnamefont {A.}~\bibnamefont
  {Vogt}},\ }\href {\doibase 10.1016/j.physletb.2005.08.067} {\bibfield
  {journal} {\bibinfo  {journal} {Phys. Lett.}\ }\textbf {\bibinfo {volume}
  {B625}},\ \bibinfo {pages} {245} (\bibinfo {year} {2005}{\natexlab{b}})},\
  \Eprint {http://arxiv.org/abs/hep-ph/0508055} {arXiv:hep-ph/0508055 [hep-ph]}
  \BibitemShut {NoStop}%
\bibitem [{\citenamefont {van Neerven}(1986)}]{vanNeerven:1985xr}%
  \BibitemOpen
  \bibfield  {author} {\bibinfo {author} {\bibfnamefont {W.~L.}\ \bibnamefont
  {van Neerven}},\ }\href {\doibase 10.1016/0550-3213(86)90165-3} {\bibfield
  {journal} {\bibinfo  {journal} {Nucl. Phys.}\ }\textbf {\bibinfo {volume}
  {B268}},\ \bibinfo {pages} {453} (\bibinfo {year} {1986})}\BibitemShut
  {NoStop}%
\bibitem [{\citenamefont {Matsuura}\ \emph {et~al.}(1989)\citenamefont
  {Matsuura}, \citenamefont {van~der Marck},\ and\ \citenamefont {van
  Neerven}}]{Matsuura:1988sm}%
  \BibitemOpen
  \bibfield  {author} {\bibinfo {author} {\bibfnamefont {T.}~\bibnamefont
  {Matsuura}}, \bibinfo {author} {\bibfnamefont {S.~C.}\ \bibnamefont {van~der
  Marck}}, \ and\ \bibinfo {author} {\bibfnamefont {W.~L.}\ \bibnamefont {van
  Neerven}},\ }\href {\doibase 10.1016/0550-3213(89)90620-2} {\bibfield
  {journal} {\bibinfo  {journal} {Nucl. Phys.}\ }\textbf {\bibinfo {volume}
  {B319}},\ \bibinfo {pages} {570} (\bibinfo {year} {1989})}\BibitemShut
  {NoStop}%
\end{thebibliography}%

\end{document}